\documentclass[structabstract]{aa}  
\usepackage{graphicx}
\usepackage{float}
\usepackage{subfig}
\usepackage{txfonts}
\usepackage{calc}

\begin{document}
   \title{On the nature of magnetic cycles at different ages of stars}

   \author{K. Ol\'ah\inst{1}, Zs. K\H ov\'ari\inst{1}, K. Petrovay,\inst{2}, W. Soon \inst{3},  S. Baliunas \inst{4 }, Z. Koll\'ath\inst{5}, \and 
               K. Vida\inst{1}
                          %\fnmsep\thanks{Just to show the usage of the elements in the author field}
          }

   \institute{Konkoly Observatory, Research Centre for Astronomy and
Earth Sciences, Hungarian Academy of Sciences, H-1121 Budapest, Konkoly Thege M. u. 15-17, Hungary\\
              \email{olah@konkoly.hu}
         \and
             E\"otv\"os University, Department of Astronomy, Pf. 32, H-1518, Budapest, Hungary
         \and
             Harvard-Smithsonian Center for Astrophysics, Cambridge, MA 02138, USA
         \and   
             No Affiliation
         \and
             University of West Hungary, Savaria Campus, Institute of Mathematics and Physics, Hungary
             }

   \date{Received ..., 2016; accepted ...}

  \abstract
   {}
   %aims heading (mandatory)
   {We study the different patterns of interannual magnetic variability in stars on or near the lower main sequence, approximately solar-type (G-K dwarf) stars in time series of 36 years from the Mount Wilson Observatory Ca\,{\sc ii}\,H\&K survey. Our main aim is to search for correlations between cycles, activity measures and ages.}
   {Time-frequency analysis has been used to discern and reveal patterns and morphology of stellar activity cycles, including multiple and changing cycles, in the datasets. Both the results from short-term Fourier transform and its refinement using the Choi-Williams distribution, with better frequency resolution, are presented in this study. Rotational periods of the stars were derived using multi-frequency Fourier analysis.}
   {From the studied 29 stars we found at least one activity cycle on 28 of them. Twelve stars, with longer rotational periods ($39.7\pm6.0$ days) have simple, smooth cycles, and the rest of the stars, with on-average much faster rotation ($18.1\pm12.2$ days) show complex and sometimes vigorously changing, multiple cycles. The cycles are longer and quite uniform in the first group ($9.7\pm1.9$ years), while are generally shorter and with greater variety in the second one ($7.6\pm4.9$). There is a clear age division between stars with smooth and complex cycles that follows the known separation between the older and younger stars at around 2 to 3~Gyr of age.}
  % conclusions heading (optional) 
  {}
  
   \keywords{Stars: activity, starspots, Stars: late-type, solar-type, Sun: activity}
   \authorrunning{Ol\'ah et al.}
   \maketitle
%
%________________________________________________________________

\section{Introduction}\label{intro}

Magnetically active stars including the Sun show multiple and variable cycles on different timescales with varying amplitudes  (see  Ol\'ah et al. \cite{olahetal}, and references therein) thereby reinforcing the paradigm of change rather than constancy for solar and stellar activity outputs. In the case of the Sun, cycles with different lengths, from the Rieger-type cycles  (i.e., with length or period less than 1 yr; see e.g., McIntosh et al. \cite{mcintosh}), through cycles from 1~yr to 22~yr including the Schwabe sunspot cycle, to the Gleissberg to deVries-Suess cycles (i.e., timescales of several decades to two centuries) are often
studied separately to explore their possible origins. With advances from the study of the production of cosmogenic isotopes and resulting, indirect solar activity proxies in Earth archives, even cycles of millennial to bi-millennial
timescales have been detected and researched (e.g., Soon et al. \cite{soon1}) enriching models of 
dynamics of solar magnetic activity variations. The interrelations among those magnetic oscillations and characteristic timescales are a subject worthy of further investigation.
The several decades long observational datasets available for stars displaying magnetic activity in most cases are too short to derive long cycles (similar to the solar Gleissberg cycle) and barely enough for finding those of several years, like the solar Schwabe cycle of 9 to 13 years.  The existence of multiple cycles of stars has long been noticed and discussed, but no special attention was paid to them owing to limitation from both observational and theoretical points of view.  

It is empirically known that the lengths of magnetic cycles are related to the stellar rotation rates: stars with slower rotation exhibit longer cycles, noticed first by Baliunas et al. (\cite{baliunas1}).  Surface magnetic activity and rotation are strongly related (cf. Reiners et al. \cite{reiners}, and references therein) and as a star evolves, its rotation slows, changing the characteristics and operation of the presumed underlying dynamo as well. This is why
 the pattern of long-term variation of active stars should reflect the working of the dynamo inside the stars. 
Such a relation has already been suggested as early as 1981 by Durney et al. (\cite{durney}), based on results of the Mount
Wilson Ca\,{\sc ii}\,H\&K survey (Wilson \cite{wilson}) which had yielded time series for only about a decade at that time. The continued datasets from the Mount Wilson Observatory Ca\,{\sc ii}\,H\&K project survey, now three times as long, allow us to study and re-examine the suggestion of Durney et al. (\cite{durney}) by using the observed patterns of the activity cycles in addition to adopting newly available basic parameters like temperatures and ages of the stars.  Our primary aim in this paper is to study if the separation between young and old chromospherically active stars, first elaborated by Soon, Baliunas and Zhang (1993a), and later revisited by Brandenburg et al.  (\cite{axel}), shows up also in the pattern or morphology of activity cycles. 

The structure of our paper is as follows: in Sect. 2. we describe the data and the time-frequency methods, in Sect. 3. we present results of the time-frequency analysis, in Sect. 4. we discuss the results, and in Sect. 5. we summarize the study.

\section{Applied data and method}\label{obs_meth}

We take advantage of the full, 36 years long Mount Wilson  (MW afterwards) Ca\,{\sc ii}\,H\&K survey (Wilson \cite{wilson}, Wilson \cite{wilson1}, Baliunas et al. \cite{baliunas}) dataset for 29 stars of spectral type G--K in order to deduce surface rotational modulation periods and activity cycles on an about 10-yr longer time-base than previously available. Wilson originally chose the sample as single (i.e., not close binaries, although several wide binary pairs were included) stars on or near the lower main sequence based on parameters then known in the mid-1960s. Modern refinements to those parameters for stars in the sample are discussed below. 

Additionally, we studied the newest version of the sunspot number dataset published by WDC--SILSO, Royal Observatory of Belgium, Brussels. The sunspot data were binned to 0.025 years ($\approx$9~days). For comparison and verification purposes, we made use of the disk-integrated solar Ca\,{\sc ii}\,K-line flux measurements from the Evans Coronal Facility at The National Solar Observatory/Sacramento Peak observed between 1976--2015 (http://lasp.colorado.edu/lisird/tss/cak.html). The data were binned (and if necessary, interpolated) to 30 days.

To study activity cycles in the sample of 29 stars from the MW survey, we used the same time-frequency analysis as in Ol\'ah et al. (\cite{olahetal}). Briefly, we use Short-Term Fourier Transform (STFT) which gives acceptable resolution both in time and in frequency domains.  The observational noise and the used spline interpolation does not alter the frequency distribution as was discussed in detail by Koll\'ath \& Ol\'ah (\cite{kollath_olah}, Sec.~4.2 and 4.3), using Monte-Carlo simulations.
We also ran the analysis using Choi-Williams distribution (CWD) which has higher frequency resolution, at the cost of the time resolution or cross terms, in order to double-check the previous cycle length determinations. In case of the Sun, we applied different amplifications for the amplitudes in different frequency intervals for better visibility of the results. Details about the time-frequency analysis program are found in Koll\'ath \& Ol\'ah (\cite{kollath_olah}). We extracted rotational periods from the data with the multiple-frequency analysis tool \emph{MuFrAn} (Koll\'ath \cite{mufran}, Csubry \& Koll\'ath \cite{mufran1}). The shorter cycles on interannual timescale are generally not always detectable in stellar activity data records owing to irregular and rather limited seasonal time sampling. 

We first discuss shorter-term cycles on the Sun which co-exist together with the Schwabe cycle \emph {simultaneously}, and may add insight to the physical manifestations of the solar dynamo operation as well as add confidence in studying and interpreting such shorter term activity oscillations should they be also found in stellar activity records of the MW H\&K survey.

\section{Results}

%-------------------------------- Fig. 1.
   \begin{figure*}[t!]
   \centering
   \includegraphics[width=5.5cm]{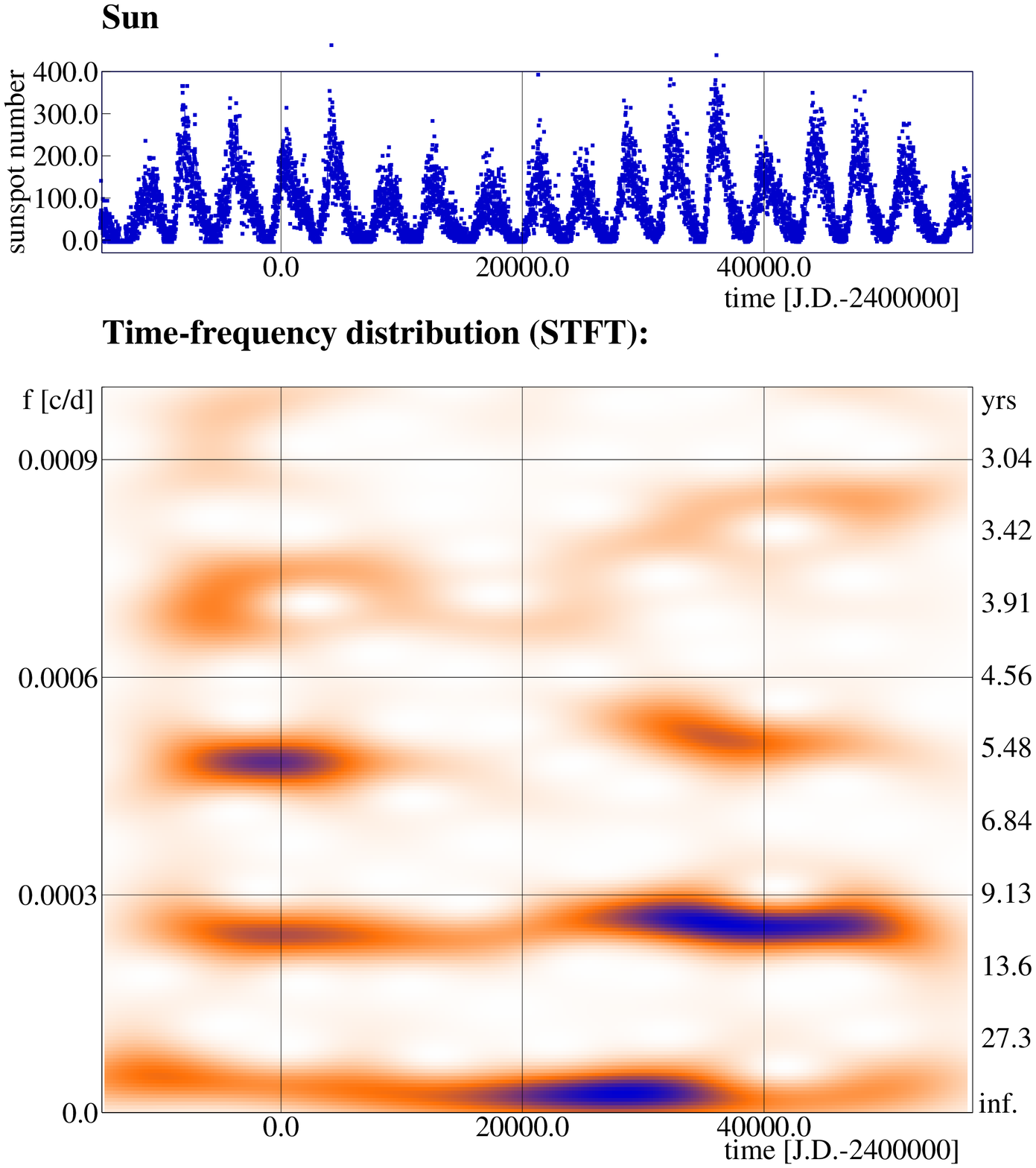}\includegraphics[width=5.5cm]{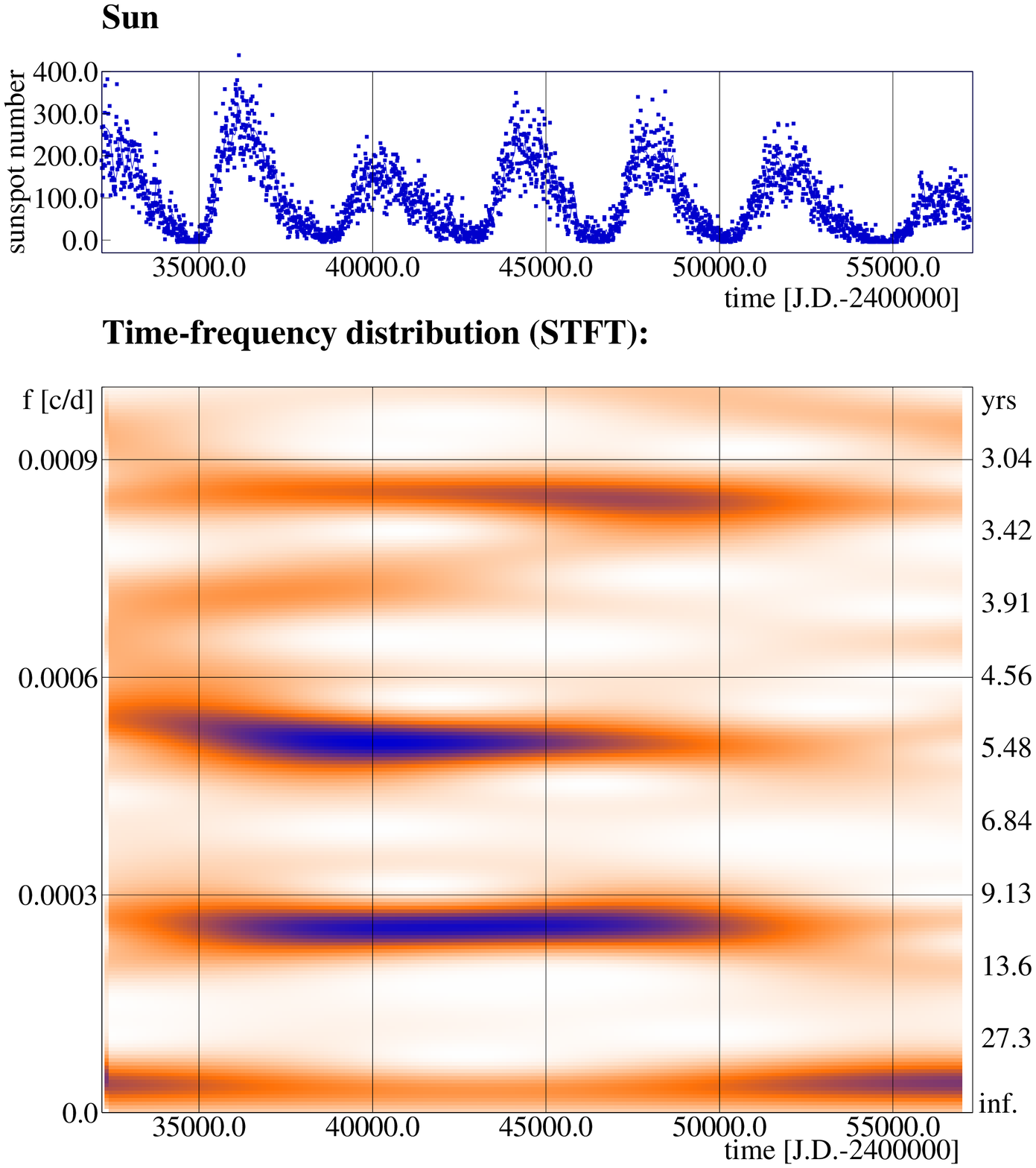}\includegraphics[width=5.5cm]{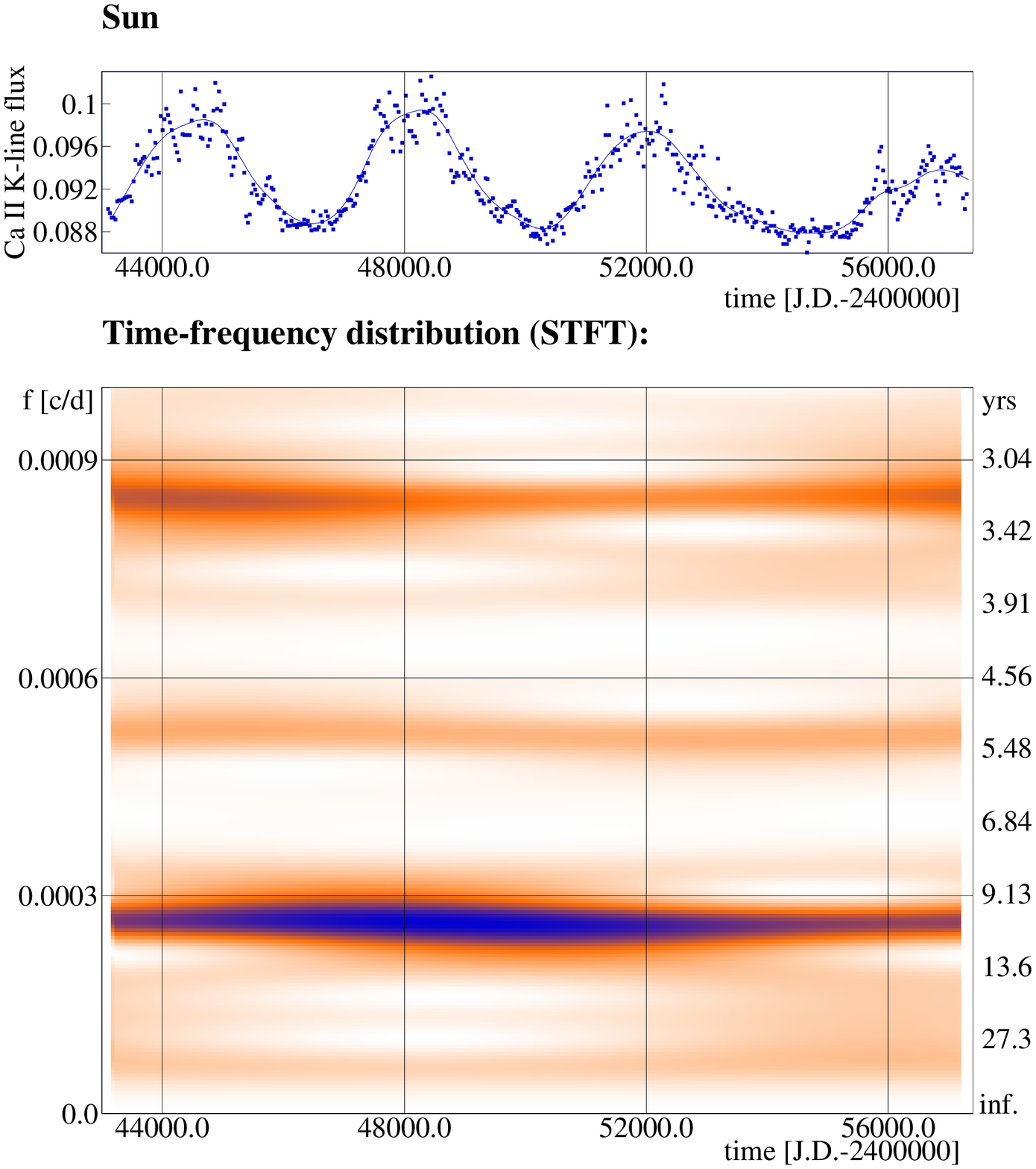}   
   \caption {Solar data (\emph {upper panels}) and STFT calculations (\emph {lower panels}). \emph {Left:} Sunspot number data from 1818 to the present and STFT. The Gleissberg cycle is seen in the bottom of the figure; above that, the Schwabe (11-yr) cycle, and above that, its half and one-third harmonic cycles are also found. Additionally, and mixed with the one-third of the Schwabe cycle, another, changing, variation of 3--4 years is continuously present in the data. \emph {Middle:} STFT of a shorter record of sunspot number (1947 to present) and \emph {right:} the same for the Ca\,{\sc ii}\,K flux of the Sun between 1976--2015. The 3--4 years long variability is well detected in both shorter datasets as well.}
         \label{qbofig}
         \end{figure*}
                 
We define "dominant cycles" of the stars as the most evident ones with high amplitude, which, at the same time, are repeated at least twice during the observational time window. High amplitude of the dominant cycle means the cycle with the highest amplitude during the observed time interval, except for some cases when the longest term trends and tendencies of the data records yield an even higher amplitude. The MW dataset are too short (36-yr) to define longer cycles like the solar Gleissberg and deVries-Suess cycles (i.e., 50--200 years), but the longest-term trends found in many stars could be parts of such long cycles. Many of the dominant cycles, however, are reminiscent of the solar Schwabe (11-yr) cycle, which is generally meant as ``solar cycle" in this paper.  

\subsection{Quasi-biennial oscillations of the Sun}\label{qbo}

Shorter cycles of the Sun are well known, ranging from the Rieger-type cycles on the timescales of less than 1 year to the so-called mid-term cycles or quasi-biennial oscillations (QBO) with typical lengths of 1--4 years. However, neither the Rieger-type cycles, nor the mid-term cycles, which are studied in detail concerning both their time-behavior and possible origin (cf. e.g., Forg\'acs-Dajka \& Borkovits \cite{emese_borko}), are \emph{continuously} present in the solar datasets. Bazilevskaya et al. (\cite{Bazil}) published an excellent review about the QBOs, and we suggest that paper and references therein, for further information. McIntosh et al. (\cite{mcintosh}) also recently added details of and insights into those characteristic variations and modeled them as overlapping, solar magnetic activity band interactions and instabilities.

We analyzed the most recent version of the sunspot number dataset between 1818--2015 (WDC-SILSO, Royal Observatory of Belgium, Brussels) with the same method as used for photometry and/or Ca-index data of stars (Ol\'ah et al. \cite{olahetal}). The result is presented in Fig.\ref{qbofig}. We see varying cycles on four different timescales in Fig.~\ref{qbofig}, left panel. Among those the longest one is the Gleissberg cycle; the next three are the Schwabe (11-yr) cycle and its half and one-third harmonic components. Mixed with the one-third component of the fundamental Schwabe cycle, an independent cycle is found, which is also variable, changing between 3 and 4 years. We only mention that cycle and consider it as a plausible upper-end limit of the QBO activity variations. However, it should be noted that Barnhart \& Eichinger (\cite{barn_eich_1}) using Empirical Mode Decomposition (EMD) combined with the Hilbert-transform found 3.4--3.6 year periods from sunspot numbers, although with high uncertainty of about 50\%, as the shortest mode from their analysis. Our independent result shows that a cyclic variability on the timescale of 3--4 years may be continuously present in solar variability manifested in sunspot number, and is not inconsistent with the recent interpretation e.g., by McIntosh et al. (2015).

The solar Ca\,{\sc ii}\,K-line flux dataset is similar to the data of the MW stars both in length (only about 3 years longer) and measure. Fig.~\ref{qbofig} (middle and right panels) shows the resulting time-frequency analysis for a shorter time-interval for the sunspot number and the solar Ca\,{\sc ii}\,K-line flux. The similarity is evident; the 3--4-yr cycle is also well detected in the Ca\,{\sc ii}\,K-line flux data and adds confidence to our interpretation.

\subsection{Cyclic changes on different time-scales of MW stars}\label{cycles}

\begin{table*}[t]
\caption{All detected cycles in the MW stars, first is the dominant cycle. In the last column the results from Baliunas et al. \cite{baliunas} are given for comparison.}             
\label{allcycles}      
\centering                          
\begin{tabular}{l l l l r l}       
\hline\hline \noalign{\smallskip} 
Star & Sp. type & P$_{rot}$ & P$_{cyc, (range)}$ & lt$^{\it c}$ &P$_{cyc}$, Baliunas et al. \cite{baliunas}\\
       &               & (days)      & (years) && (years) \\
\hline

\multicolumn{5}{c}{simple cycles}\\
\hline
HD 3651    & K0V$^{\it a}$  &  44$^{\it b}$      &      11.6(9.57--13.7)-lt & yes & 13.8$\pm$0.4\\
HD 4628    & K2.5V$^{\it a}$  & 41.6     &    8.94(8.30--9.5) & no &8.37$\pm$0.08\\
HD 10476  & K1V$^{\it a}$  & 33.7      &   9.8(9.01--9.85) & no &9.6$\pm$0.1\\
HD 16160   & K3V$^{\it a}$  & 57        &   12.1 & no &13.2$\pm$0.2 \\
HD 26965    & K1V$^{\it a}$  & 43$^{\it b}$ &   10.0(9.57--10.5)  & no &10.1$\pm$0.1\\
HD 32147   & K3+V$^{\it a}$  & 39.3      &   10.6(9.85--11.3), lt & yes &11.1$\pm$0.2\\
HD 81809   & G5V$^{\it a}$  & 39.3     &    8.69(8.10--9.28) & no &8.17$\pm$0.08\\
HD 103095  & K1V$^{\it a}$  & 36.5     &    6.95(6.9--7.0), lt & yes &7.30$\pm$0.08\\
HD 160346  & K2.5V$^{\it a}$   & 35.3       &  7.35(7.2--7.5), lt & yes &7.00$\pm$0.08\\      
HD 166620  & K2V$^{\it c}$  & 41.5       &  13.6(9.6--17.6) & no &15.8$\pm$0.3\\
HD 201091  & K5V  & 37.1      &   6.95(6.7--7.2), lt & yes &7.3$\pm$0.1\\
HD 219834B & K2V  & 34.0    &     9.29(9.01--9.57) & no &10.0$\pm$0.2\\
\noalign{\smallskip}
Sun & G2V & 27.275 & 11(9--14), 3.65(3.3--4.0)$^{\it d}$ & yes &\\
\hline

\multicolumn{5}{c}{complicated cycles}\\
\hline
HD 1835    & G2V$^{\it a}$  & 7.84     &   7.6(7.3--7.9), 2.4(2.50--2.28),  3.97(4.85--3.09), lt & yes &9.1$\pm$0.3\\
HD 20630   &G5V  & 9.08      &   5.32(5.36--5.27) & no &5.6$\pm$0.1\\
HD 76151   & G3V$^{\it a}$  & 15.2     &    5.32(6.07--4.56), lt & yes &2.52$\pm$0.02\\
HD 78366   & G0V  & 9.7       &   13.45(12.6--14.3)-lt,  4.0(3.85--4.15) & yes &$12.2\pm$0.4,  5.9$\pm$0.1 \\
HD 95735   & M2V  & 54        &   3.90(3.75-4.04), 12.7(10.3-15.0)-lt & yes &\\
HD 100180  & F9.5V$^{\it a}$  & 14.6     &   13.2(16.6--9.85), 3.63  & no &3.56$\pm$0.04, 12.9$\pm$0.5\\
HD 114710  & G0V  & 12.9      &   16.6(17.6--13.7--18.6), 7.7(7.2--8.2), 5.25(5.1--5.4) & no &16.6$\pm$0.6, 9.6$\pm$0.3 \\
HD 115404  & K2.5V$^{\it a}$  & 18.8       &  10.8(9.57--12.1), 5.08(5.83--4.33), 3.4(3.96--2.84) lt & yes &12.4$\pm$0.4\\
HD 131156A & G8V  & 6.25      &   3.7(3.6--3.8), 12, lt &yes & \\
HD 131156B & K4V  & 11.05     &   4.2(4.64--3.76), 2.28, lt &yes & \\
HD 149661  & K0V$^{\it a}$  & 21.3      &   4.0, 6.2(5.6--6.8), 12.25(11.3--13.2)-lt & yes &17.4$\pm$0.7, 4.00$\pm$0.04 \\    
HD 152391  & G8+V$^{\it a}$  & 11.4      &   9.9(12.6--7.17), 2.8(2.5--3.1) & no &10.9$\pm$0.2\\
HD 156026  & K5V  & 29.2      &  4.33(3.95--4.71), 8.1, lt($\approx$20)& yes &21.0$\pm$0.9\\
HD 165341A & K0V$^{\it a}$  & 18.9     &  11.4(9--13.8), 5.17, lt & yes &5.1$\pm$0.1\\
HD 165341B & K6V  &  ---       &     --- && Var\\
HD 190406  & G0V$^{\it a}$  & 15.5      &   15.8(15.0--17.6)-lt, 8.35(7.69--9.01), 4.65(4.44--4.86), 2.33, 3.01 & yes &2.60$\pm$0.02, 16.9$\pm$0.8\\
HD 201092  & K7V  & 34.1     &    13.4(10.1--16.6)-lt, 4.45(4.71--4.18), (2.07:) & yes &11.7$\pm$0.4\\
\noalign{\smallskip}\hline                  
\end{tabular}
\tablefoot{ $^{\it a}$spectral types are from Gray et al. (\cite{gray1}, \cite{gray2}), otherwise from Baliunas et al. \cite{baliunas}, $^{\it b}$period form Barnes et al. (\cite{barnes}), $^{\it c}$lt: long-term trend in the data, $^{\it d}$determined in the present paper.} 
\end{table*}

We next analyzed Ca-index measurements from Mount Wilson Observatory for 29 stars. The time series of MW stars have about the same length of $\approx36$ years and serves as a source in studying relations among stellar rotation, cycles and other parameters. We found multiple cycles on the stars in most cases, of which several are also variable in time. Some of the cycles are only temporarily seen, while others are present during the entire observed timespan, similarly to the solar case. We did not use any special treatment to get the cycle lengths, which should be considered rather as characteristic timescales of certain physical mechanisms or processes and therefore are not strictly periodic. Most of these cycles and even their variabilities are well visible on the datasets themselves, and the numerical values were taken interactively from the time-frequency diagrams.

\begin{figure*}[tbp]
 \centering
\includegraphics[width=4.3cm]{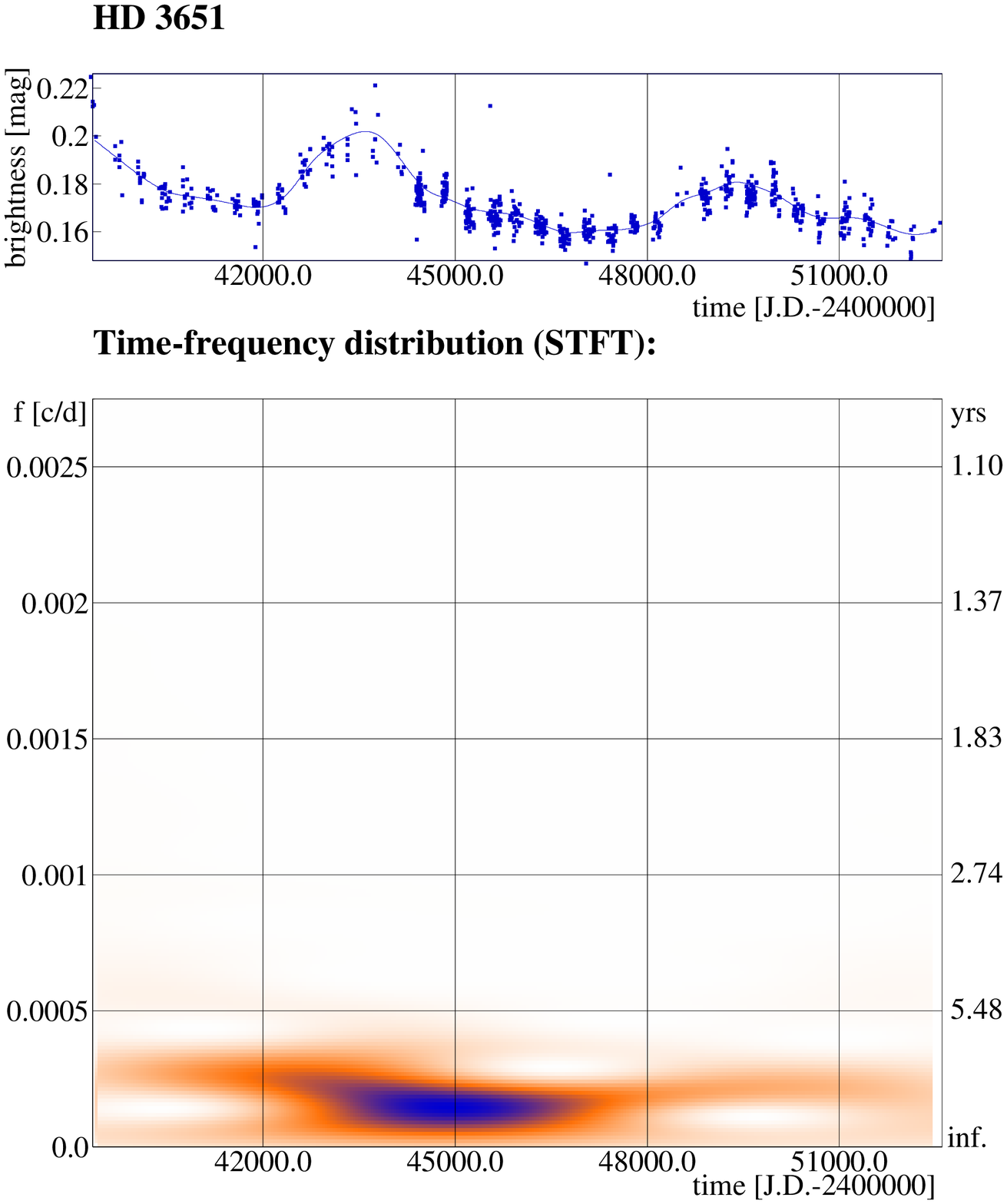}\includegraphics[width=4.3cm]{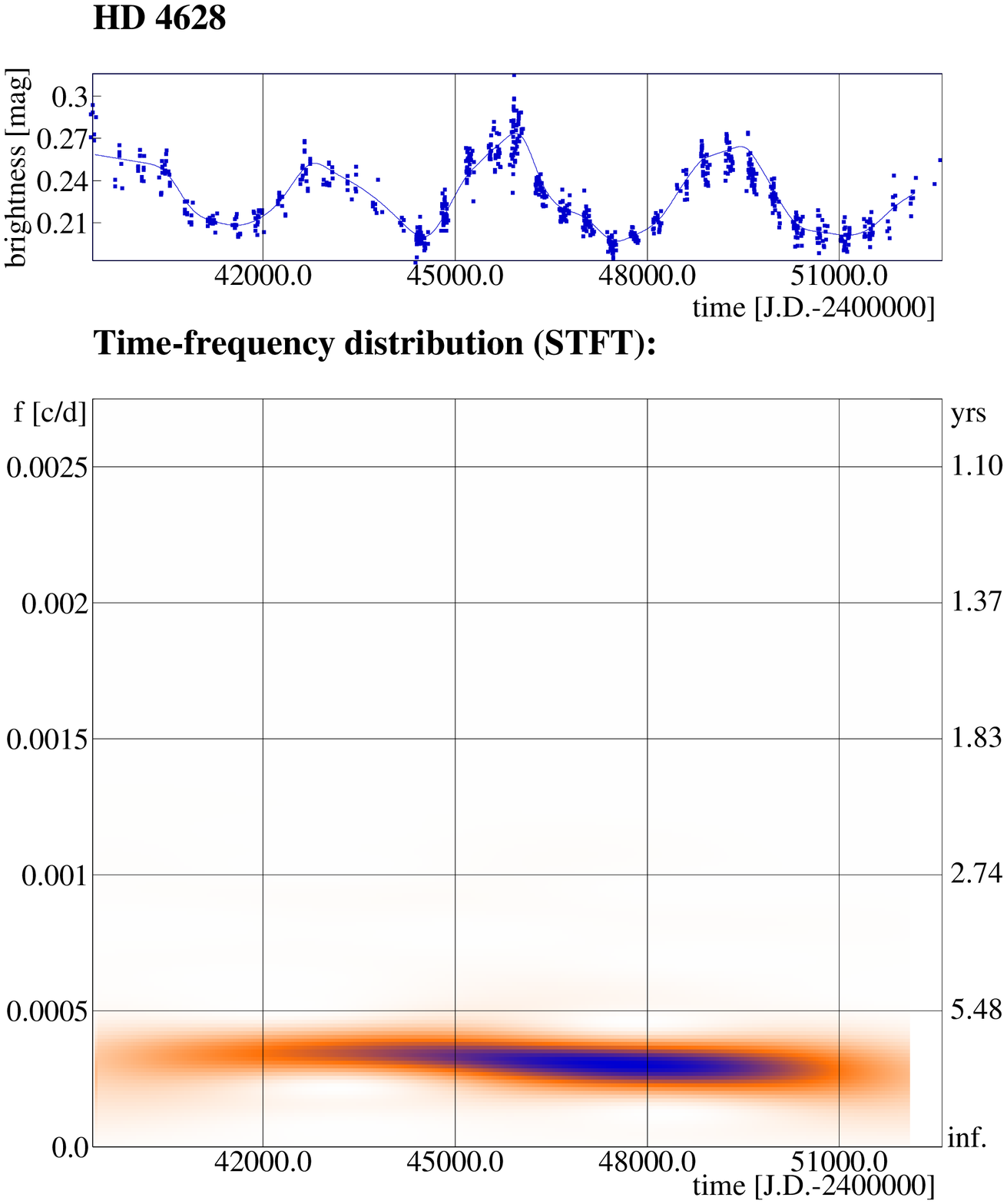}\includegraphics[width=4.3cm]{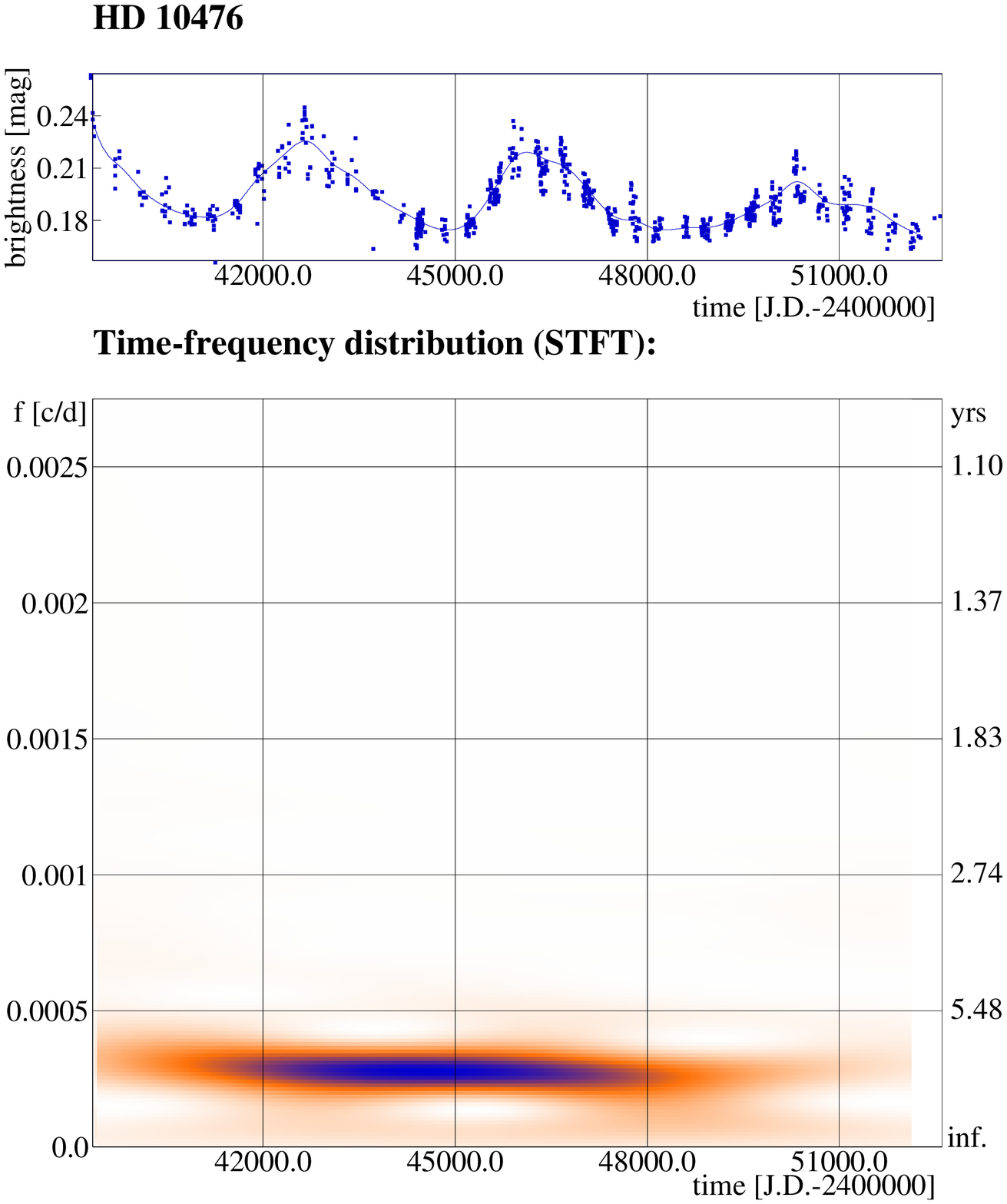}\includegraphics[width=4.3cm]{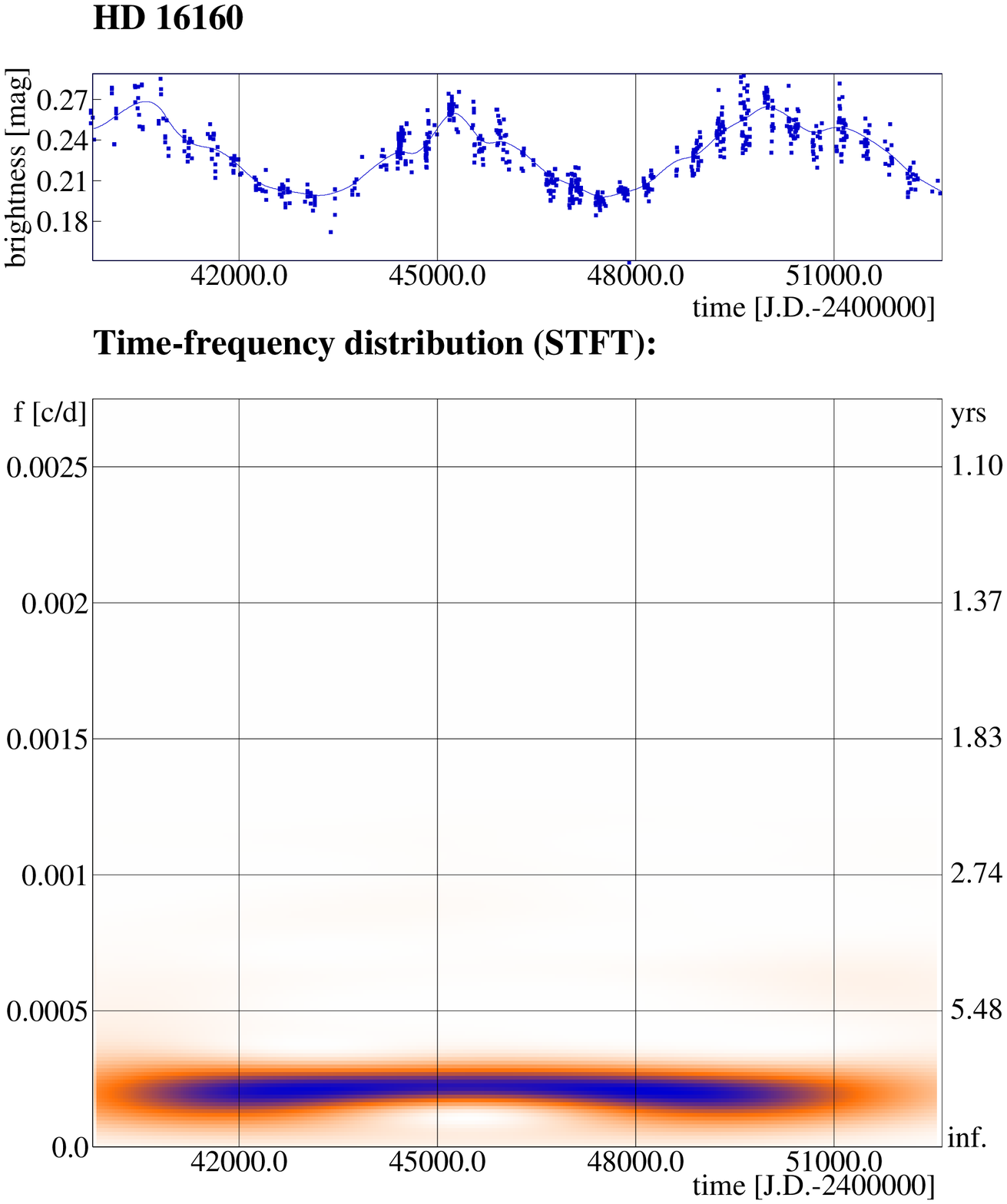}
\includegraphics[width=4.3cm]{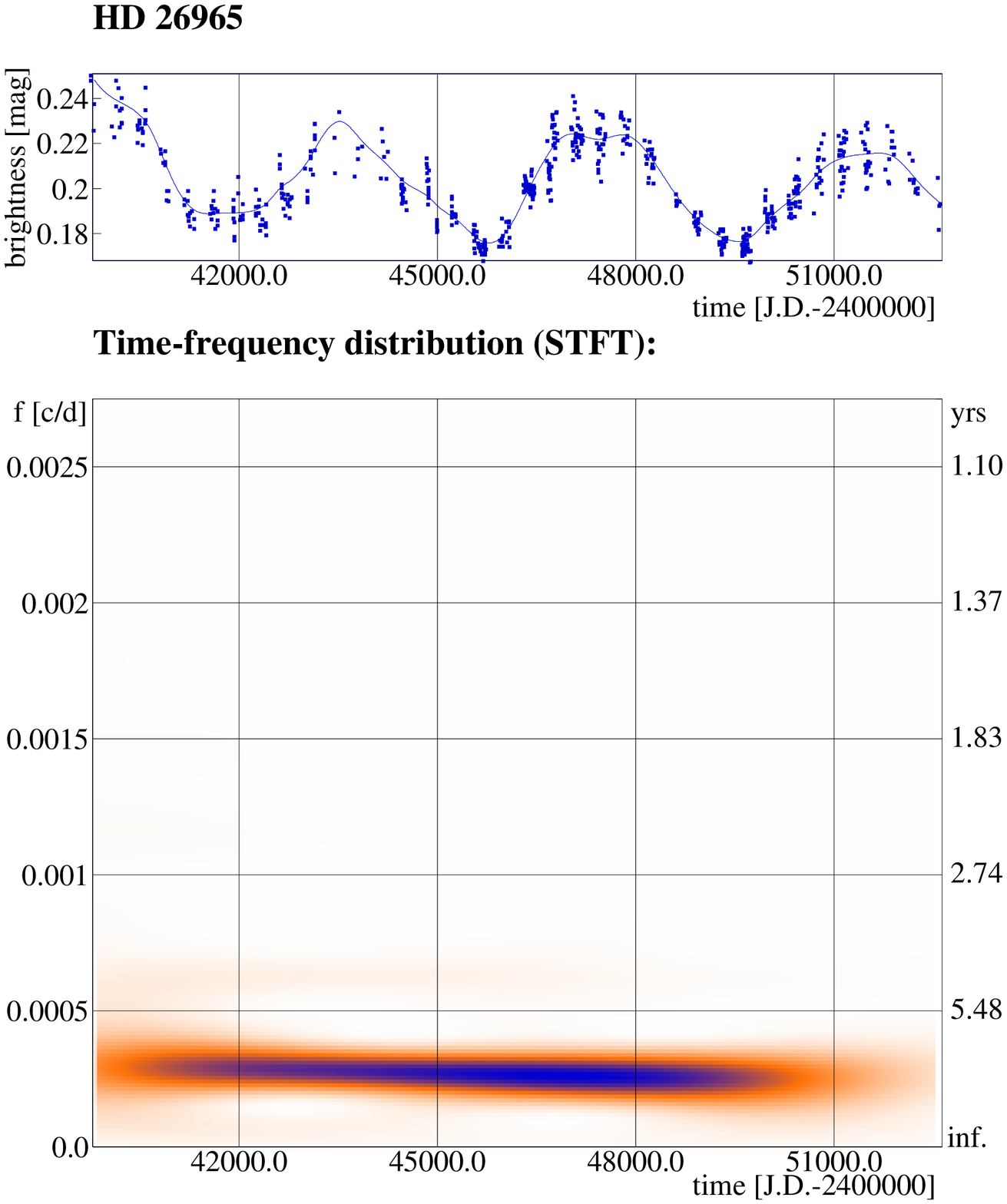}\includegraphics[width=4.3cm]{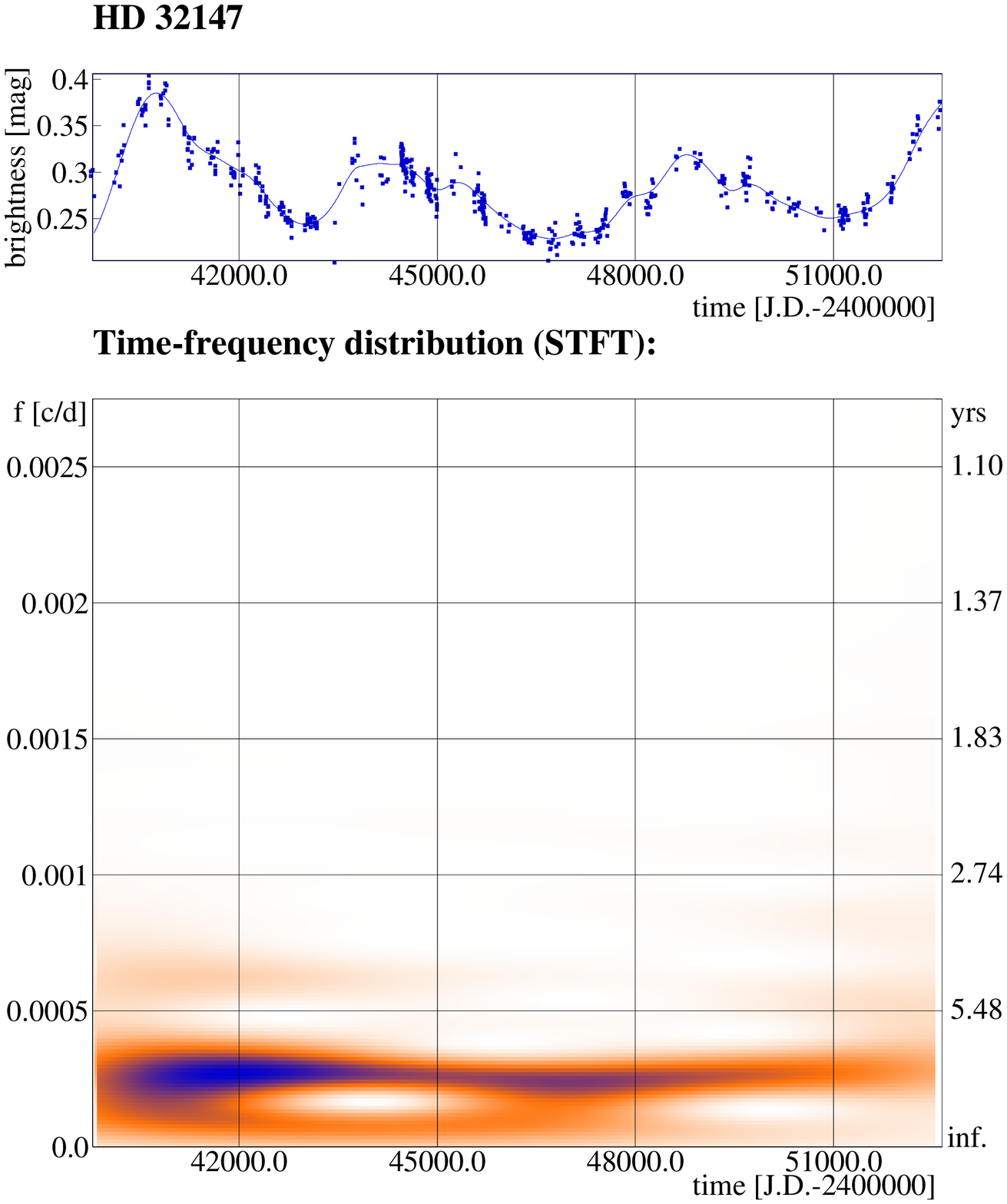}\includegraphics[width=4.3cm]{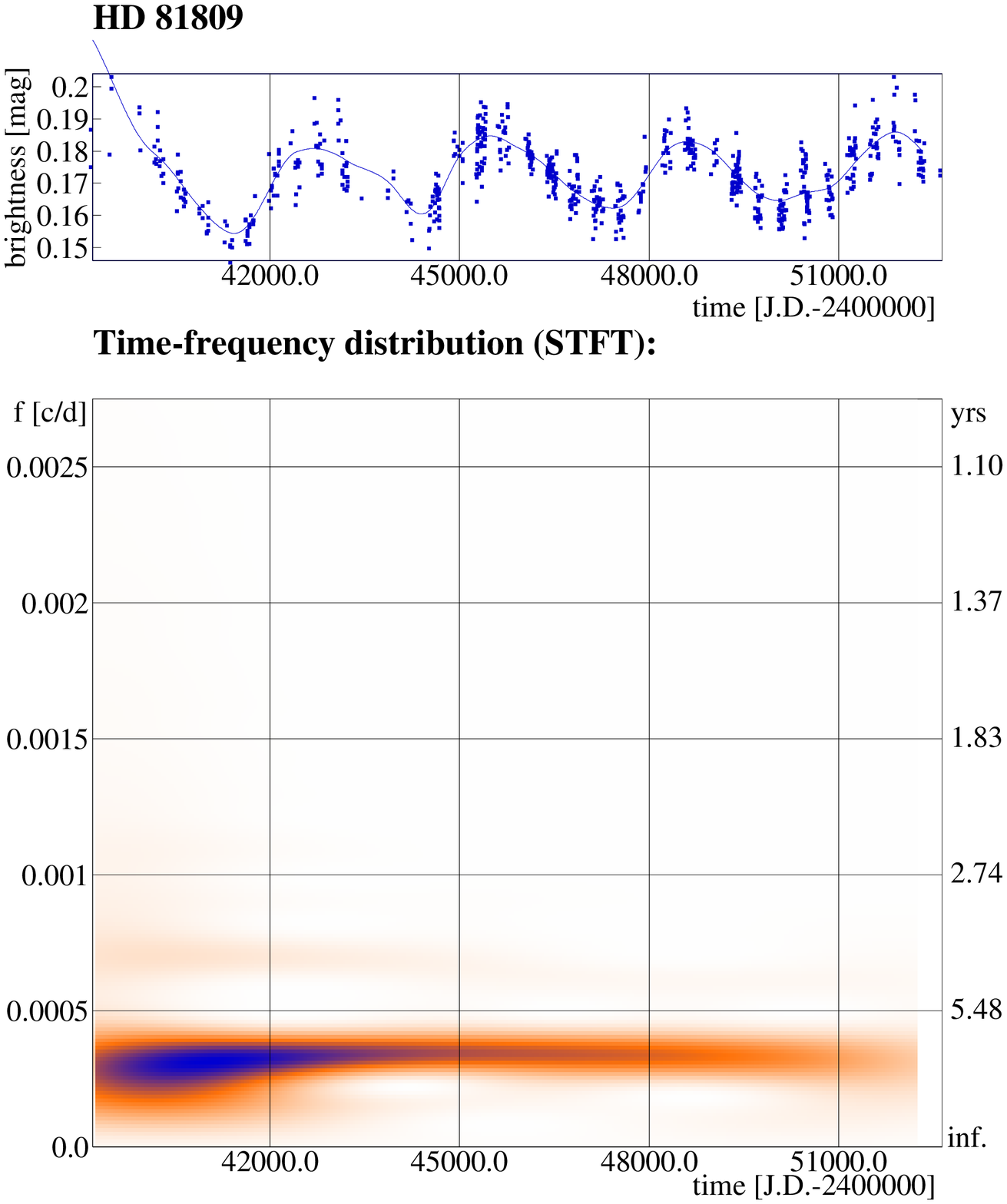}\includegraphics[width=4.3cm]{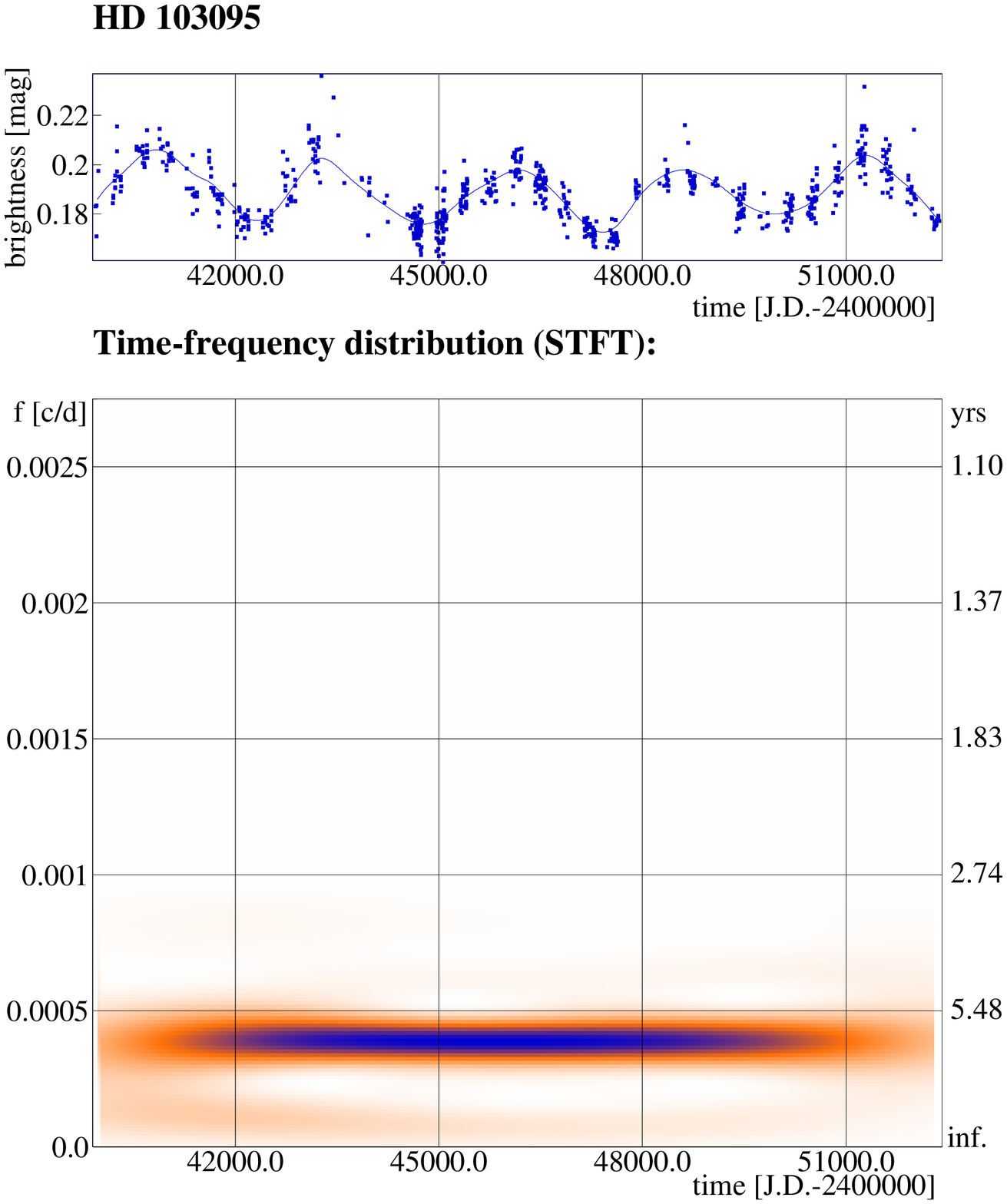}
\includegraphics[width=4.3cm]{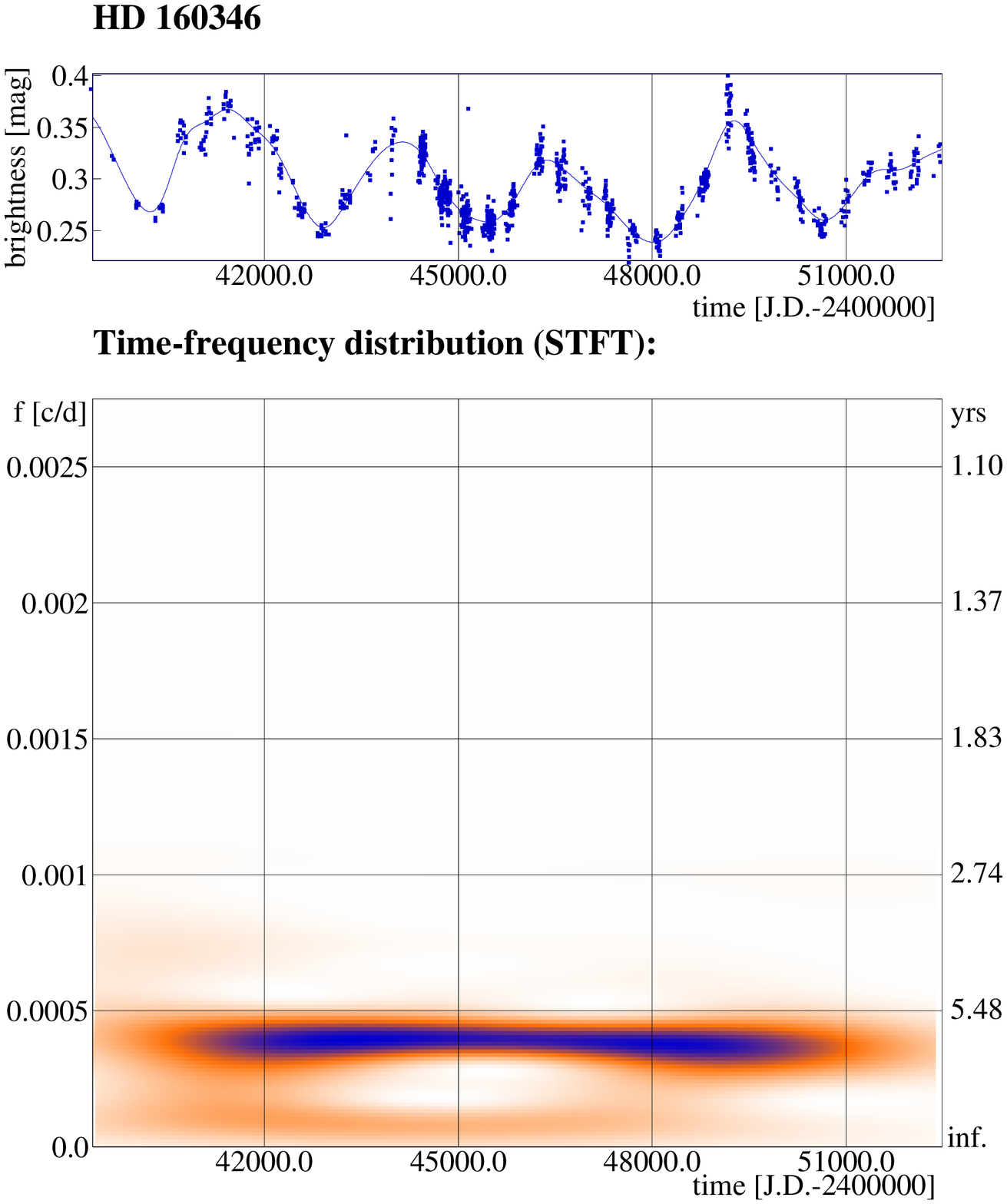}\includegraphics[width=4.3cm]{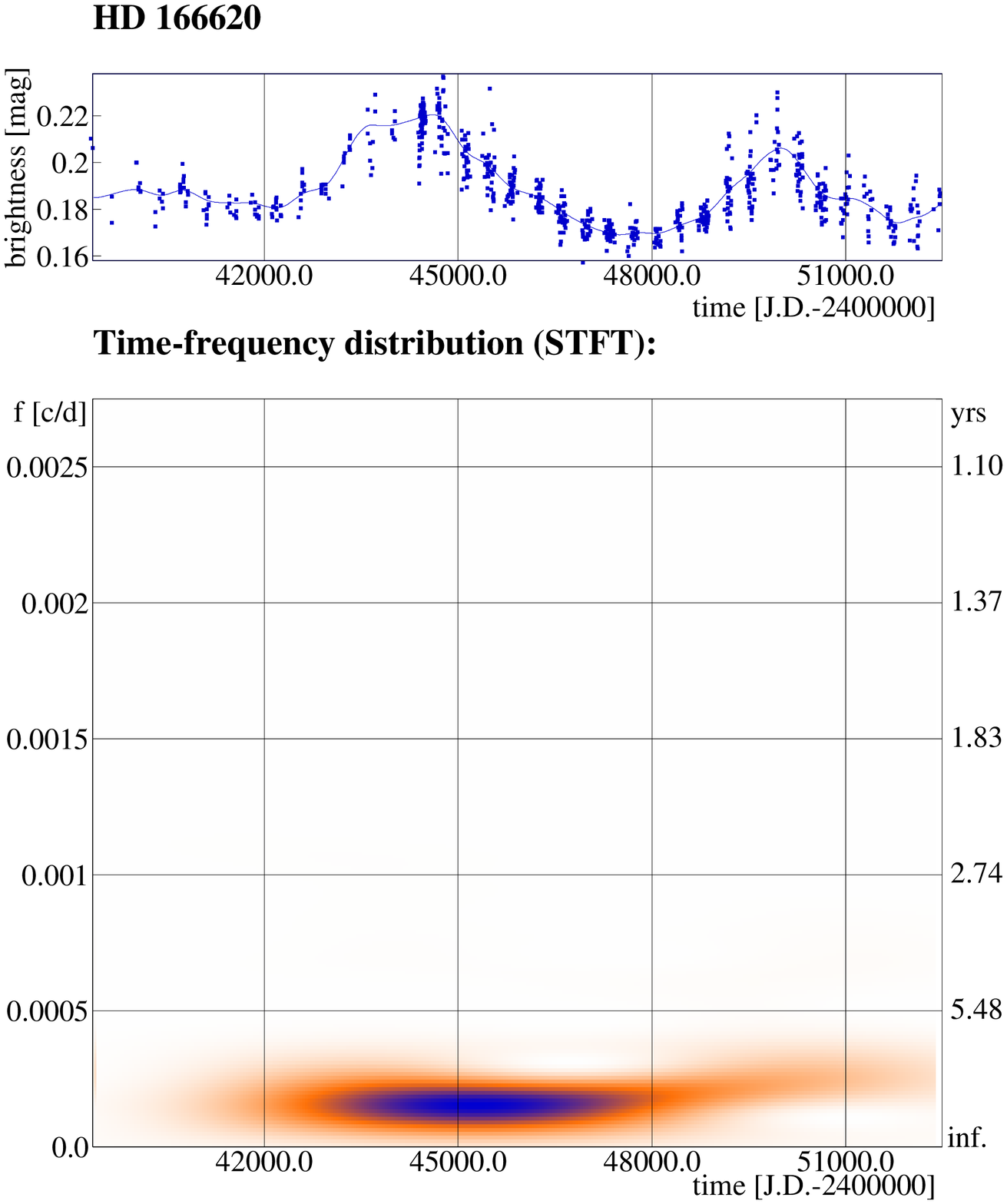}\includegraphics[width=4.3cm]{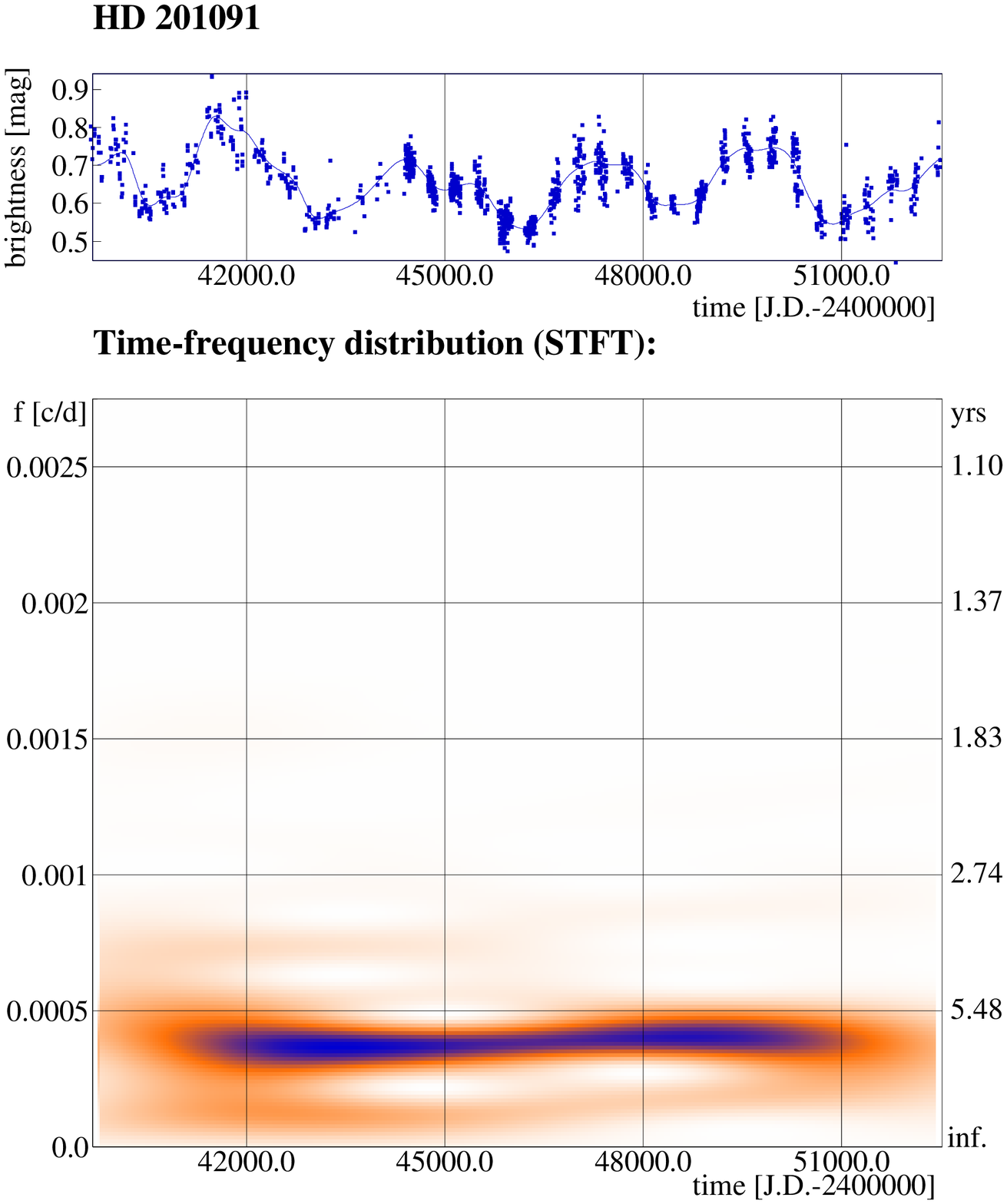}\includegraphics[width=4.3cm]{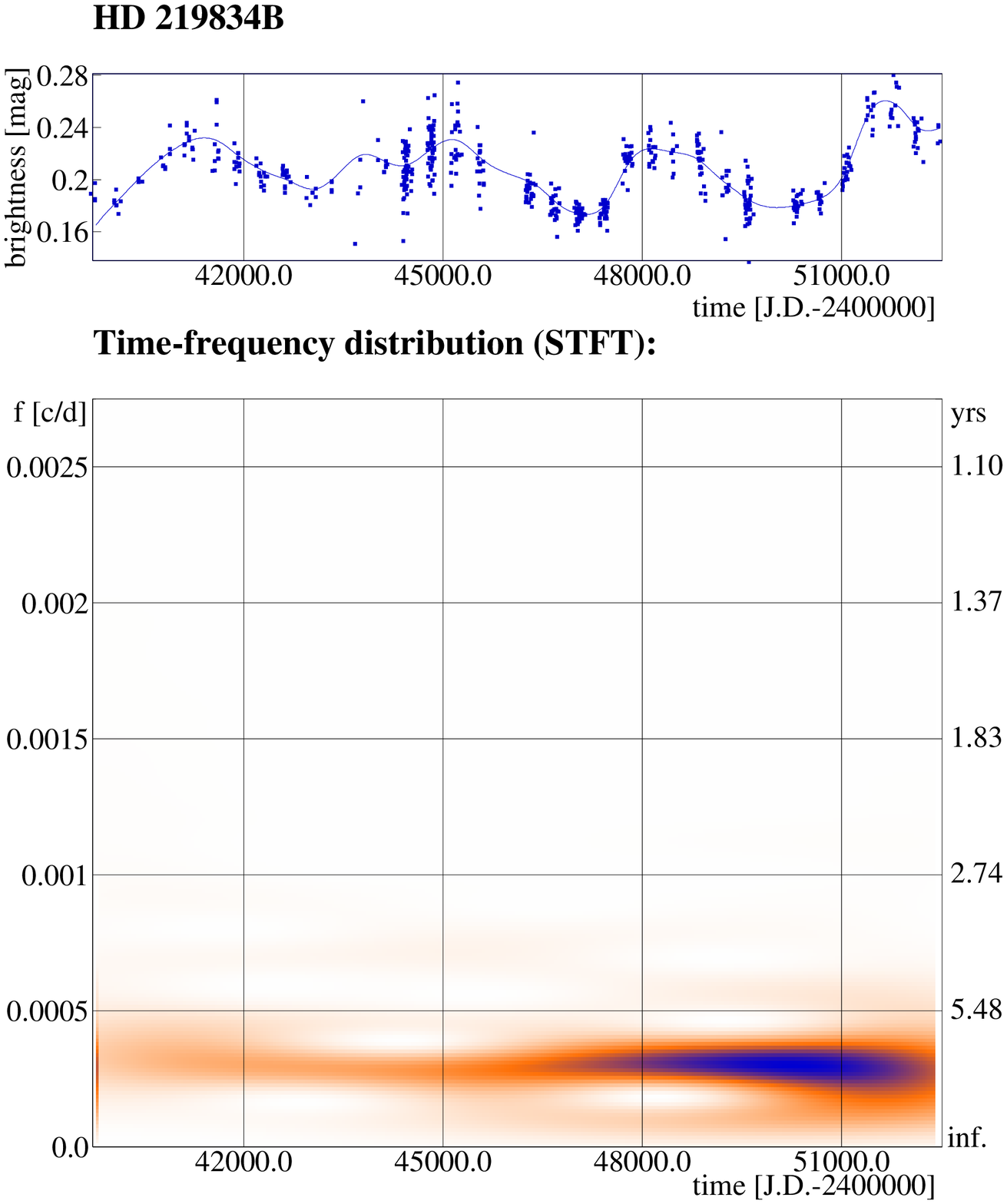}
\caption{STFT of  MW stars with simple cycles. In the upper panels, the datasets and spline interpolation we used are plotted; the lower panels show the time-frequency diagrams. We applied no amplification, so the strengths of the low amplitude secondary cycles can directly be compared. (However, in a few cases the low amplitude signals are too faint in the prints).}
 \label{stft_simple}
 \end{figure*}

\begin{figure*}[tbp]
\centering
 \includegraphics[width=4.3cm]{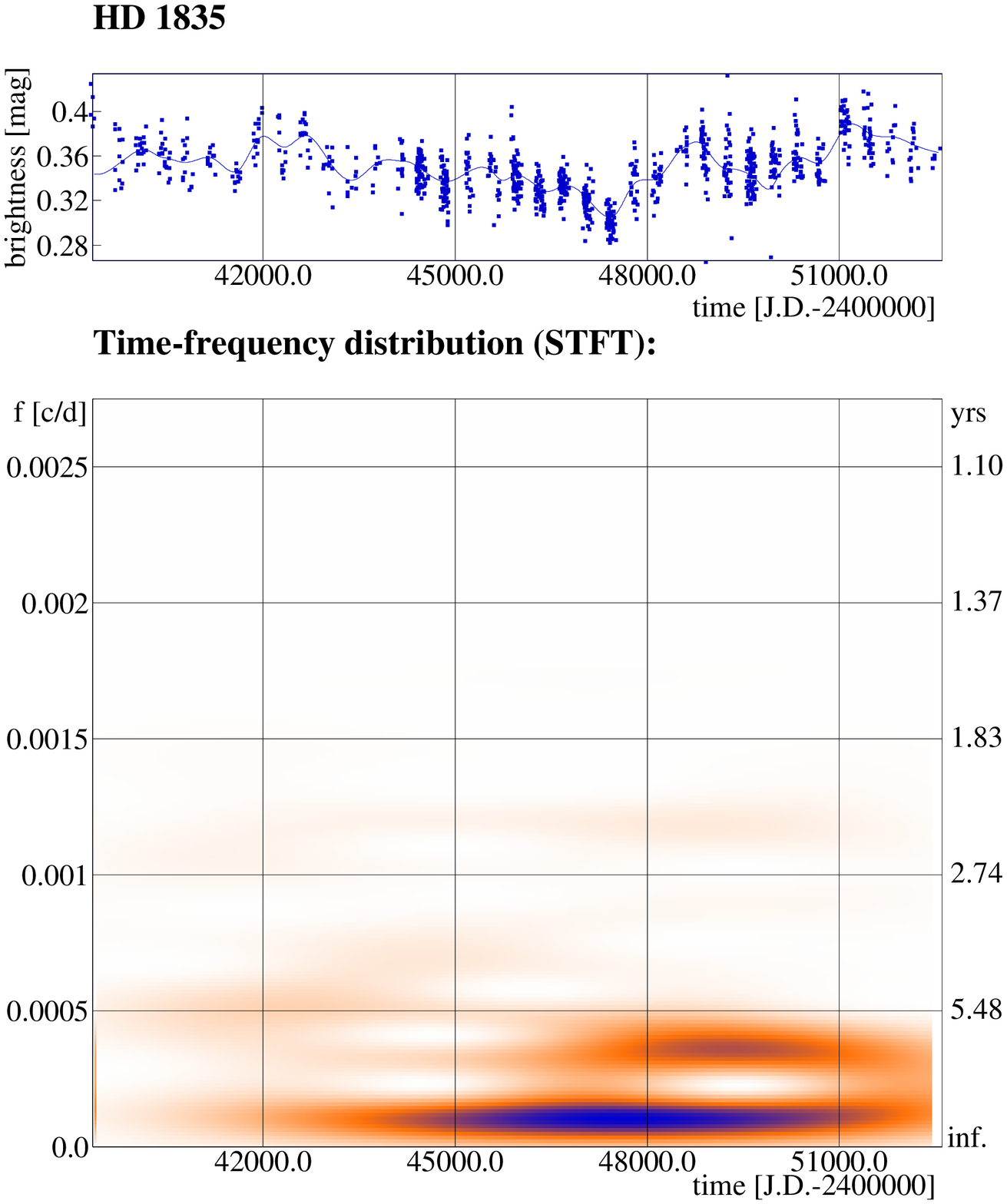}\includegraphics[width=4.3cm]{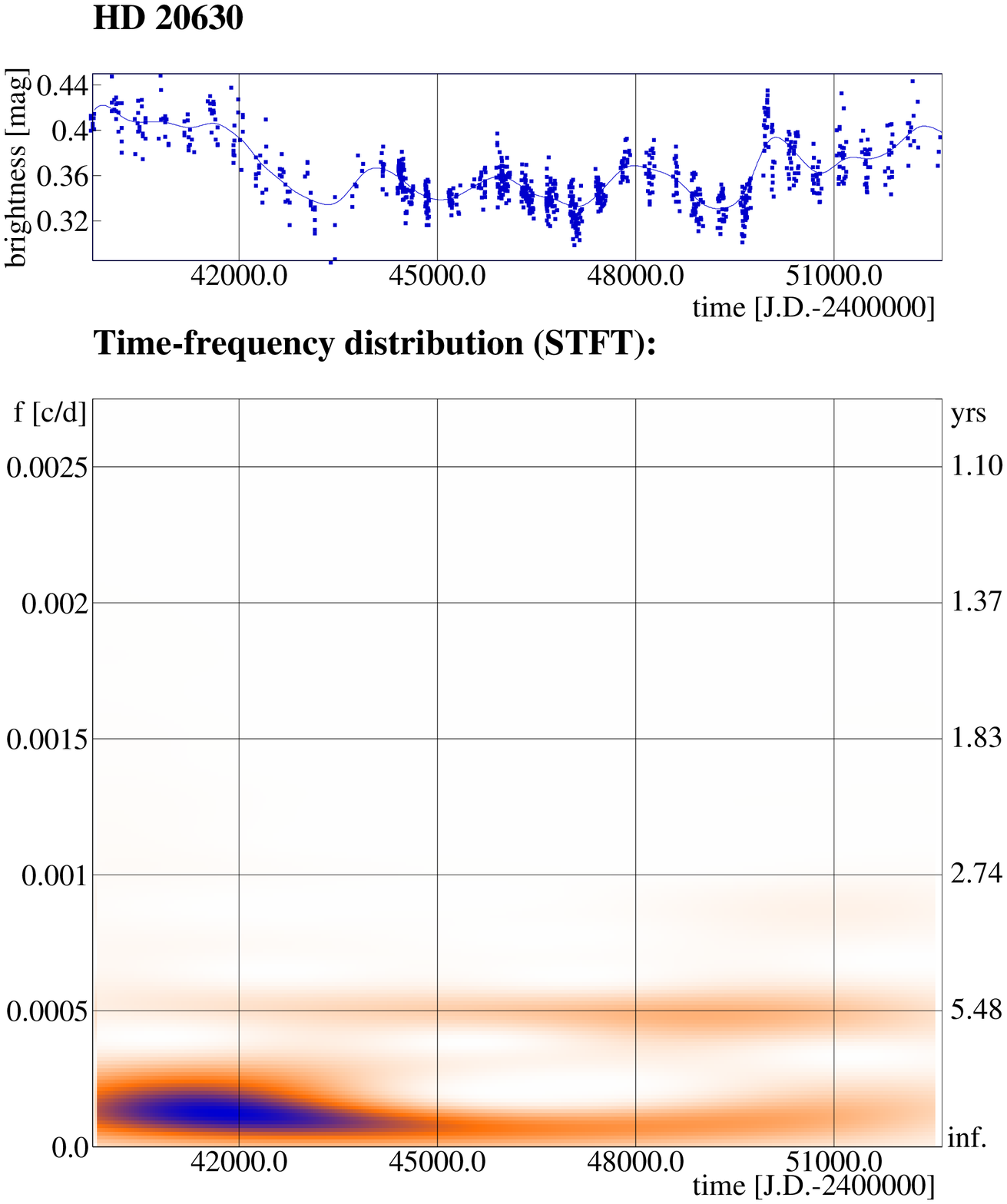}\includegraphics[width=4.3cm]{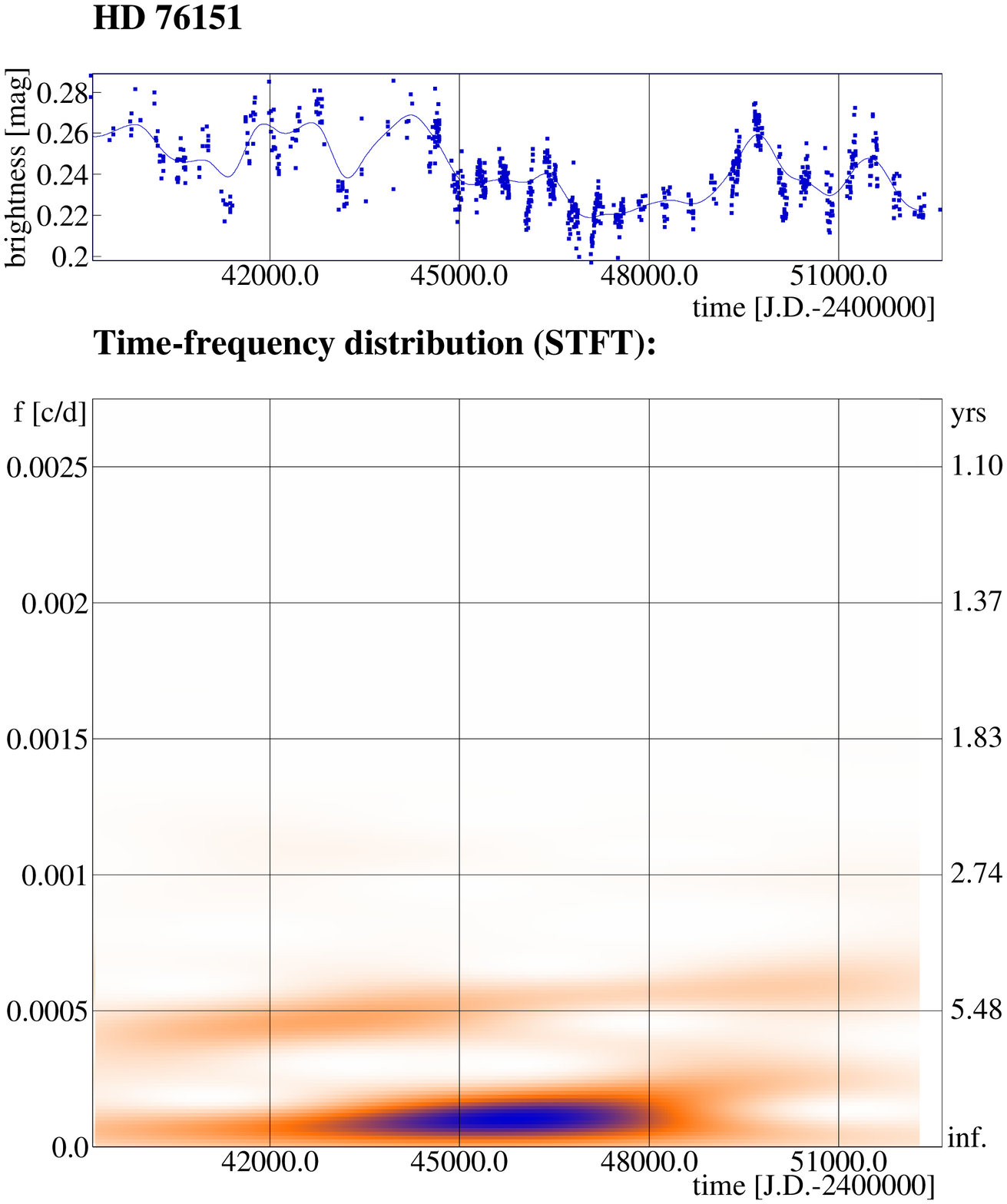}\includegraphics[width=4.3cm]{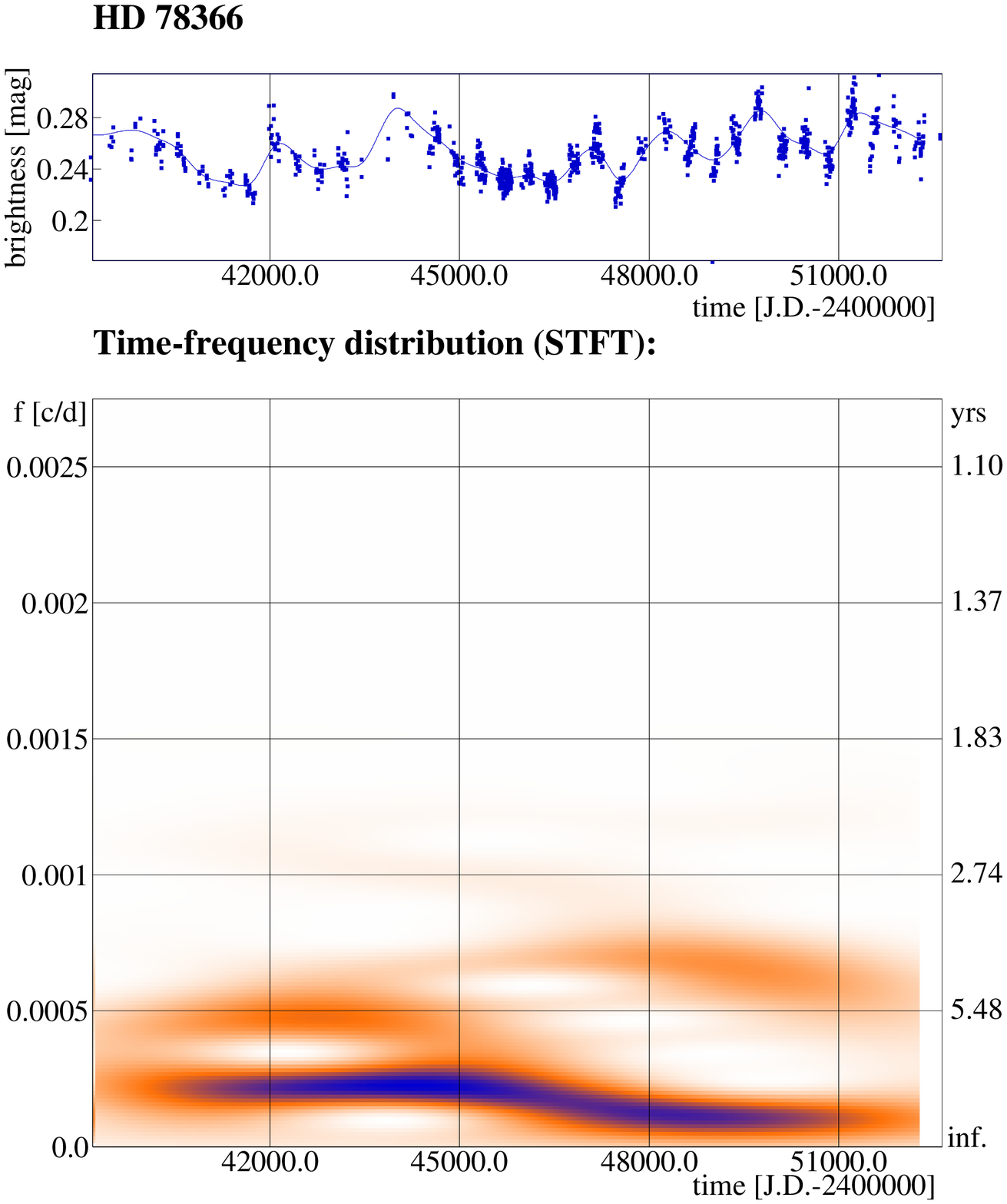}
\includegraphics[width=4.3cm]{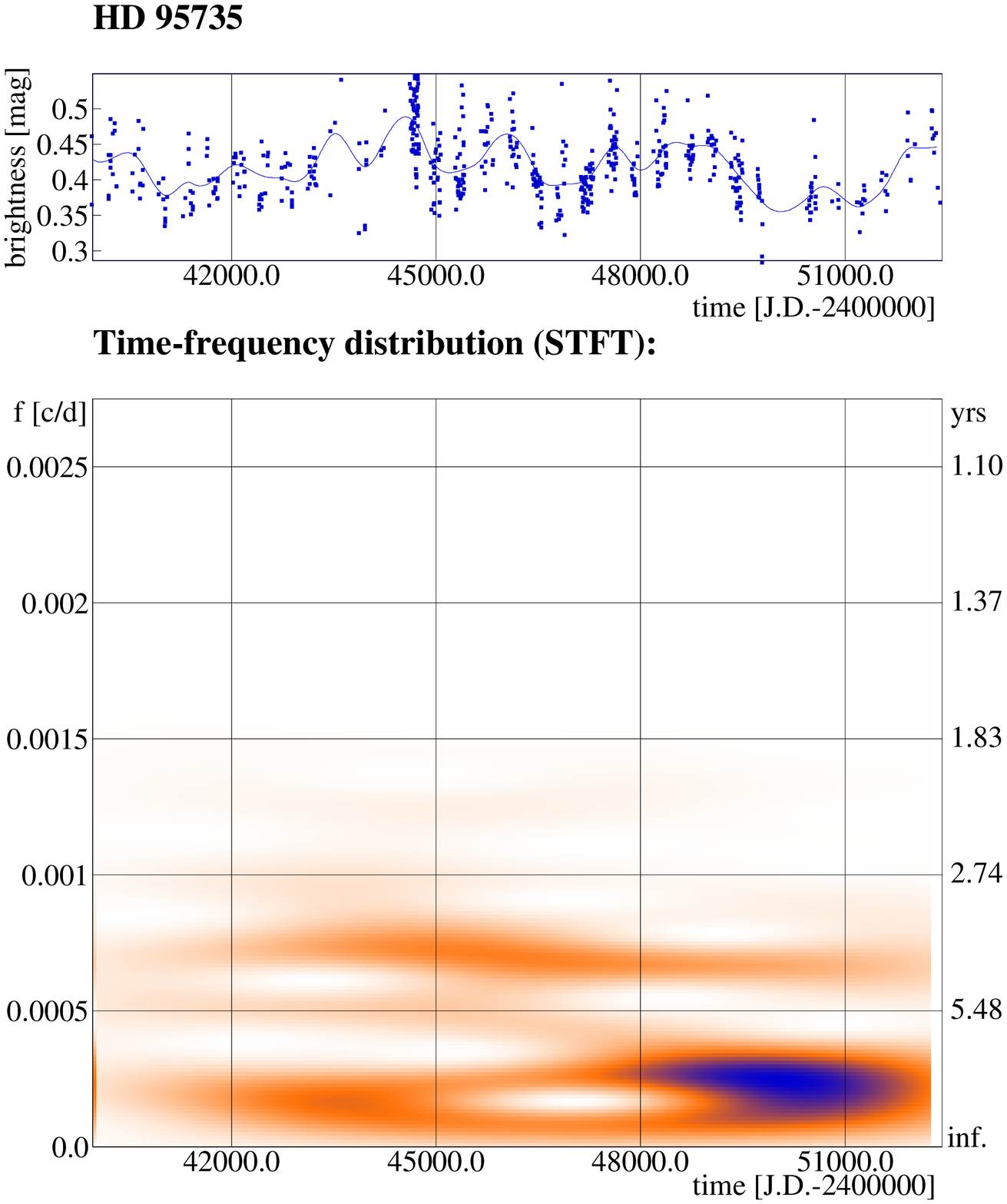}\includegraphics[width=4.3cm]{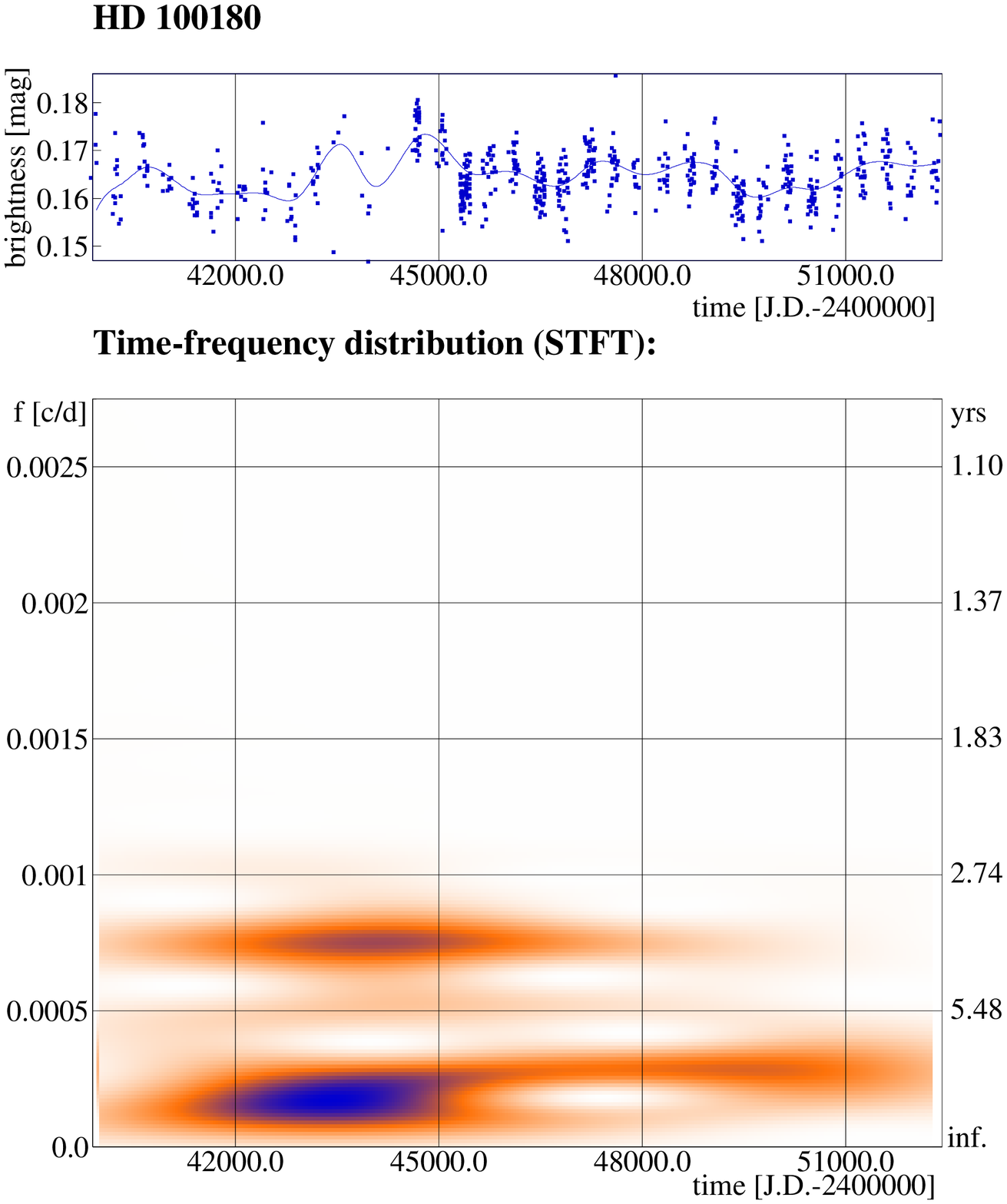}\includegraphics[width=4.3cm]{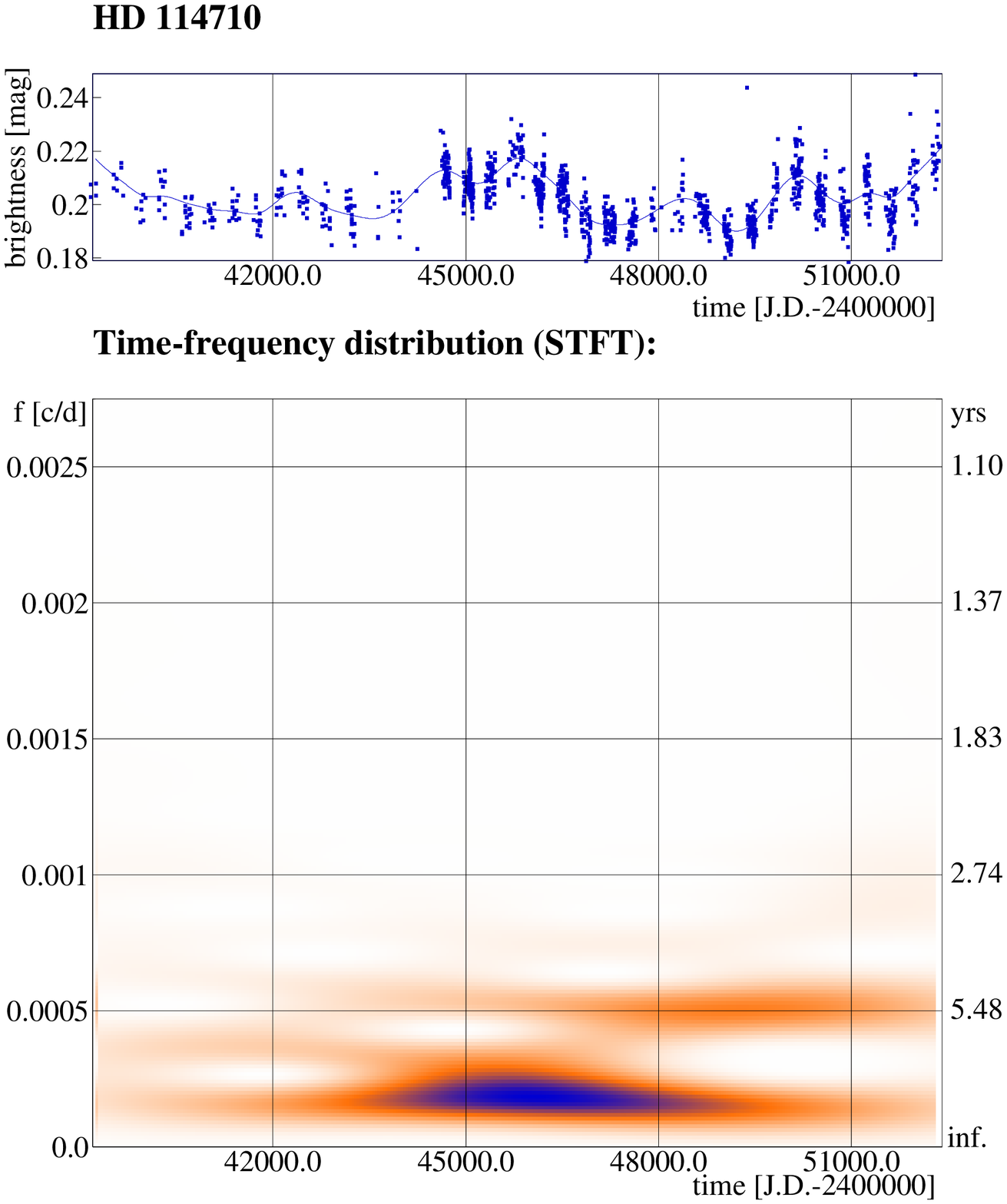}\includegraphics[width=4.3cm]{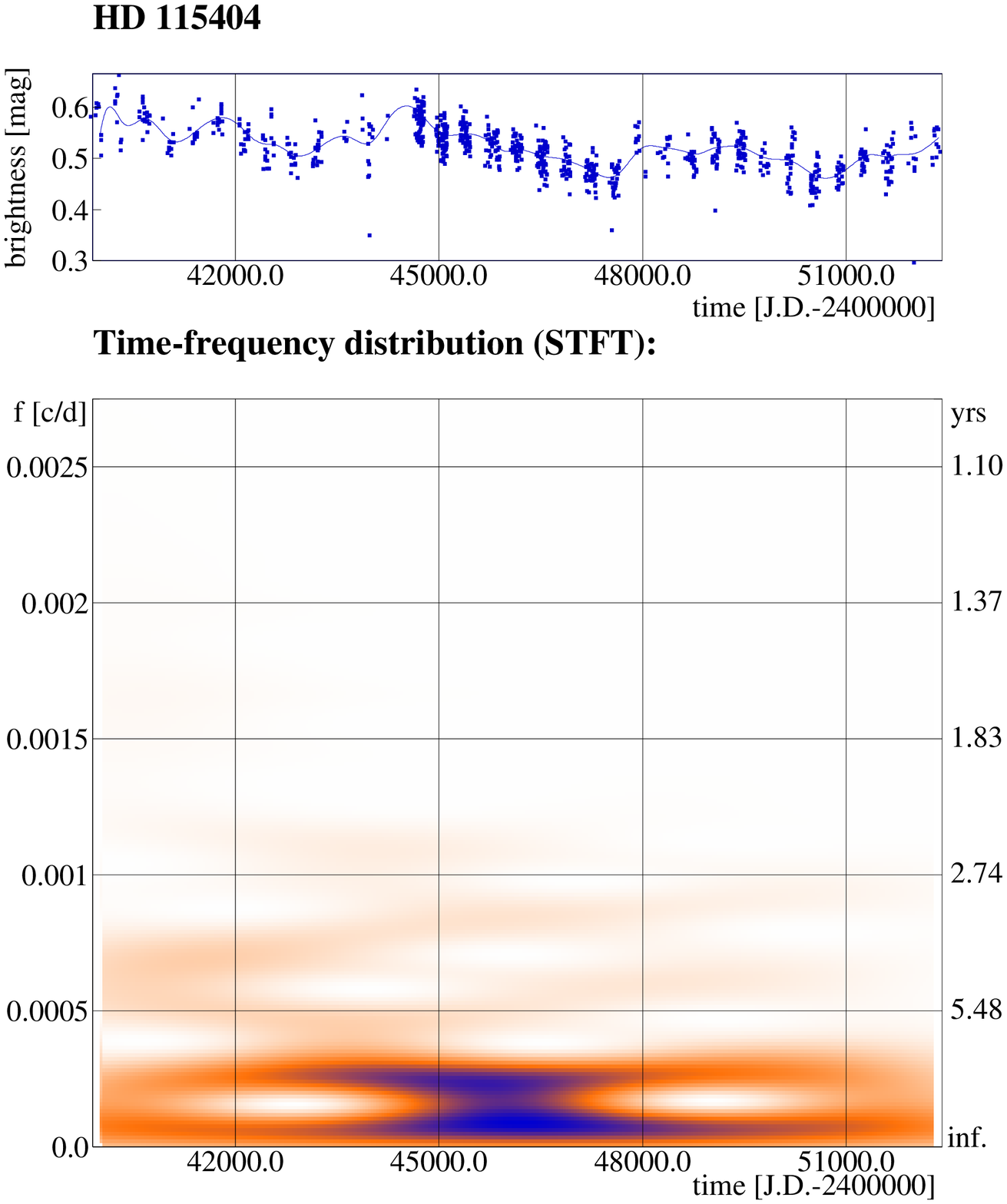}
\includegraphics[width=4.3cm]{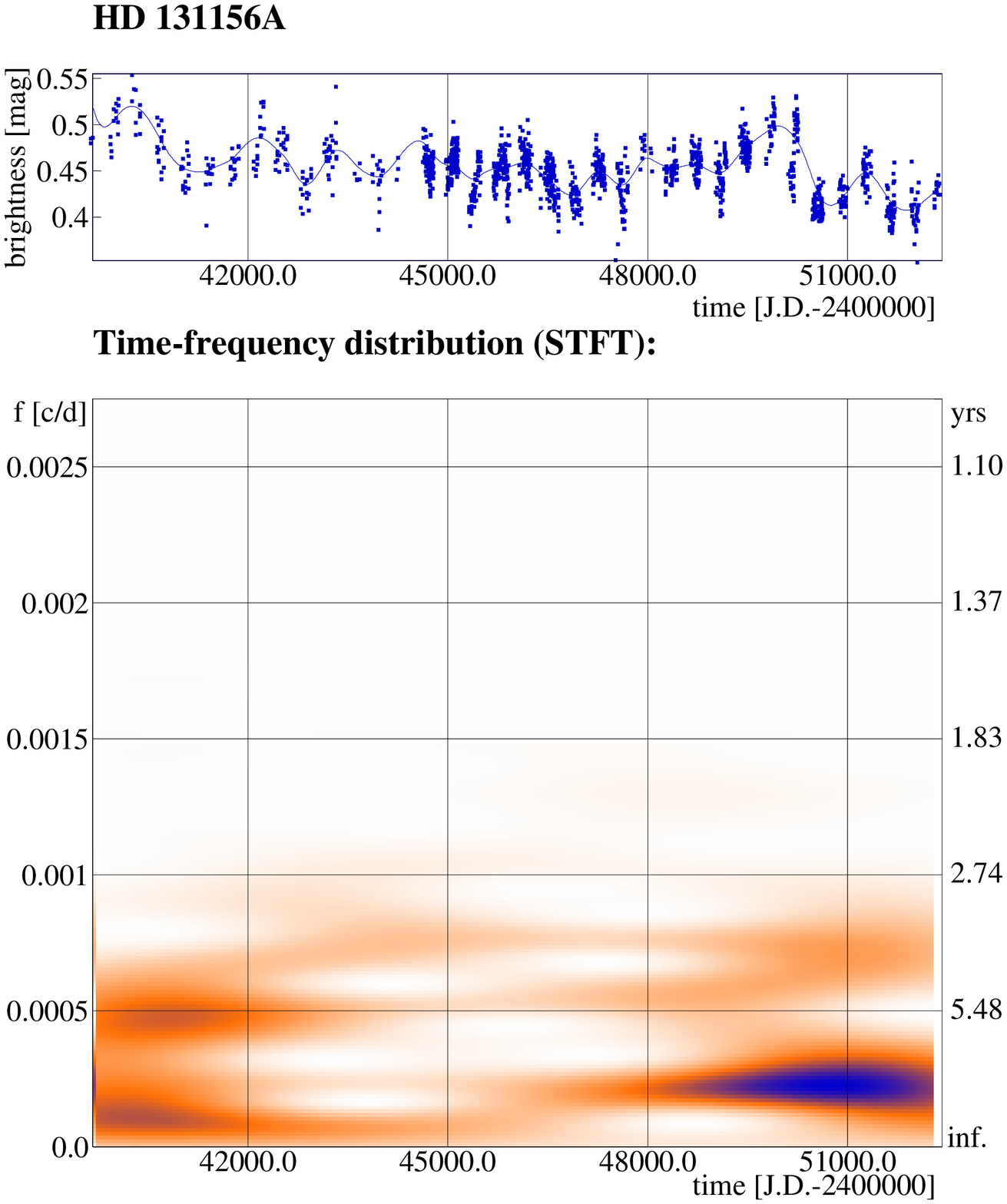}\includegraphics[width=4.3cm]{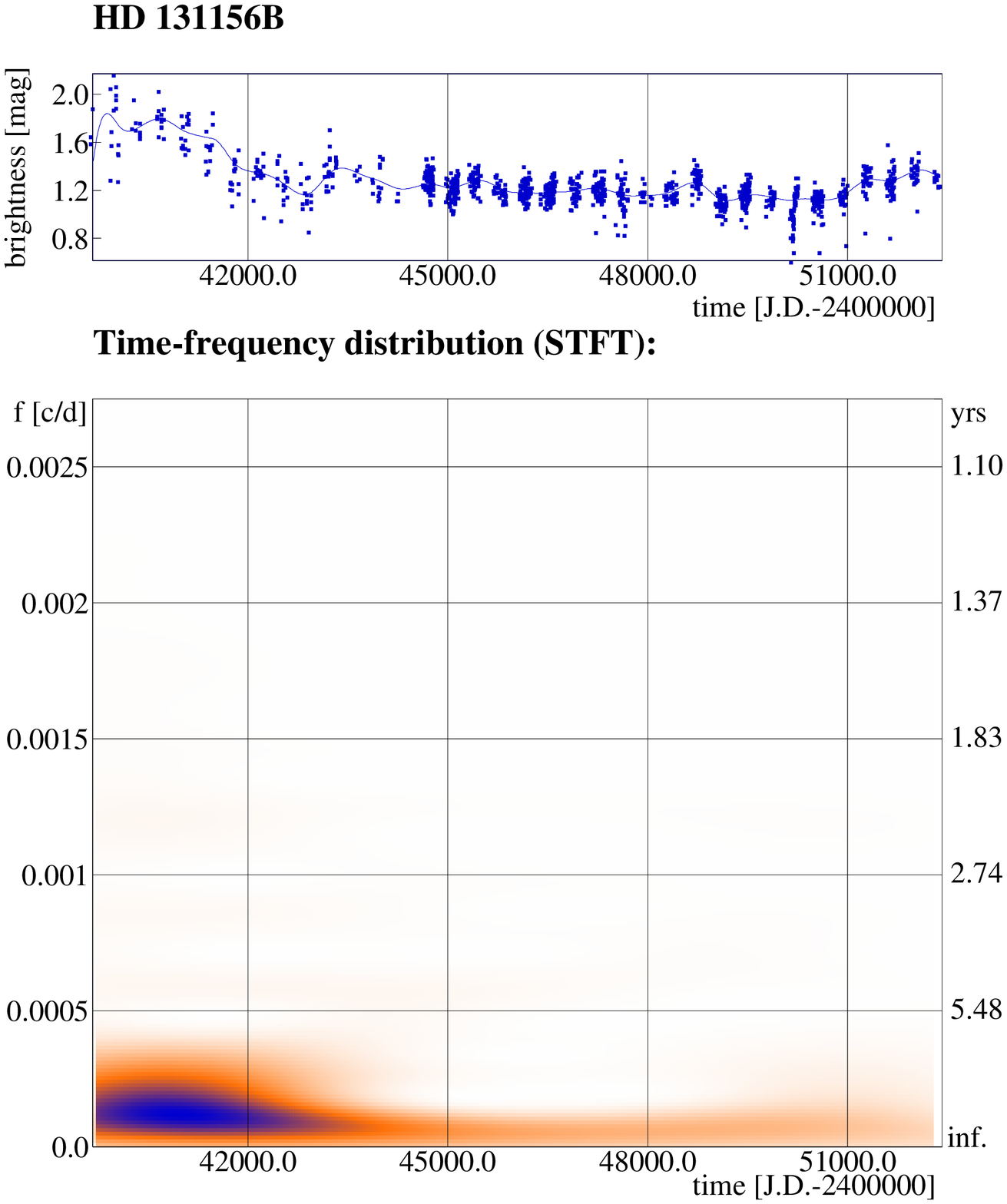}\includegraphics[width=4.3cm]{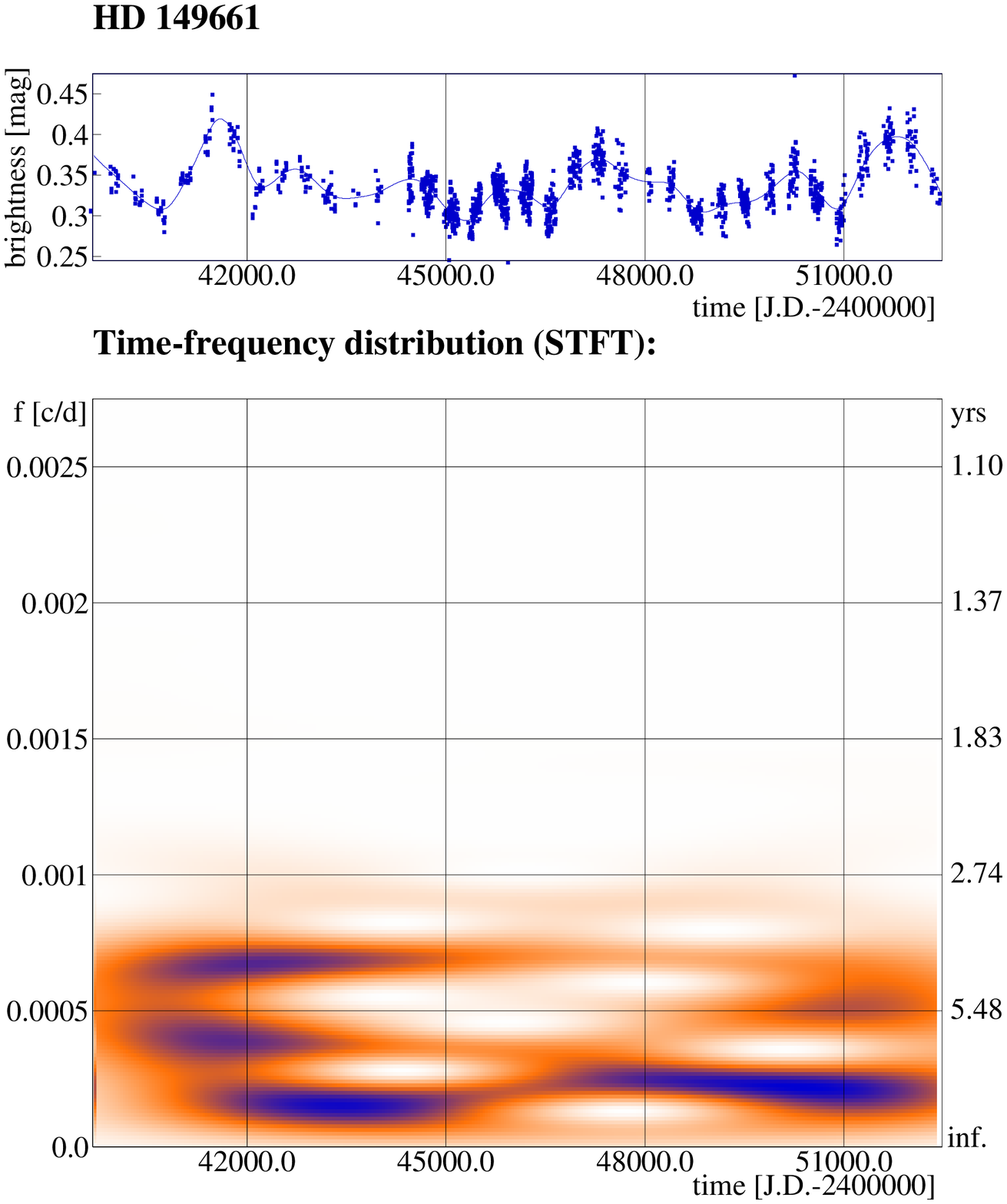}\includegraphics[width=4.3cm]{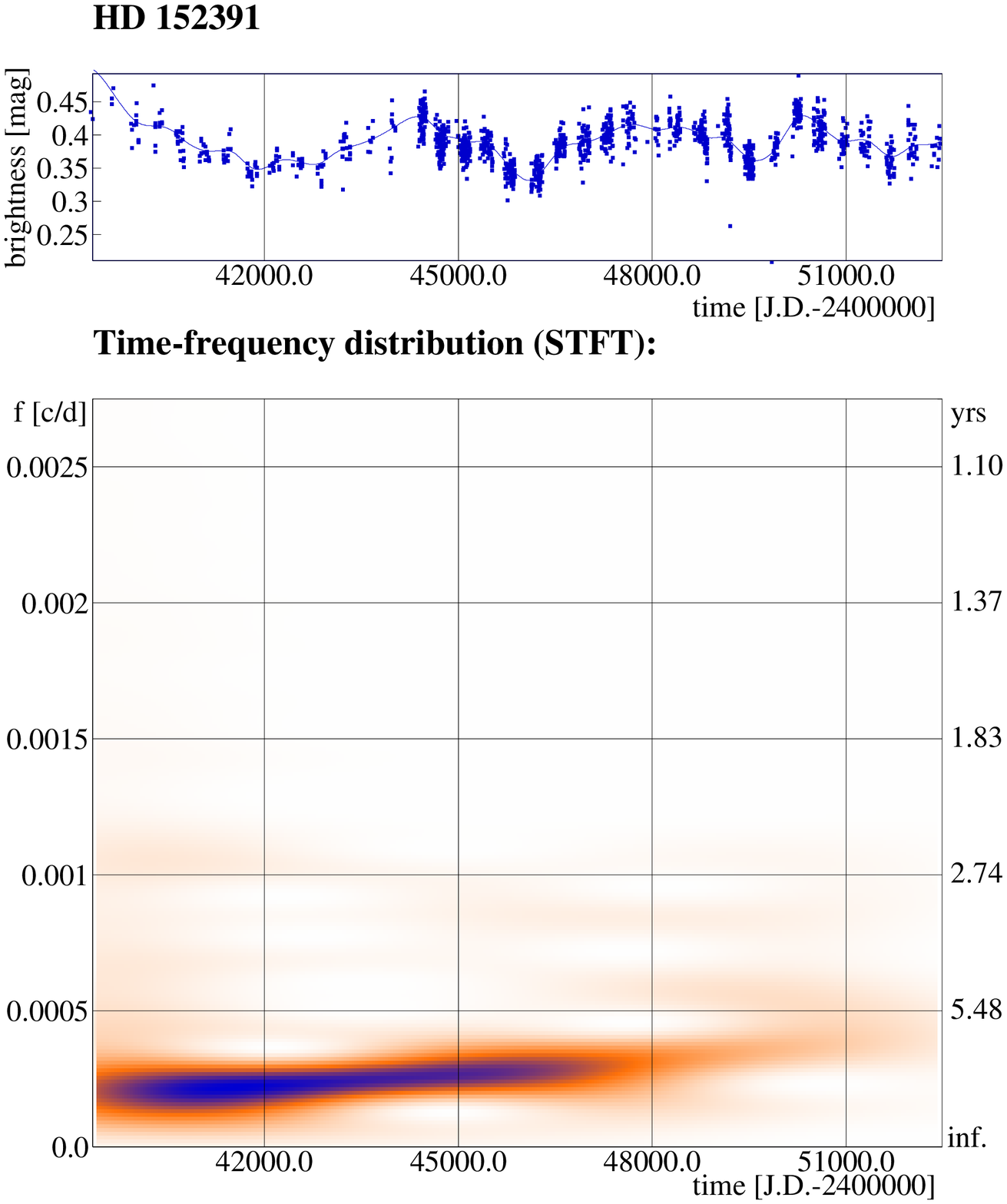}
\includegraphics[width=4.3cm]{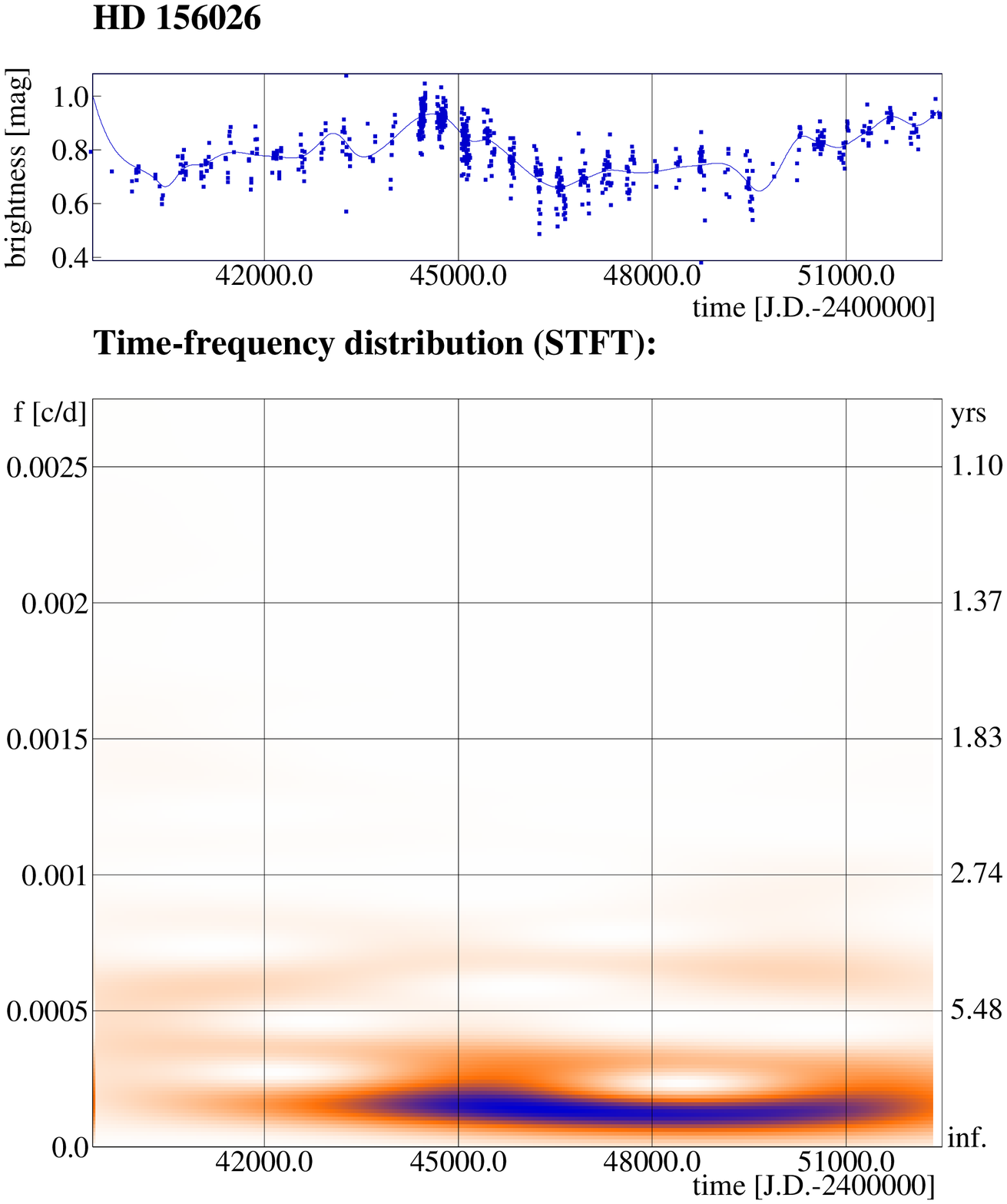}\includegraphics[width=4.3cm]{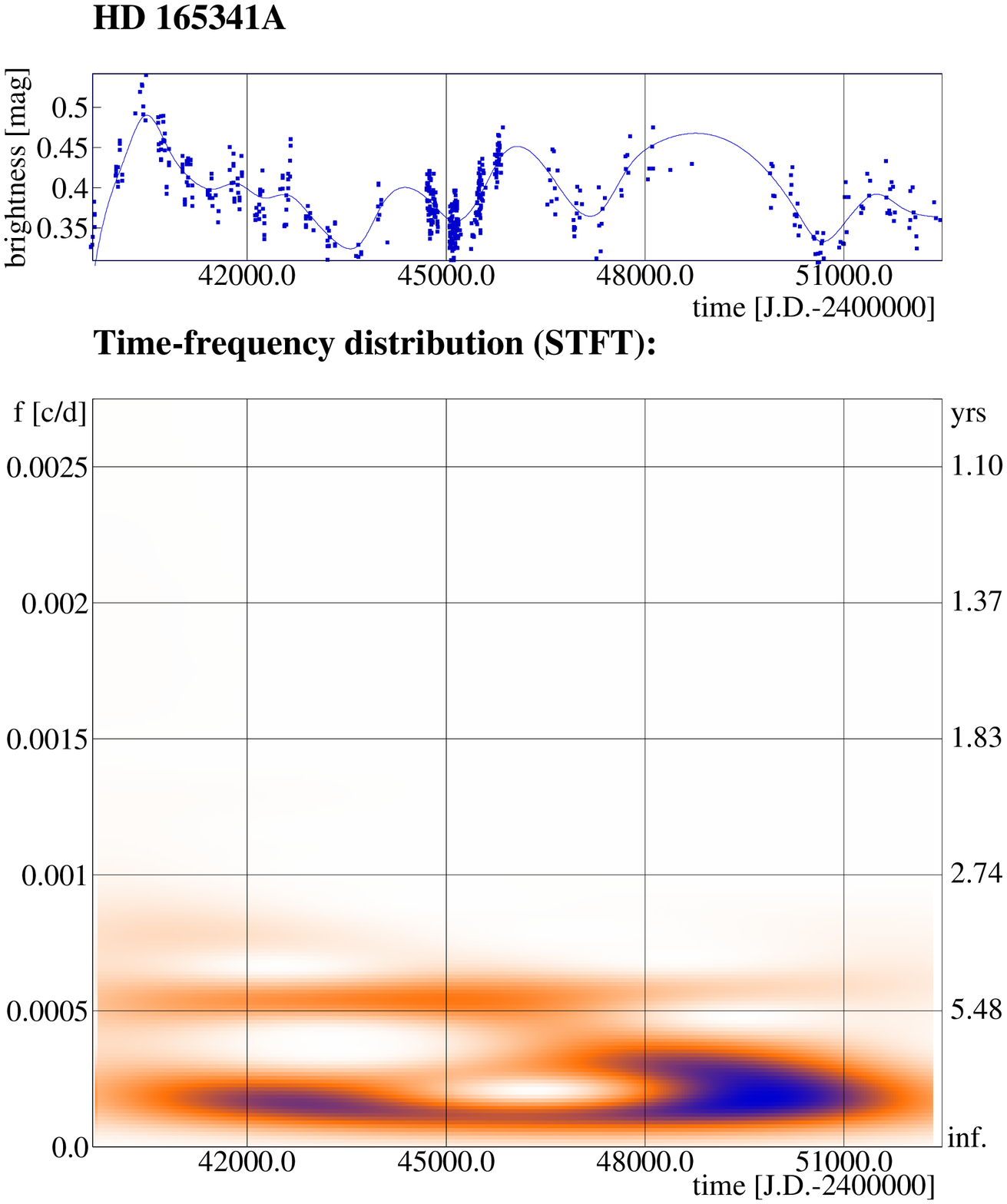}\includegraphics[width=4.3cm]{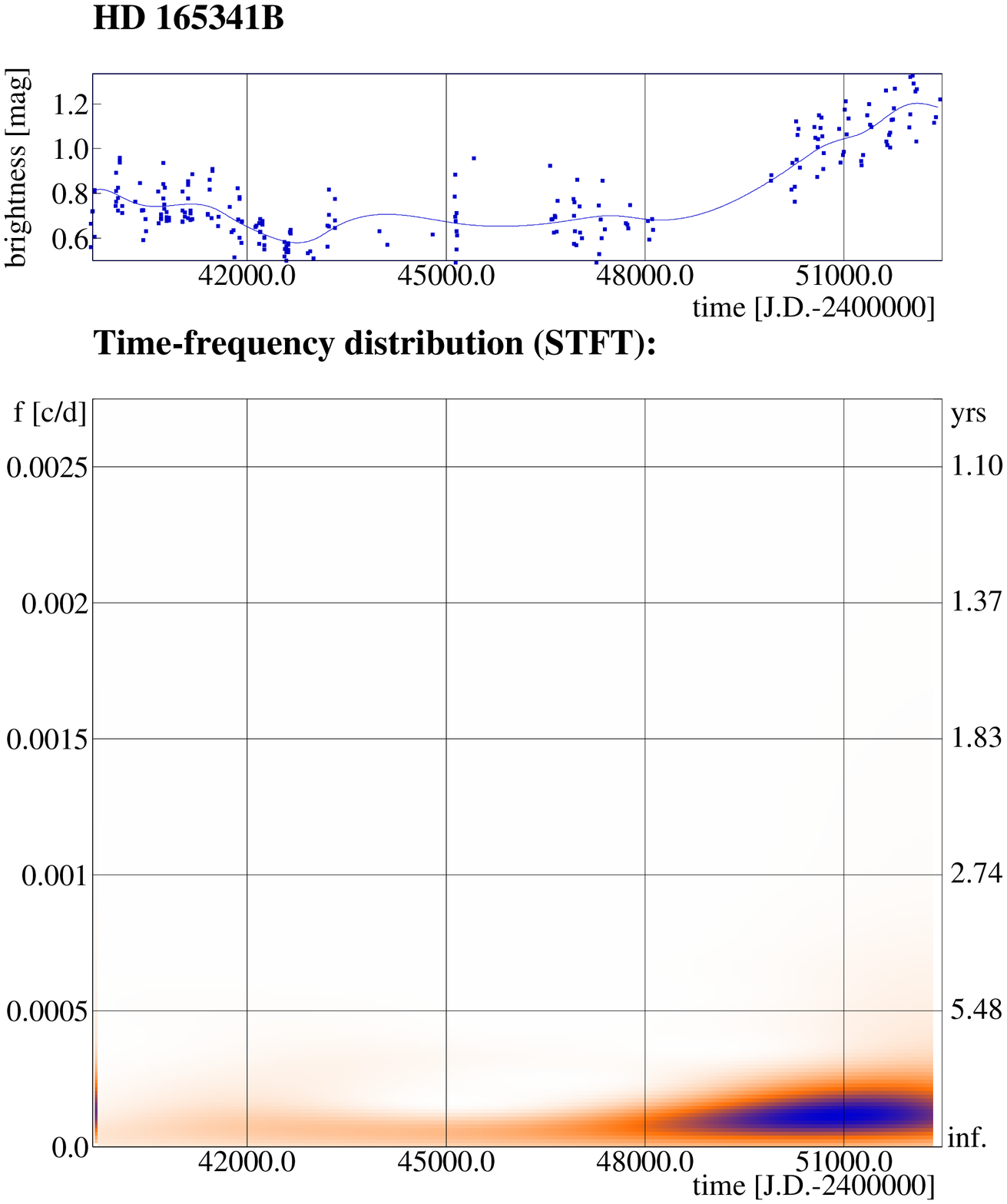}\includegraphics[width=4.3cm]{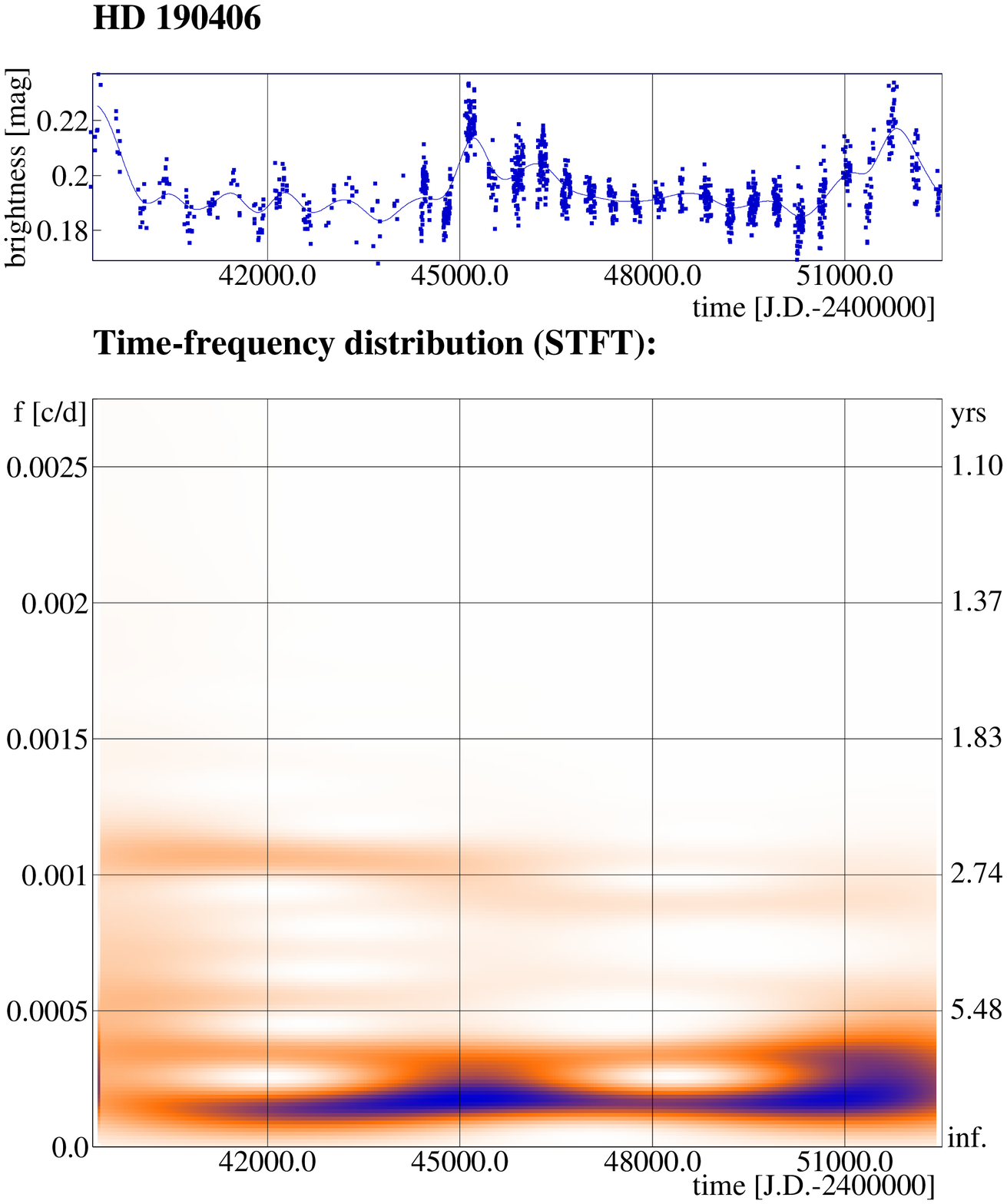}

  \hspace*{-12.8cm}\includegraphics[width=4.3cm]{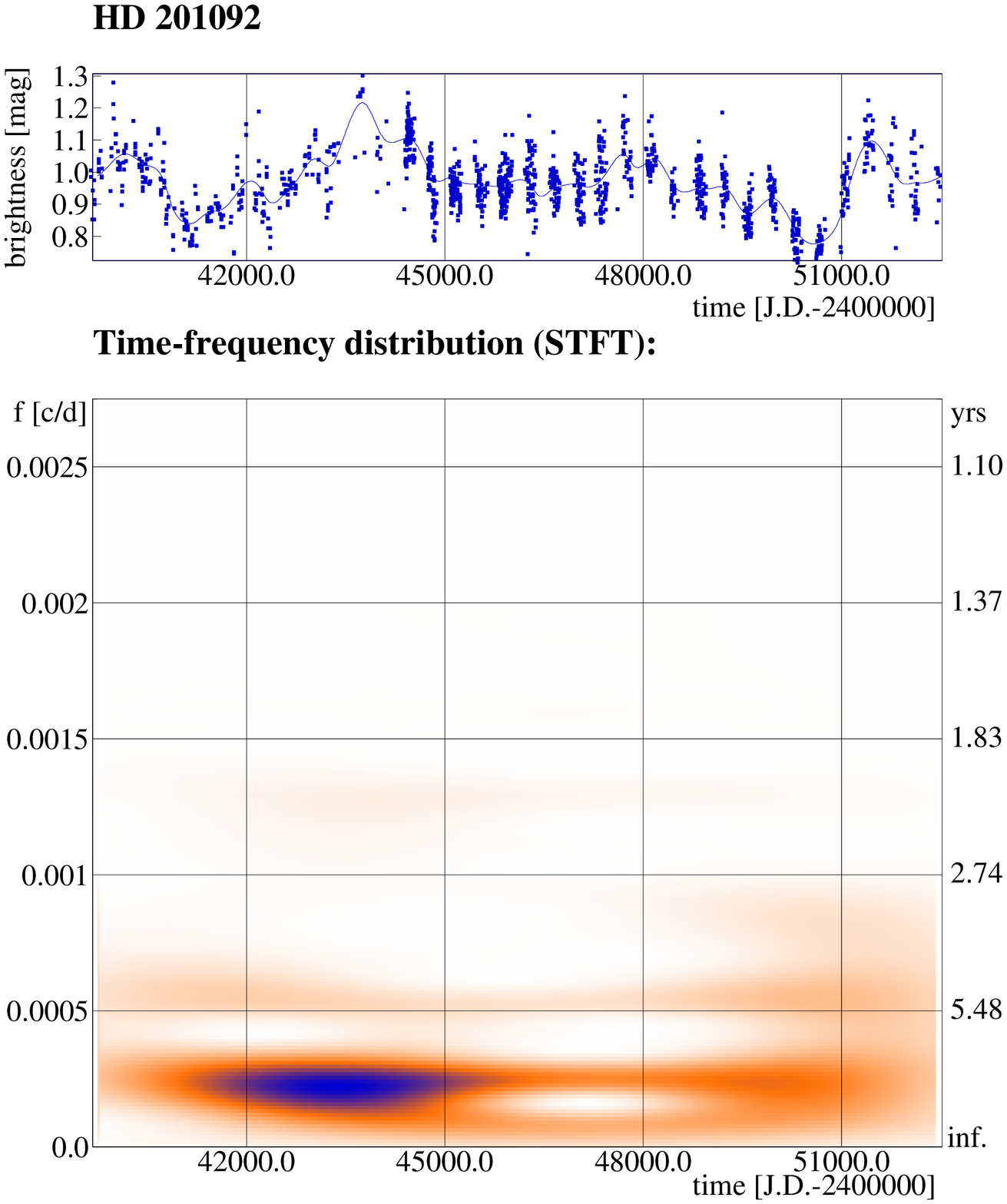}
  \caption{STFT of  MW stars with complex cycles. In the upper panels, the datasets and the spline interpolation we used are plotted; the lower panels show the time-frequency diagrams. We applied no amplification so the cycle amplitudes can directly been compared.}
    \label{stft_compl}
    \end{figure*}

The type of the long-term variability of the MW stars divides the sample in two classes by the simple or complex nature of their dominant cycles. Twelve stars show high-amplitude, slow variability with well-seen cycles, and sixteen have more erratic variations. The STFT maps are plotted in Figs.~\ref{stft_simple} and ~\ref{stft_compl} for the two types, see also the Appendix for the confirmatory CWD maps. The numerical results are given separately for the two groups in Table~\ref{allcycles}. In the table, spectral types from different sources (Gray et al. \cite{gray1}, \cite{gray2}, and Baliunas et al. \cite{baliunas}), rotational periods and mean cycle lengths, determined in this paper, are given. In cases of changing cycles, the extreme values or limits of the cycle lengths are also listed, in parentheses. When available, in the last column, cycle lengths from Baliunas et al. (\cite{baliunas}) are given for comparison. Only one star (HD~165341B) does not have any cycle; its time series displays only a slow, long-term variation in the course of the observations. Altogether, 14 stars show more than 1 cycle, and 27 stars (except HD~16160) have at least one changing cycle. 

Comparing the results of the derived cycles with those published by Baliunas et al. (\cite{baliunas}) a very good general agreement is found. Baliunas et al. (\cite{baliunas}) used Lomb-Scargle periodograms for the datasets, which were about 10 years shorter than in the present paper. Their published errors of the cycle lengths could reflect changing cycles: e.g., for HD~100180, we find two cycles of 3.63 years and a variable one between 16.6--9.85 years while Baliunas et al. (\cite{baliunas}) got 3.56$\pm$0.04 and 12.9$\pm$0.5 years. Our newly determined rotational periods from the longer dataset are within the minimum and maximum periods from the seasonal results of Donahue et al. (\cite{donahue}) for the stars in common with the present MW sample. For 17 MW stars, we find long-term changes which could be parts of other variability on long timescales, e.g., similar to the solar Gleissberg cycle.

Looking at Table~\ref{allcycles}, it is already evident that the 12 stars with simple cycle pattern have generally longer rotational periods of $39.7\pm6.0$ days (median: 39.3 d), and corresponding cycle lengths of $9.7\pm1.9$ years (median: 9.5 yr), respectively. Some of those stars have shorter, secondary cycles with low amplitudes of a few percents of the main cycle amplitude on timescales of around 3.5--4 years. The rotation and cycle lengths show a much higher dispersion for 16 stars with complex cycles: their rotation of $18.1\pm12.2$ days (median: 14.9 d) is on average half as long as that of the other group, and the scatter is on the same order as the value itself. The average cycle length is $7.6\pm4.9$ years (median: 5.2 yr), i.e., it is also shorter and more scattered than for the stars with simple cycles. Moreover, the stars with complex cycle variations show secondary, and sometimes tertiary cycles that can be shorter or longer than the dominant one, with amplitudes ranging between 5--65\% of the dominant cycles. Figs.~\ref{stft_simple} and ~\ref{stft_compl} show the striking difference between the long-term changes of the two groups.

The recent detection of dual magnetic activity cycles at 1.7-yr and about 12-yr in the rapidly rotation solar analog star HD~30495, 
with spectral type G 1.5V (P$_{rot}=11$ days), and estimated age of about 1 Gyr, by Egeland et al. (\cite{egeland}) fits well with our results. Furthermore, because of the longer observational records available, those authors were also able to discern the individual, decadal-long cycles to have durations ranging from 9.6 to 15.5 years. In addition, Metcalfe et al. (\cite{metcalfe}) reported also the detection of simultaneous operation of two stellar activity cycles of 2.95$\pm$0.03 years and 12.7$\pm$0.3 years for the young star, HD~22049 ($\epsilon$~Eridani) which hosts a planet with absolute mass of 15 times the mass of Jupiter. Findings of multiperiodicity by other authors are consistent with similar results presented in our Figs.~\ref{stft_simple} and ~\ref{stft_compl}.
     
 %-------------------------------- Fig. 2.
 
%-------------------------------- Fig. 3.
 
 \section{Discussion} 
 
The rotational and cycle periods are two important observables of magnetically active stars. Their ratio, $P_\mathrm{cyc}/P_\mathrm{rot}$ can be related to the dynamo number $N_D$, giving a directly measurable parameter closely related to the magnetic activity of the stars, first proposed by Soon et al. (\cite{soon}). For a detailed discussion of this topic, see Baliunas et al. (\cite{baliunas1}) who, based on the observations of MW stars with a shorter time-base ($\approx25$ years) at that time, stressed the importance of the \emph{measured quantities} in examining stellar dynamos.

\begin{figure}[tbp]
   \centering
      \includegraphics[width=8cm]{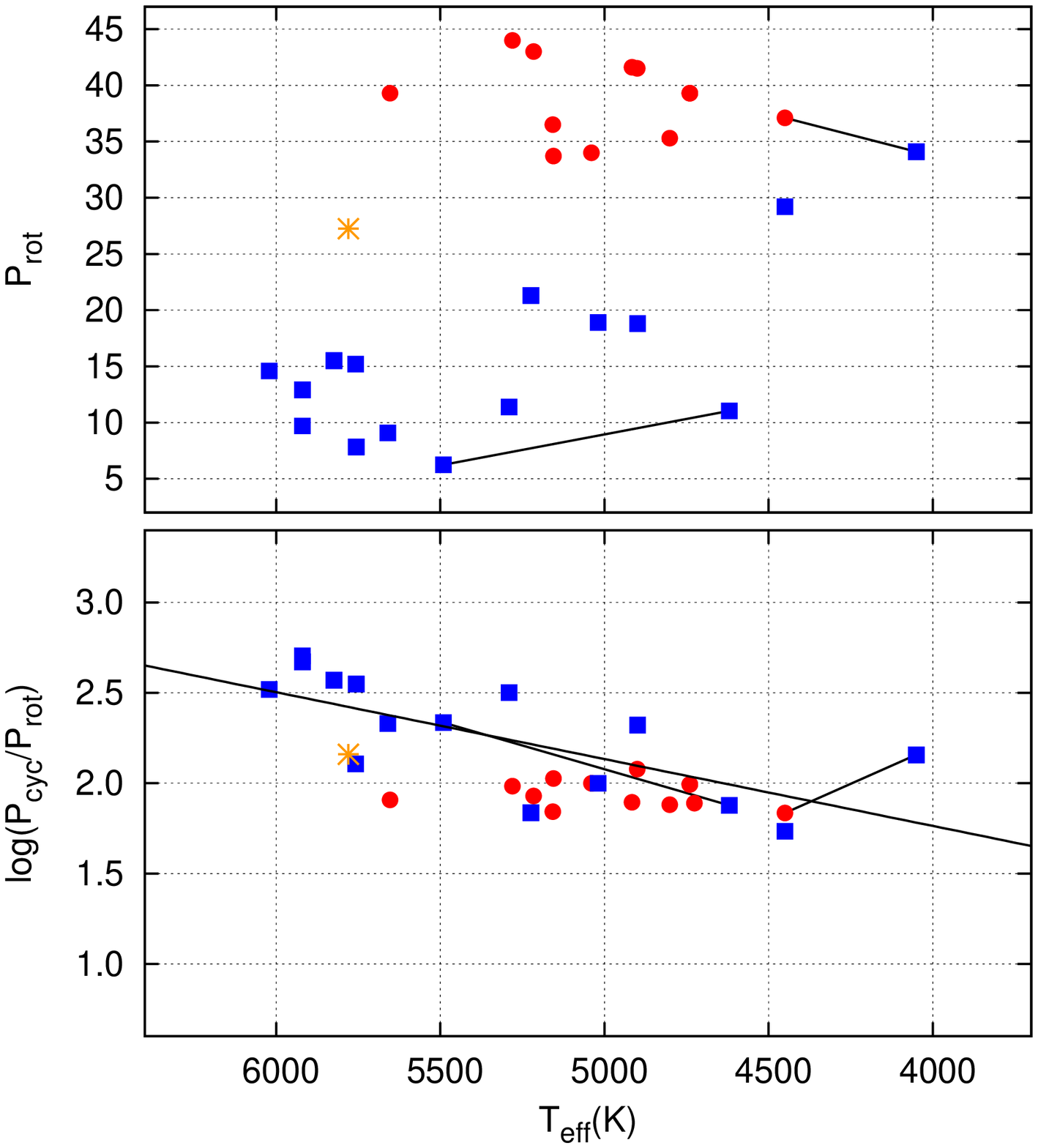}
   \caption{Relations of rotational and cycle periods vs. $T_\mathrm{eff}$. Stars with simple cycles are plotted with red dots, while complex ones with blue squares and the Sun with an orange star. The components of the wide binaries HD~131156A/B and HD~201091/HD~201092 are connected.  \emph{Top}: Rotational periods versus effective temperatures. \emph{Bottom}: $P_\mathrm{cyc}/P_\mathrm{rot}$ (related to the dynamo number) versus effective temperature. The fit for the stars with complex cycles shows a relation being higher at hotter temperatures. }
   \label{rot_cyc_teff}
 \end{figure}

Durney et al. (\cite{durney}), and more recently B\"ohm-Vitense (\cite{bohm}) studied the rotations and cycles of the MW stars in the function of their  temperatures. We re-examine the relation for the stars of our sample plotted in Fig.~\ref{rot_cyc_teff}. Effective temperatures, when available, are from Gray et al. (\cite{gray1}, \cite{gray2}). For stars having no temperatures in those two papers, we get spectral types from Baliunas et al. (\cite{baliunas}),  and use the color-temperature calibration of Pecaut and Mamajek (\cite{pec_ma}, Table 5.) to calculate effective stellar temperatures. Comparing the temperatures of the 18 stars from Gray et al. (\cite{gray1}, \cite{gray2}) with color-temperature calibration of the same stars from Pecaut and Mamajek (\cite{pec_ma}) we find, on average, temperature differences less than 60\,K with the latter calibration being hotter. However the small difference in temperature calibration methods basically does not affect our result depicted in Fig.~\ref{rot_cyc_teff}.

The rotational periods show a bimodal distribution between the stars with simple and complex cycles as a function of effective temperature, as shown in the top panel of Fig.~\ref{rot_cyc_teff}. The picture is the same as in B\"ohm-Vitense (\cite{bohm}), her Figure 4., showing the difference between the ``A'' (active) and ``I'' (inactive) stars; different dynamos working in the stars of the two groups was given as the possible reason of the bimodality  in that paper. In the bottom panel of Fig.~\ref{rot_cyc_teff}, no strong correlation is seen for the stars with simple cycles, and a possible slight increase of $P_\mathrm{cyc}/P_\mathrm{rot}$ is found for the stars with complex cycles towards higher temperatures, i.e., earlier type stars. Note, that the coolest star, HD~95735, which is an M-dwarf, is excluded from both relations of Fig.~\ref{rot_cyc_teff}, since its internal structure may differ significantly from the rest of the stars. The MW survey was actually able to detect this convective M-dwarf flaring in real time during the 1981 observing season; consult the data points plotted in Fig.~1g (right panel, bottom) of  Baliunas et al.  (\cite{baliunas}).

%-------------------------------- Fig. 4.
 
%-------------------------------- Fig. 5.
  \begin{figure}[h!!!]
   \centering
   \includegraphics[width=8cm]{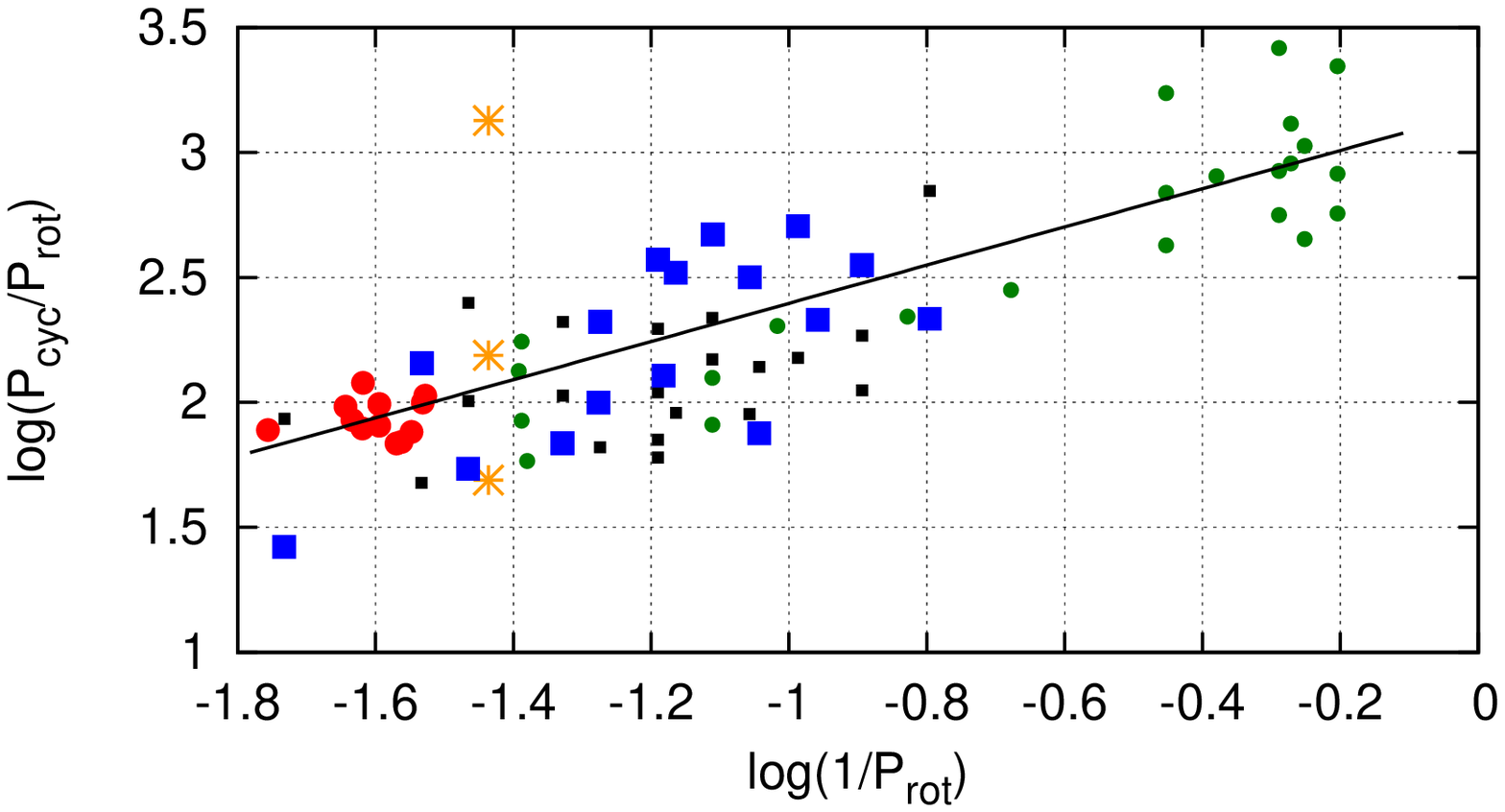}
   \caption{Relation between the observed rotational- and cycle periods. Stars with simple cycles are plotted with red dots, while complex ones with blue squares and the Sun with orange stars (Gleissberg-, Schwabe-, and 3--4-yr cycles). Small black squares denote additional cycles, and small green dots are the results from Ol\'ah et al. (\cite{olahetal}). The fit is only done for the dominant cycles derived in this paper. It goes through the region of the fast rotating K-dwarf stars (singles and binaries as well) in the sample of Ol\'ah et al. (\cite{olahetal}). See the text for further discussion. }
 \label{rot_cyc}
 \end{figure}
 
We plot  $P_\mathrm{cyc}/P_\mathrm{rot}$ in the function of $1/P_\mathrm{rot}$ on Fig.~\ref{rot_cyc} like that first shown by Baliunas et al. (\cite{baliunas1}), displaying all cycles derived in this paper and given in Ol\'ah et al. (\cite{olahetal}) using the same time-frequency analysis method. Apart from the MW stars, the plot contains fast rotating stars, both singles and binaries, and also evolved stars. The fit, with a slope of $0.76\pm0.15$, was done only for the dominant periods of MW stars derived in this paper (big symbols), based on the Ca-index measurements, i.e., essentially the same as in Baliunas et al. (\cite{baliunas1}). The relation matches the region of the fast rotating K-dwarf stars ($P_\mathrm{rot}$ is between about 1.5--3 days) and does not contradict with the results for evolved stars; note, that all results from Ol\'ah et al. (\cite{olahetal}) are plotted with small green dots. 
However, the samples of the cycles derived from photometry (i.e., photospheric variability) for fast rotating stars of different types, i.e., single stars, binaries, dwarfs and giants, are too small to see if there is any basic difference between the cycle patterns of these active stars concerning their binarity and/or evolutionary status. The general trend between the rotation and cycle length plotted on Fig.~\ref{rot_cyc} looks quite similar for all kind of active stars, also regardless of the source of the data (chromosphere or photosphere). Values arising from the secondary and tertiary cycles of the MW stars fall within the scatter of the relation.

Several attempts have already been made to connect rotation, activity cycles and ages with the observed properties of the MW stars. The closest study to ours is from Soon et al. (\cite{soon}), their Figure~2, which shows clear separation in log($P_\mathrm{cyc}/P_\mathrm{rot})^2$ between older and younger stars, and is interpreted as a consequence of the change of the dynamo efficiency as stars evolve. To construct a similar diagram, we used the pattern of the long-term variability of the stars (simple-complex cycles and cycle lengths) and their estimated gyro-ages from Barnes (\cite{barnes}). The errors of the gyro-ages are between 13--20\%, and are not plotted, in order to maintain clarity of the figures.  Barnes (\cite{barnes}) in his Table~3 gives the ages of the MW stars of our sample derived from isochrone fitting, activity-rotation relation, and his own technique. The 25\% discrepancy between the chromospheric and gyro ages of the MW stars arises in stars with $B-V<0.6$, i.e., from the bluer stars earlier than G0-1V with  $T_\mathrm{eff}$ hotter than about 6000~K. We consider such a discrepancy to be not so serious as to render the determined age to be incorrect. 

\begin{figure}[tbp]
   \centering
   \includegraphics[width=8cm]{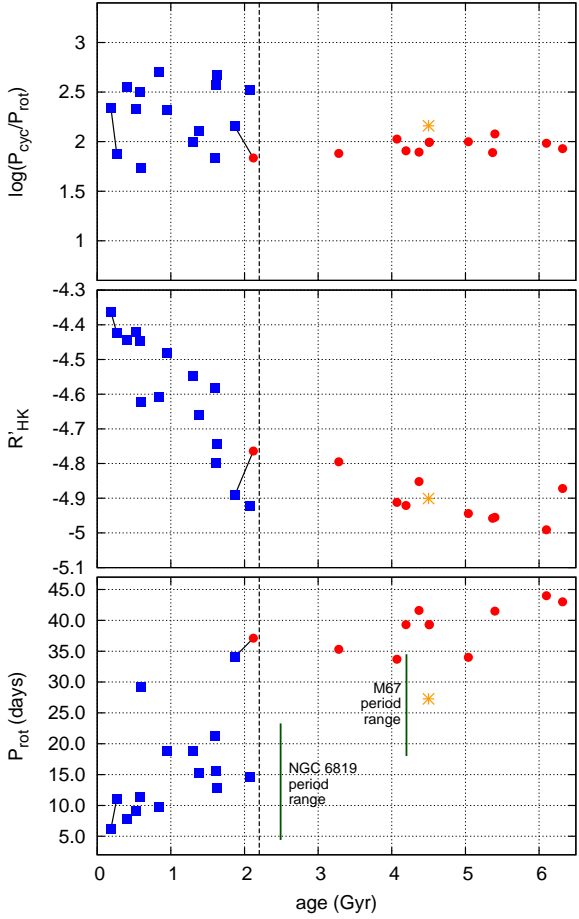}
   \caption{Relations between the observed rotational- and cycle periods, and $R'_{HK}$ versus age. Stars with simple cycles are plotted with red dots, while complex ones with blue squares and the Sun with an orange star. The components of the wide binaries HD~131156A/B and HD~201091/HD~201092 are connected. \emph{Top: } $P_\mathrm{cyc}/P_\mathrm{rot}$ show a clear separation between stars with simple and complex cycles in the sense that the simple cycles have about the same value while the complex ones are more scattered. The division occurs near the age of the Vaughan-Preston gap (Donahue \cite{donahue}). \emph{Middle: }Average $R'_{HK}$ of the stars from Barnes (\cite{barnes}); the younger ones, especially below 1.5~Gyr have higher average activity.  \emph{Bottom:} $P_\mathrm{rot}$ of the stars in the function of age, showing a clear separation, also near the Vaughan-Preston gap. The range of the rotational periods in NGC~6819 and M67 at their ages of 2.5~Gyr and 4.2~Gyr are shown by green lines, from Meibom et al. (\cite{meibom}) and Barnes et al. (\cite{barnes_etal}).}
         \label{relations}
         \end{figure}

We find a clear separation between the stars with simple and complex cycle patterns as a function of age in the sense that stars with simple cycles are older, as seen on Fig.~\ref{relations}, top panel. The separation occurs near the age of the Vaughan-Preston gap (i.e., a bifurcation in the activity owing to the lack of stars with intermediate activity, Vaughan \& Preston \cite{v-p}) at about 2.2~Gyr (Donahue \cite{donahue}) and indicating change in the dynamo. (The M-dwarf HD~95735, slightly older than 3~Gyr, is not part of the relation and is not plotted.) The morphology of stellar activity cycles appears to undergo a characteristic switch at a transitional age, and clearly requires a theoretical explanation.

Fig.~\ref{relations}, middle panel, shows chromospheric emission as a fraction of luminosity, the dimensionless quantity $R'_{HK}$ (Noyes et al. \cite{noyes}) of the stars given in Barnes (\cite{barnes}) as a function of gyro-age. The youngest and fastest rotating stars, younger than $\approx$1.5~Gyr show much higher time-averaged activity than the rest of the sample. After that, a slow downward trend of the chromospheric activity with age is suggested by the data.  

At first sight, the spread of data in the $P_\mathrm{rot}$ and gyro-ages in Fig.~\ref{relations} seems too large to be physically meaningful. But we consider the possibility and reality of radial migration of individual stars, including the Sun and stars in the contemporary solar neighborhood, or even a whole cluster within the age of the Milky Way under the near-resonance interaction between the co-rotating star's angular momentum with the transient spiral density wave studied e.g., recently by Roskar et al. (\cite{roskar}). The spread of metallicity that may result from the migration of the stars in the solar neighborhood, is  more than 1.0~dex, without any reliable trend until at least 6--7~Gyr, as plotted by Sellwood and Binney (\cite{sel-bin}) in their Figure 3. That picture is similar to ours for MW stars with available Fe/H values (Gray et al. \cite{gray1}, \cite{gray2}) and from the metallicity indicator $< C_{RV} >$ (Soon et al. \cite{soon2}), which we plot in Fig.~\ref{metal} in the Appendix. Such a large metallicity difference could indicate different evolutionary paths,  with the full inclusion of the likely scenarios of radial migration within the Galaxy, even for stars of very similar mass as of our MW sample, thereby accounting for at least some of the spread in Fig.~\ref{relations}.

A plausible explanation for the apparent dichotomy between stars of higher and lower level of activity 
(``active'' vs. ``inactive'' stars) was already suggested by Durney et al. (\cite{durney}) as an explanation 
for the Vaughan-Preston gap. In their scenario, in stars above a critical rotation rate, several dynamo modes are excited, leading to a higher overall level of activity.  In slower rotating stars, in contrast, only the fundamental mode is excited. Our results seem to agree with this general scenario, a difference in the overall activity level ($R'_{HK}$) across the V-P gap is quite obvious from our data. In addition, in  Fig.~\ref{relations}, bottom panel, an indication of a non-uniform evolution of the rotation period may be seen.  

Another possibility is the switching of the nonlinear dynamo from one saturated state to another. Such a possibility arises in the bimodal interface dynamo proposed by Petrovay (\cite{kristof1}). In this dynamo, under fairly general assumptions on the nonlinearity (some of which were confirmed by Petrovay et al. \cite{kristof2}), two alternative solutions exist: a strong-field, short-period solution with a 
thin tachocline and a weak-field, long-period solution with a thicker tachocline. Transition from the latter state to the former state would be expected to lead to a shortening of the dominant dynamo period and to an increase in the level of activity compared to what would be
expected from the extension of the ``active branch'' towards lower rotation rate.  Some support for the bimodal interface dynamo may be present, as seen in Fig.~\ref{relations}, middle panel. (In this sense, using ``inactive branch'' for the regular group may be misleading.) In addition, the reorganization of the tachocline may potentially lead to a change in the overall rotation rate of the convective envelope via altered torque balance. On the other hand, in its present form, this scenario offers no explanation to the complex, multiperiodic variation seen in the presumed weak-field dynamo.

%-------------------------------- Fig. 6. 
                  
\subsection{The validity of gyro-ages for the MW stars}

%-------------------------------- Fig. 7. 

   \begin{figure}[tbp]
   \centering
   \includegraphics[width=8cm]{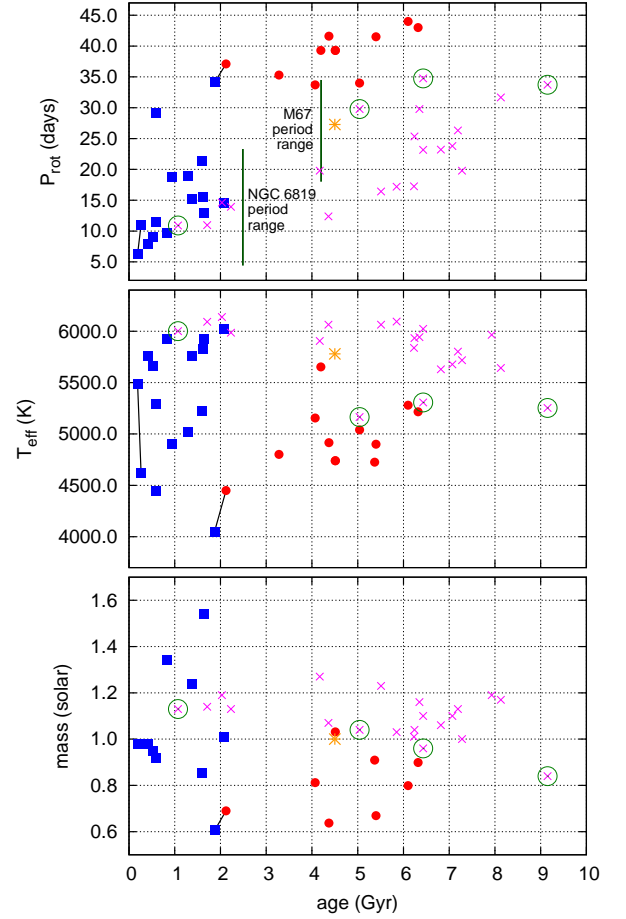}
   \caption{Rotational periods and effective temperatures of the stars as a function of age. Stars with simple cycles are plotted with red dots, while complex ones with blue squares and the Sun with an orange star.  Stars from van Saders et al. (\cite{saders}) are plotted with magenta crosses, four stars with $\log g>4.4$ are marked with green circles. The components of the wide binaries HD~131156A/B and HD~201091/HD~201092 are connected. \emph {Top}: rotational periods. The range of the rotational periods in NGC~6819 and M67 at their ages of 2.5~Gyr and 4.2~Gyr are shown by green lines, from Meibom et al. (\cite{meibom}) and Barnes et al. (\cite{barnes_etal}). \emph {Middle and bottom}: effective temperatures and masses in the function of age, respectively. See text for further discussion.}
         \label{age}
         \end{figure}

Meibom et al. (\cite{meibom}) selected 30 stars in NGC~6819, an open cluster of about 2.5~Gyr old, to study their rotation with high precision. They found a clear relationship between the rotational periods and stellar masses at the age of the cluster supporting the validity of the rotational evolution model of Barnes (\cite{barnes1}) with the best fit of the rotational period versus $(B-V)_0$ color index. More recently, Barnes et al. (\cite{barnes_etal}) derived rotational periods of 20 cool stars in the 4.2~Gyr old open cluster M67 based on data from the Kepler K2 mission and found also good relation between the rotational periods and $(B-V)_0$ color indices for stars with masses between 0.8--1.15~$M_{\odot}$.

Fig.~\ref{relations}, bottom panel, shows the rotational periods of MW stars as a function of age. The Vaughan-Preston gap is present, and the range of rotation periods is shown for the 30 stars in NGC~6819 by Meibom et al. (\cite{meibom}) at the cluster age of 2.5~Gyr, and for the 20 stars in M67 by Barnes et al. (\cite{barnes_etal}) at the cluster age of 4.2~Gyr.  The period range of the NGC~6819 stars fits well the range of the younger MW stars in the field.  The M67 results of Barnes et al. (\cite{barnes_etal}) shows that the longest period stars of this sample have $(B-V)_0$ of about 1.0 corresponding to about 0.8~$M_{\odot}$ matching our results for the older MW stars which have similar masses (see below), near the age of the Sun. The gyro-ages of the MW stars and that of NGC~6819 and M67 clusters were derived based on the same background, using the expressions of Barnes (\cite{barnes}, \cite{barnes1}). 

Almost all stars of our sample (cf. Fig.~\ref{rot_cyc_teff}) fall into the temperature region where chromospheric and gyro-ages agree well with each other. van Saders et al. (\cite{saders}) found anomalously rapid rotation among old, field stars based on rotational periods from Kepler data and astroseismically derived ages, which generally do not agree with the position of the older group of MW stars. In Fig.~\ref{age}, top panel, data from van Saders et al. (\cite{saders}) are plotted with magenta crosses over our data. Four stars with $\log g>4.4$ from van Saders et al. (\cite{saders}), marked with green circles in Fig.~\ref{age} fall near the positions of MW stars in both panels, showing that the relations given by the MW stars are valid in a narrow $\log g$ range around 4.5. Note that the average $\log g=4.50\pm0.17$ for 23 MW stars with available values.  In this average, we did not take into account HD~81809 ($\log g=3.84$, Gray et al. \cite{gray2}), since it is a spectroscopic binary with components of 1.7/1.0 solar masses (Favata et al. \cite{favata}); Barnes (\cite{barnes}) also notes that the gyro-age of this object (10.6~Gyr) is uncertain because of its binarity; therefore, we used the chromo-age (4.19~Gyr) instead, in the plots.

The middle panel of Fig.~\ref{age} shows temperature vs. age, for the same stars as in the top panel. Older than 4~Gyr, the stars from van Saders et al. (\cite{saders}) are about 1000~K hotter than the MW stars of similar ages. The masses of the older MW stars, on average is $0.79\pm0.13~M_{\odot}$ (masses are from Valenti \& Fischer \cite{val-fisch}), whereas all stars from van Saders et al. (\cite{saders}) are in a narrow mass range of $1.10\pm0.10M_{\odot}$, see the bottom panel of Fig.~\ref{age}. This difference suggests that the stars in the sample of van Saders et al. (\cite{saders}) were originally F-type stars with higher initial mass and different internal structure than those of MW stars of similar age. The weak, if any, magnetic field of F-stars results in less effective magnetic braking leaving the stars with faster rotation at advanced ages. The masses of the younger MW stars on average are about 1.0~$M_{\odot}\pm0.3$ with quite a large standard deviation.
         
\subsection{The wide binaries HD~131156A/B and HD~201091/HD~201092 in the sample}

The stars of the wide binary system HD~131156A/B are the youngest stars (about 0.2~Gyr) among the MW stars in this paper, with fairly similar dominant cycles, around 4 years. The dominant cycles of both components have small amplitudes compared to the long-term variability; thus, those dominant cycles can hardly be seen on the maps, which have no artificial amplifications (cf. Sec.\ref{obs_meth}). Those dominant cycles are, however, evident in the time series themselves. Also, Morgenthaler et al. (\cite{morgenthaler}) from magnetic field monitoring of HD~131156A, found that the topologies of the field were similar in 2007.59 and 2011.07, i.e., about 3.5 years apart, and suggested a possible relation with a cycle of this length, which supports our cycle length of 3.7-yr. 

The stars of HD~201091/HD~201092 are near the Vaughan-Preston gap in the age-$P_\mathrm{cyc}/P_\mathrm{rot}$ diagram (Fig.~\ref{relations}, top panel) indicating slightly different evolutionary stages, possibly owing to the difference in mass between the stars (0.690/0.605~$M_{\odot}$, Kervella et al. \cite{kervella}). While HD~201091 has smooth cycles of about 7-yr, which slightly changes with time, its later and less massive companion HD~201092 has more erratic, long-term variability (cf. Figs.~\ref{stft_simple} and ~\ref{stft_compl}). This different pattern is confirmed by the coronal variations of the stars. Robrade et al. (\cite{robrade}), using X-ray observations from XMM-Newton, sampled  
 HD 201091 over the 9 years interval between 2002 and 2011, confirms the 7 years cycle found from the MW chromospheric record. HD~201092 showed irregular X-ray fluctuations around intermediate state with slightly declining overall trend during the 9-yr XMM observations that are well in line with those observed for chromospheric activity record.

\subsection{The place of the Sun among the cycling stars}

Except for the slightly faster rotation of the Sun than the MW stars of similar ages, the solar parameters and the results of the time-frequency analyses of solar datasets (sunspot number, Ca\,{\sc ii}\,K-line flux) fit well the different relations drawn for the MW stars plotted in Figs.~\ref{rot_cyc_teff},~\ref{rot_cyc} and ~\ref{relations}. One possible reason for Sun's slightly faster rotation in the sample is, that according to Sellwood and Binney (\cite{sel-bin}) the Sun is slightly metal-rich comparing with the stars of similar age, which may affected its rotational evolution via less effective magnetic braking.

Another explanation for the Sun's slightly faster rotation could come from the planetary system of the Sun. Recently, Maxted, Serenelli \& Southworth (\cite{maxted}) found evidence that in some cases, host stars of exoplanets rotate faster than expected owing to their tidal spin-up. Apart from the Sun, among the studied 29 MW stars, only HD~3651 is known to have a planet, with an orbital period of 62.2 days on a very eccentric orbit ($e=0.63$), discovered by Fischer et al. \cite{fischer} from radial velocity measurements. HD~3651 also has a distant T-dwarf companion star (e.g., Liu et al. \cite{liu}). The rotational period of HD~3651 is about 44 days, i.e., quite slow. Wittenmyer et al. (\cite{wittenmyer}) showed that in case of HD~3651 (and in 8 other cases outside of MW chromospheric activity samples) the radial velocity data can be better modeled with a planetary system of 2 planets on nearly circular orbits than with just one planet on an eccentric orbit. The second possible planet is closer to HD~3651, with a 31 days orbit. However, the two inferred planets are too small to affect the rotation of the central star.

\section{Summary}

We studied activity cycles of stars on or near the lower main sequence by analyzing observations of 29 MW stars observed for 36 years. The sample of stars covers a range of ages, with an average $\log g=4.5\pm0.17$. The older stars, older than 2--3~Gyr, form a homogeneous, slowly rotating group of 12 stars with spectral types between G5V--K5V. On average, they have similar rotational periods, of $39.7\pm6.0$ days. Their cycles look simple; the average dominant cycle length is $9.7\pm1.9$ years. In some cases, secondary, low-amplitude cycles of about 3.5--4 years also appear. The Sun itself rotates somewhat faster than the sample of 12 older stars with simple cycles. The universality of the sunspot Schwabe cycle of about 11-yr for the Sun together with the average dominant cycle of about 10-yr of the 12 MW stars fits well with the recent popular attribution of the sunspot cycle as ``Nature's Third Cycle" by Choudhuri (\cite{choudhuri}). 

In contrast, the younger 16 stars have, on average, much faster rotation of $18.1\pm12.2$ days with a large dispersion, and shorter, complex cycles of $7.6\pm4.9$ years, with additional, sometimes multiple, smaller amplitude cycles. The range of the spectral types for this group is somewhat wider (from F9.6V to K7V) than that of the older counterparts. There is a clear age separation between the two groups near the Vaughan-Preston gap (i.e., around 2--3~Gyr). This picture suggests that stars arriving on the lower main sequence are magnetically active, and show fast rotation, high mean levels of magnetic activity, and complex interannual variations in activity. As they age, magnetic braking slows the rotation of the stars, leading to simpler, cycling variations, similar to the contemporary solar Schwabe cycle.  

The change in patterns outlined may owe to a higher number of dynamo modes being excited in the fast rotating younger stars or to a more profound switch in the operation mode of the dynamo, as in a bimodal interface dynamo.

A clear implication of our results is that monitoring stellar activity variations promises a rewarding and affordable investment, demonstrated by the presence of interannual-scale activity cycles. Observational efforts targeting activity cycles need at least a few decades to get reliable results, like the Mount Wilson project that spanned 36 years, which, more than 10 years after its termination, is still an extremely valuable source of information.

\begin{acknowledgements}
Thanks are due to the anonymous referee for the useful advices and suggestions.
This work has been supported by the Hungarian Science Research Program OTKA-109276 and the Lend\"ulet-2012 Young Researchers' Programs of the Hungarian Academy of Sciences. K.O. is grateful to G. Kov\'acs for enlightening discussions about stellar ages, and to A. Mo\'or for useful advices. W.S. works were partially supported by SAO grant SAO-504514-4210-
40504514SS5000 and SAO proposal 000000000003010-V101.
\end{acknowledgements}

\begin{appendix}
\section{Additional figures}\label{appA}
 
 \begin{figure*}[t!!!]
 \centering
\includegraphics[width=4.3cm]{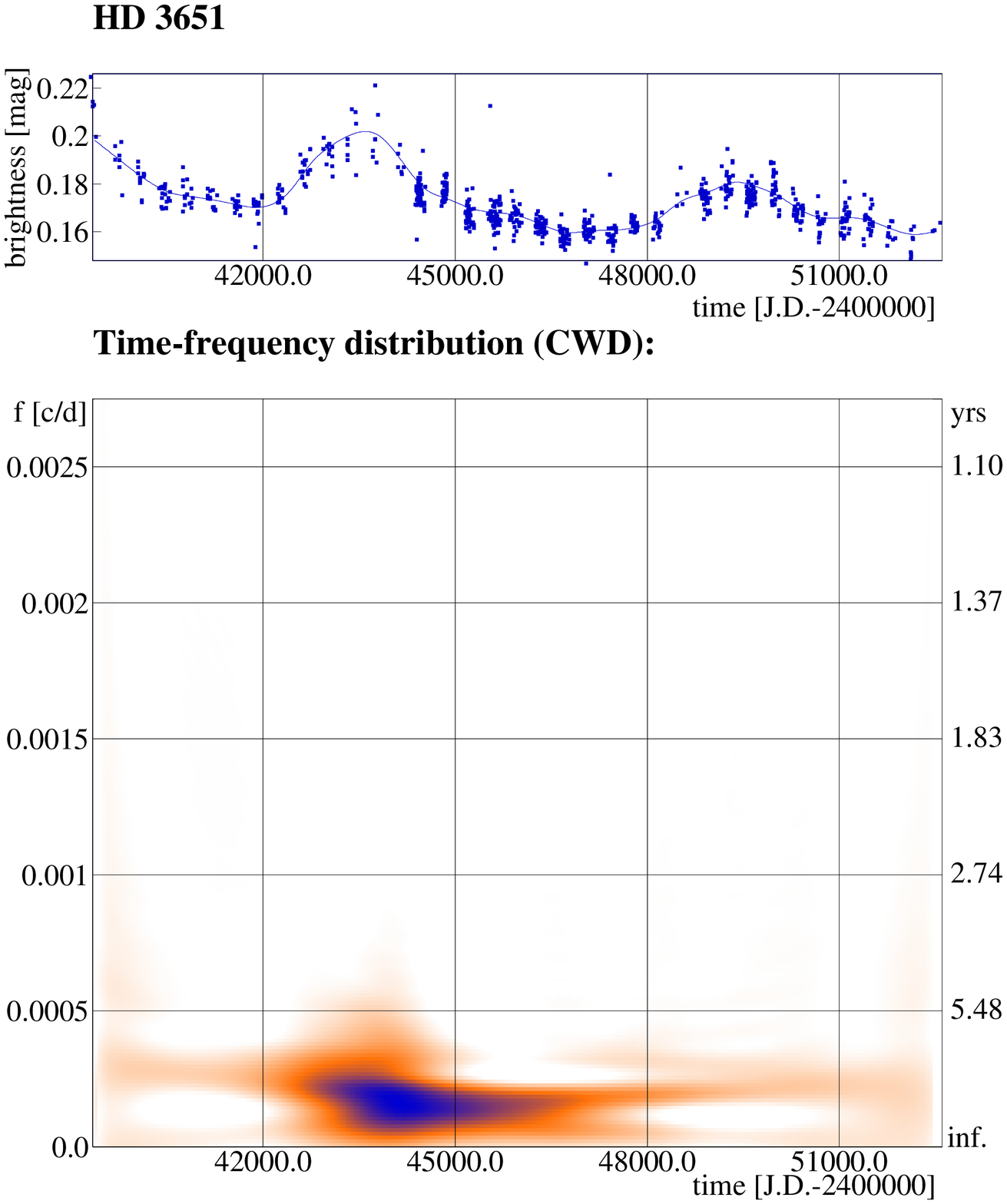}\includegraphics[width=4.3cm]{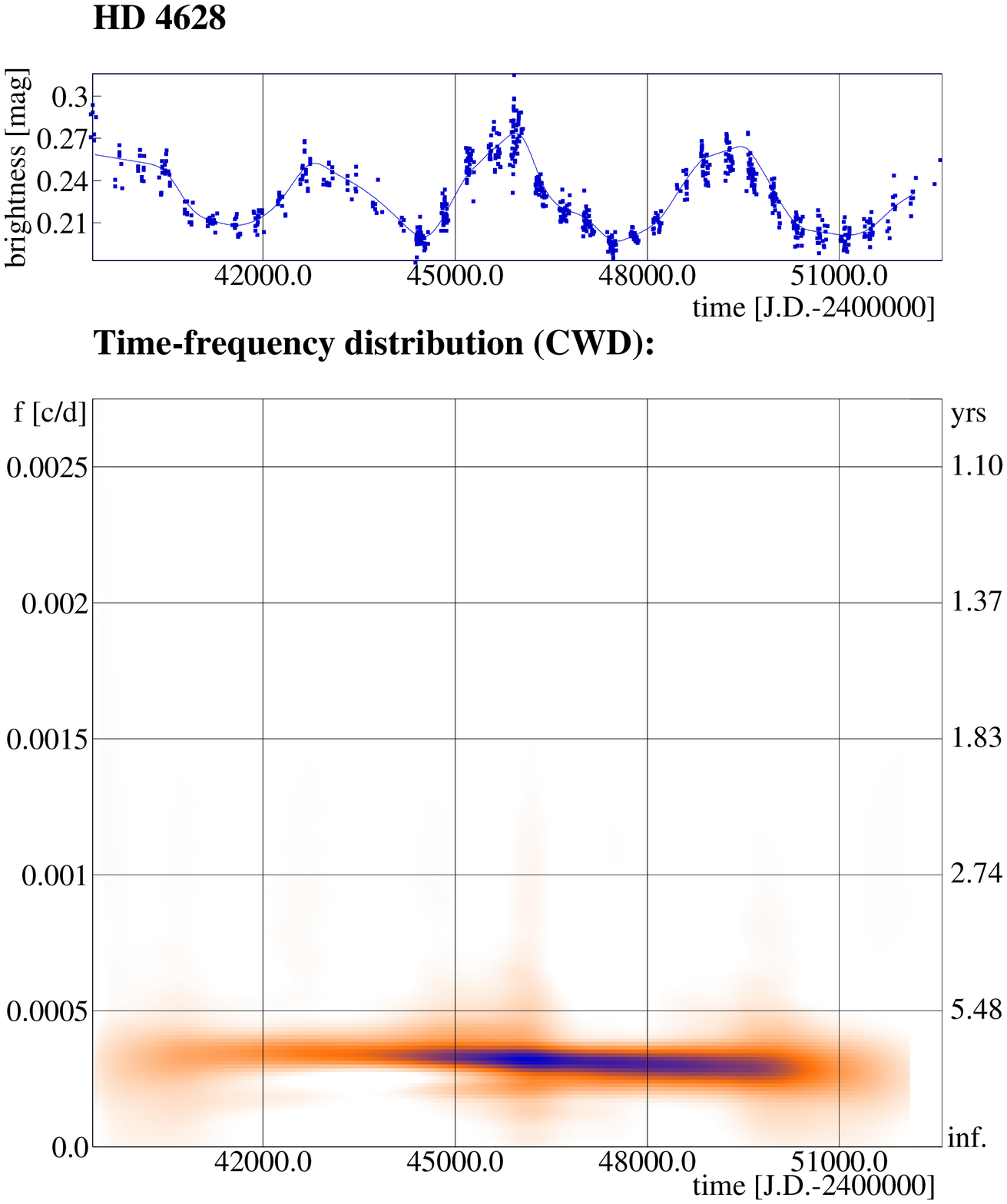}\includegraphics[width=4.3cm]{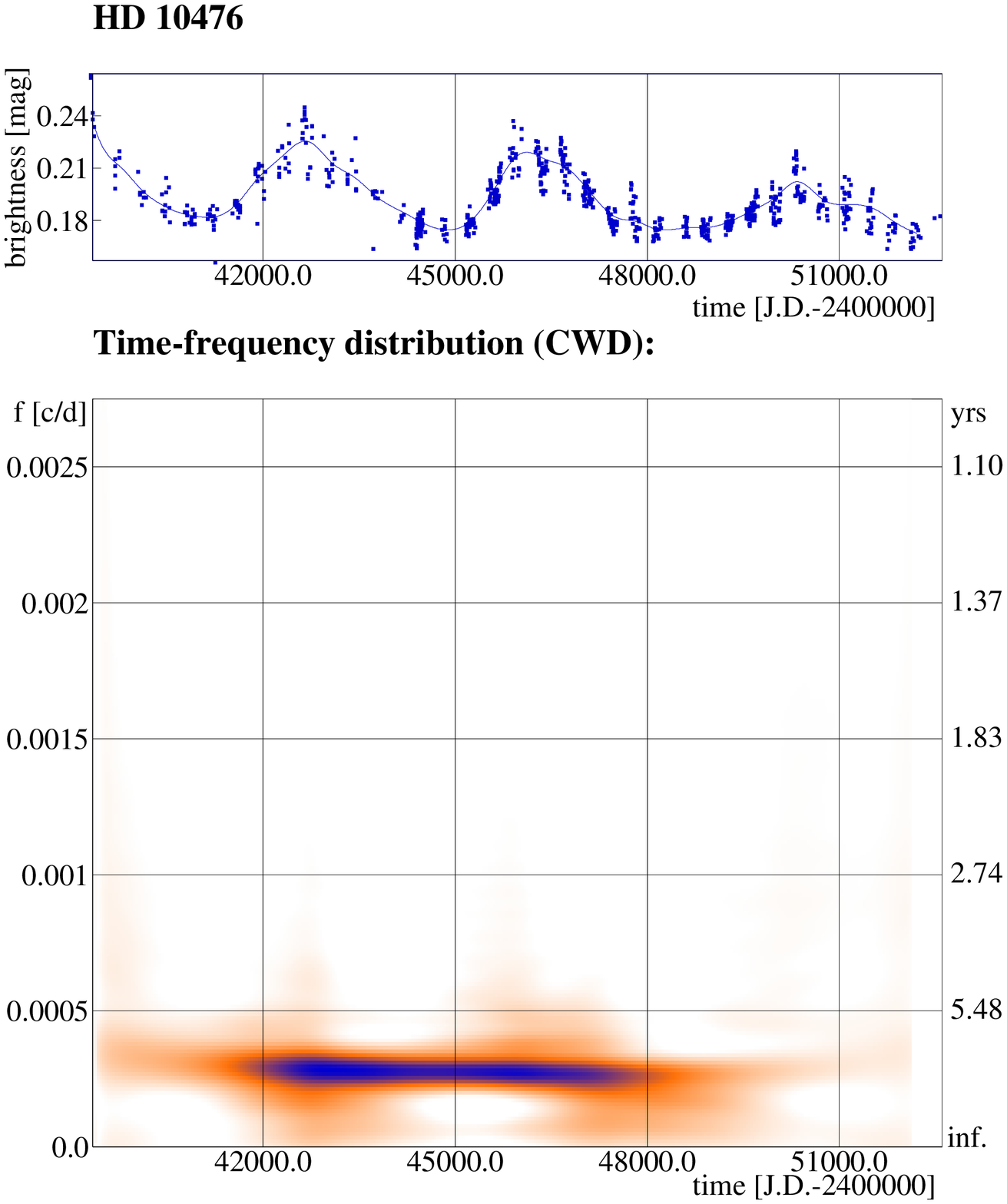}\includegraphics[width=4.3cm]{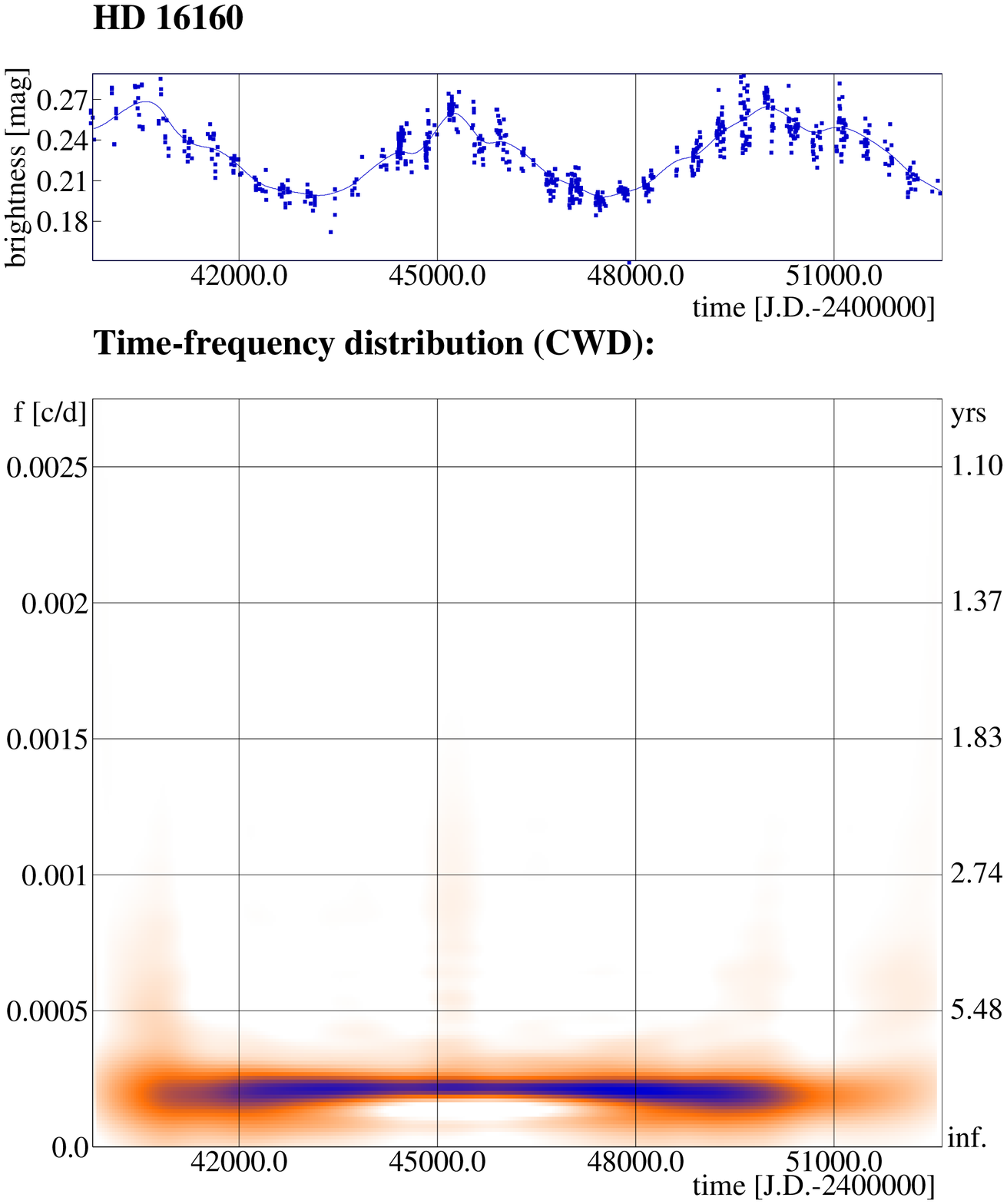}
\includegraphics[width=4.3cm]{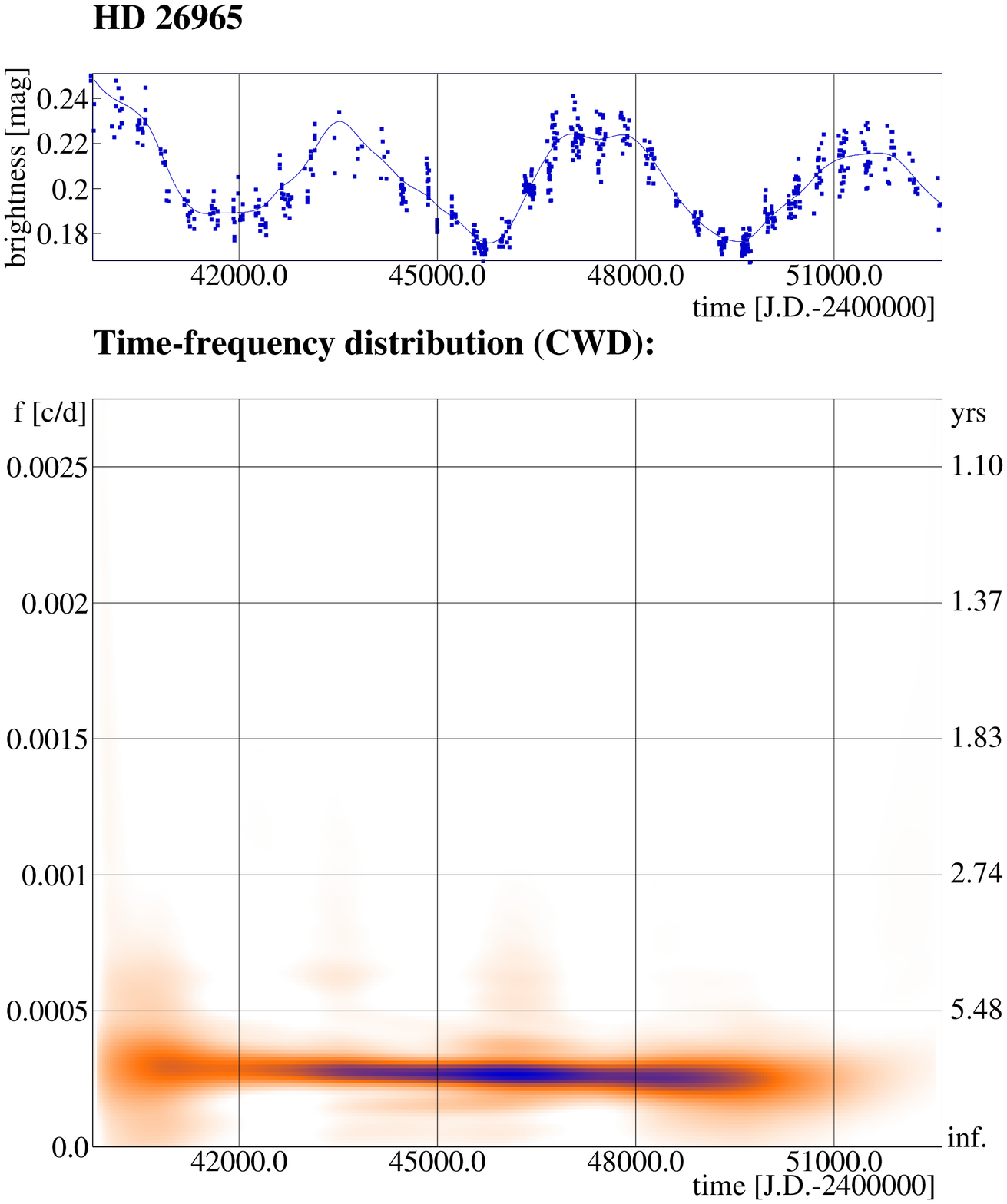}\includegraphics[width=4.3cm]{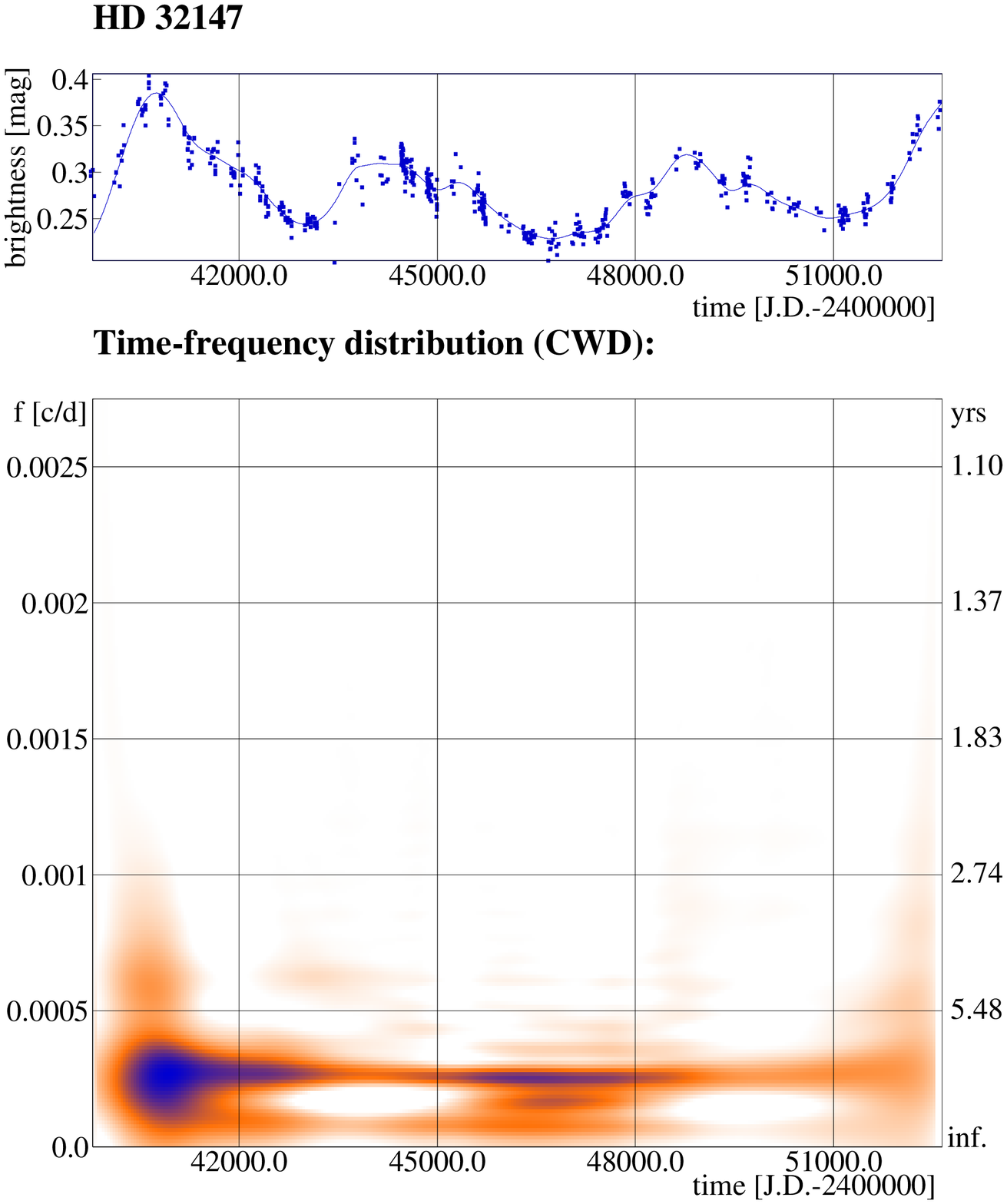}\includegraphics[width=4.3cm]{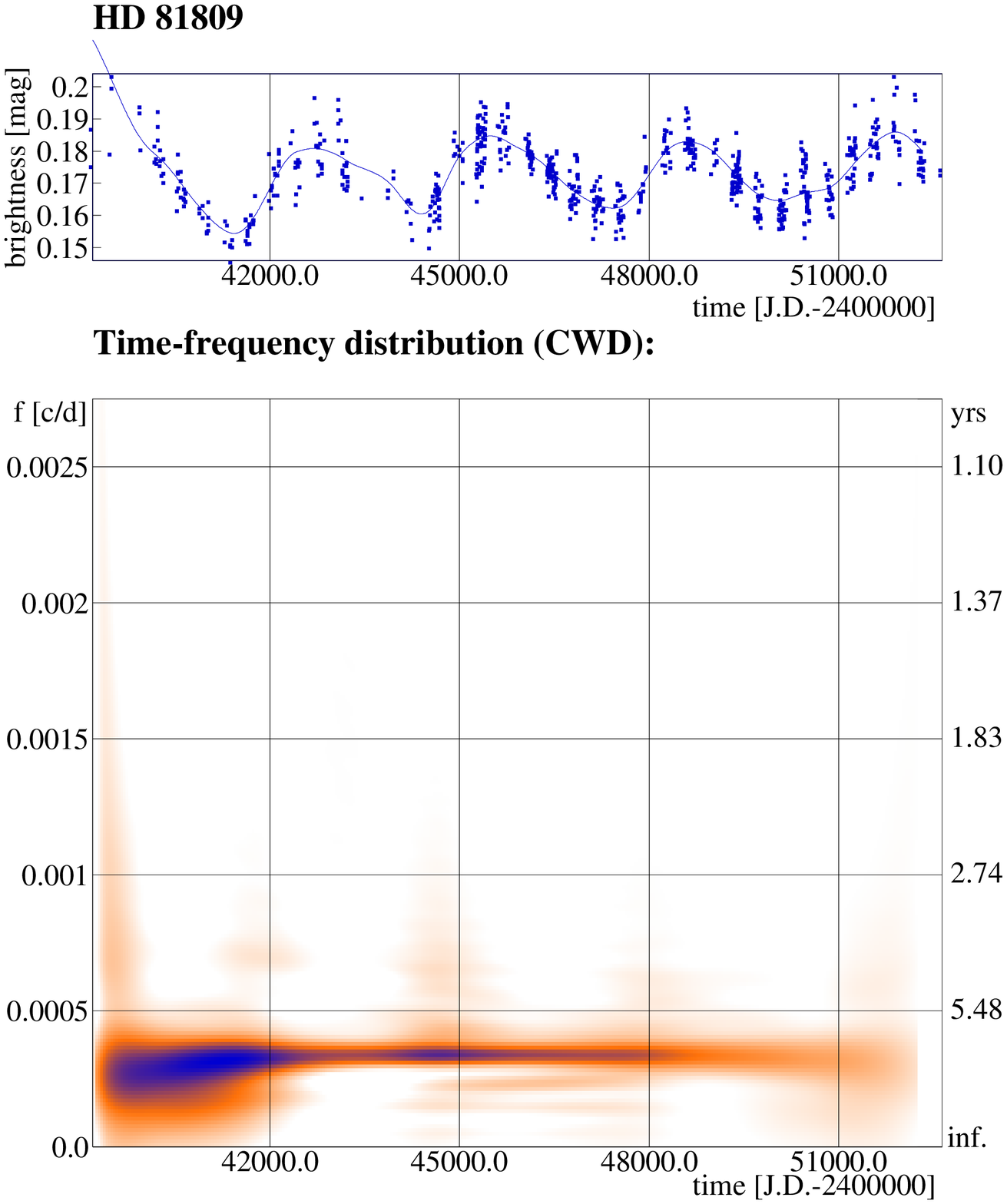}\includegraphics[width=4.3cm]{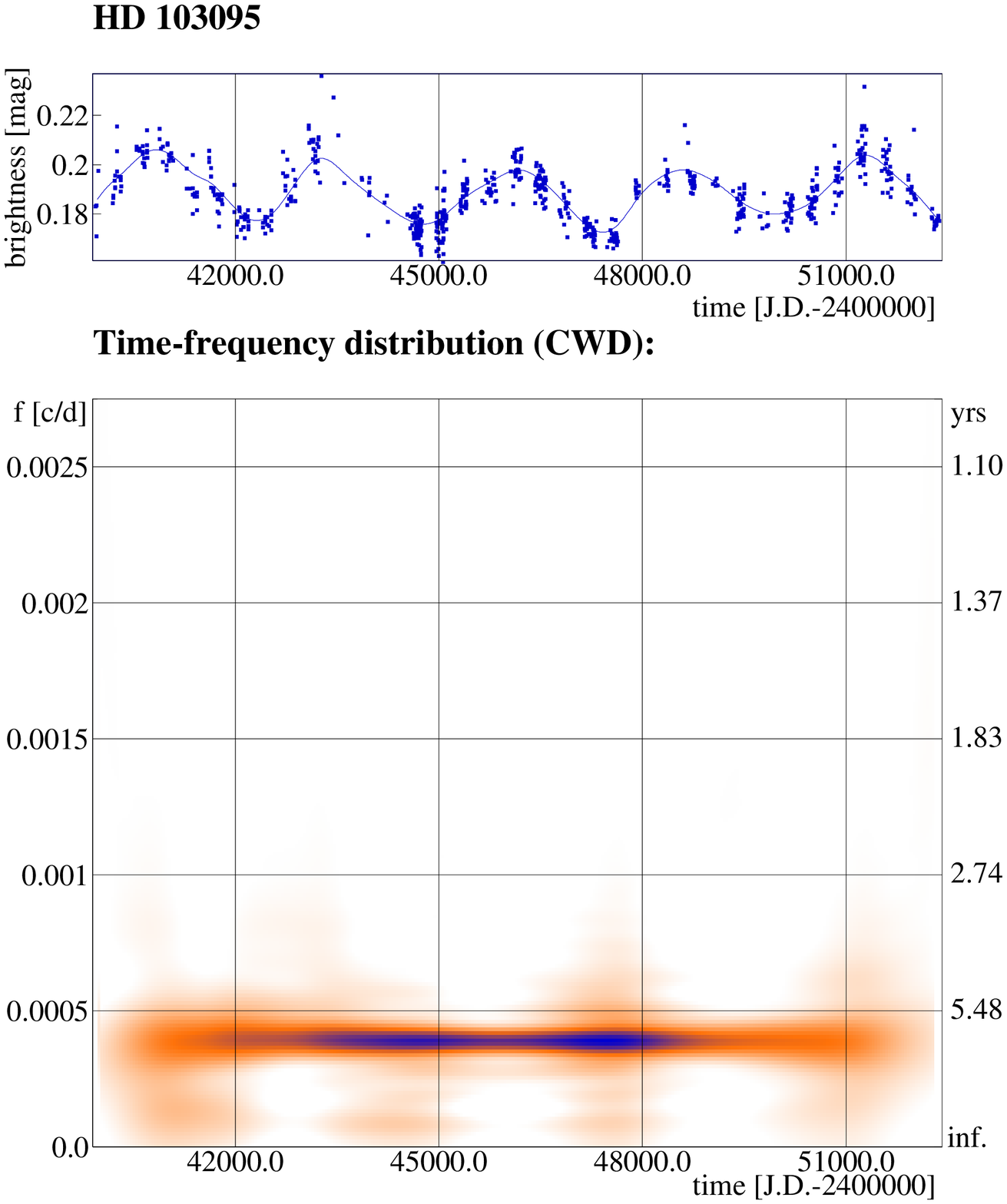}
\includegraphics[width=4.3cm]{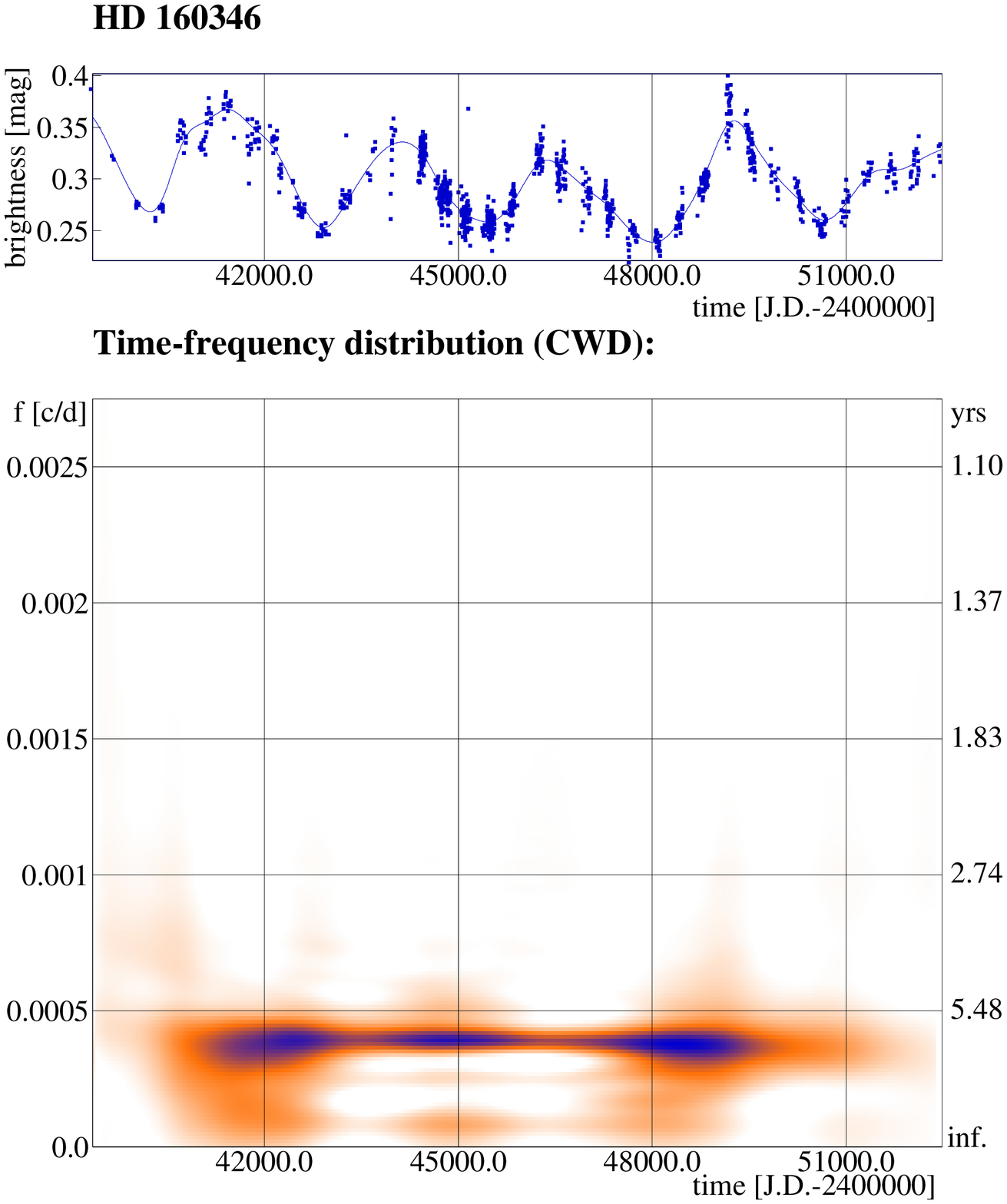}\includegraphics[width=4.3cm]{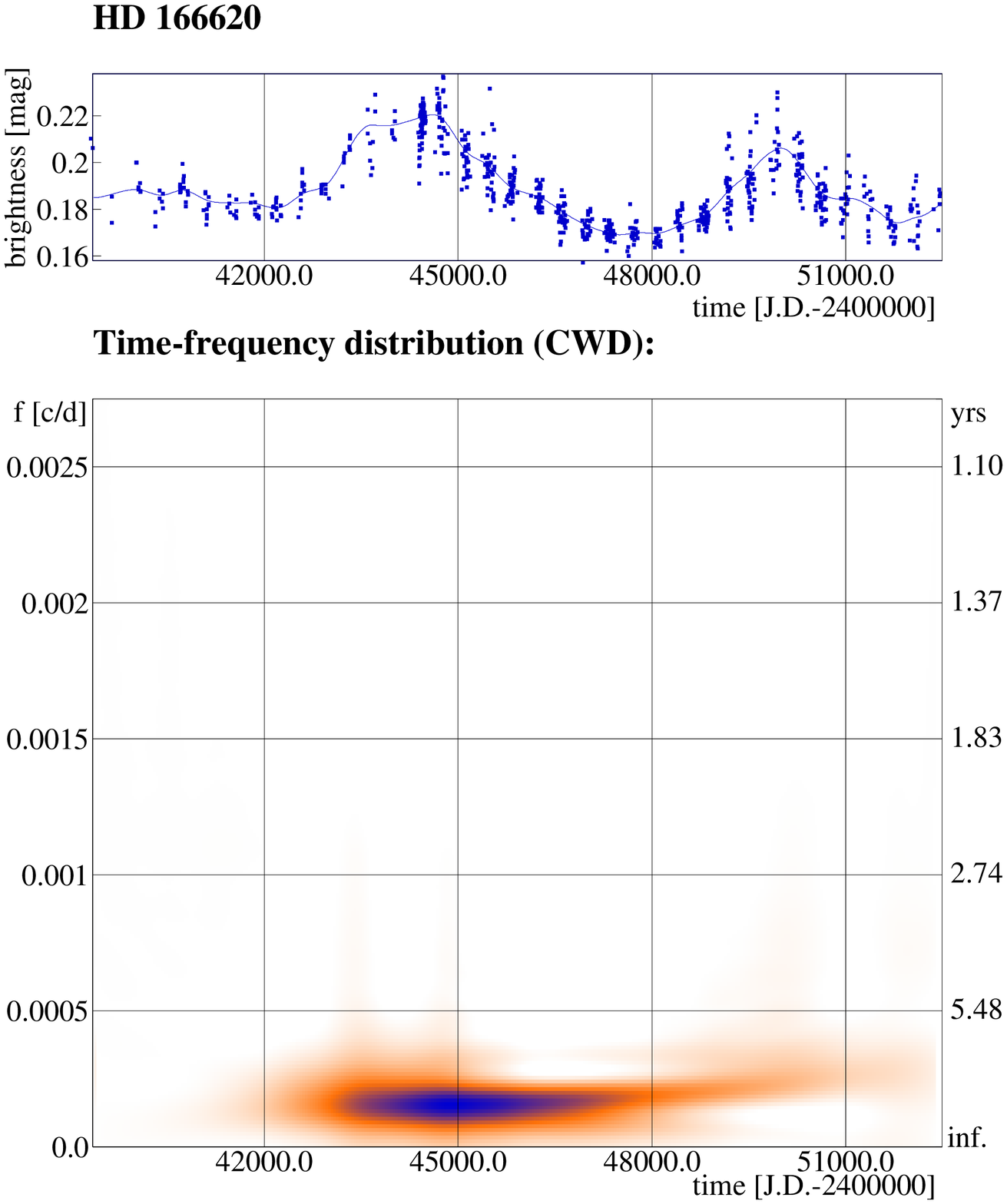}\includegraphics[width=4.3cm]{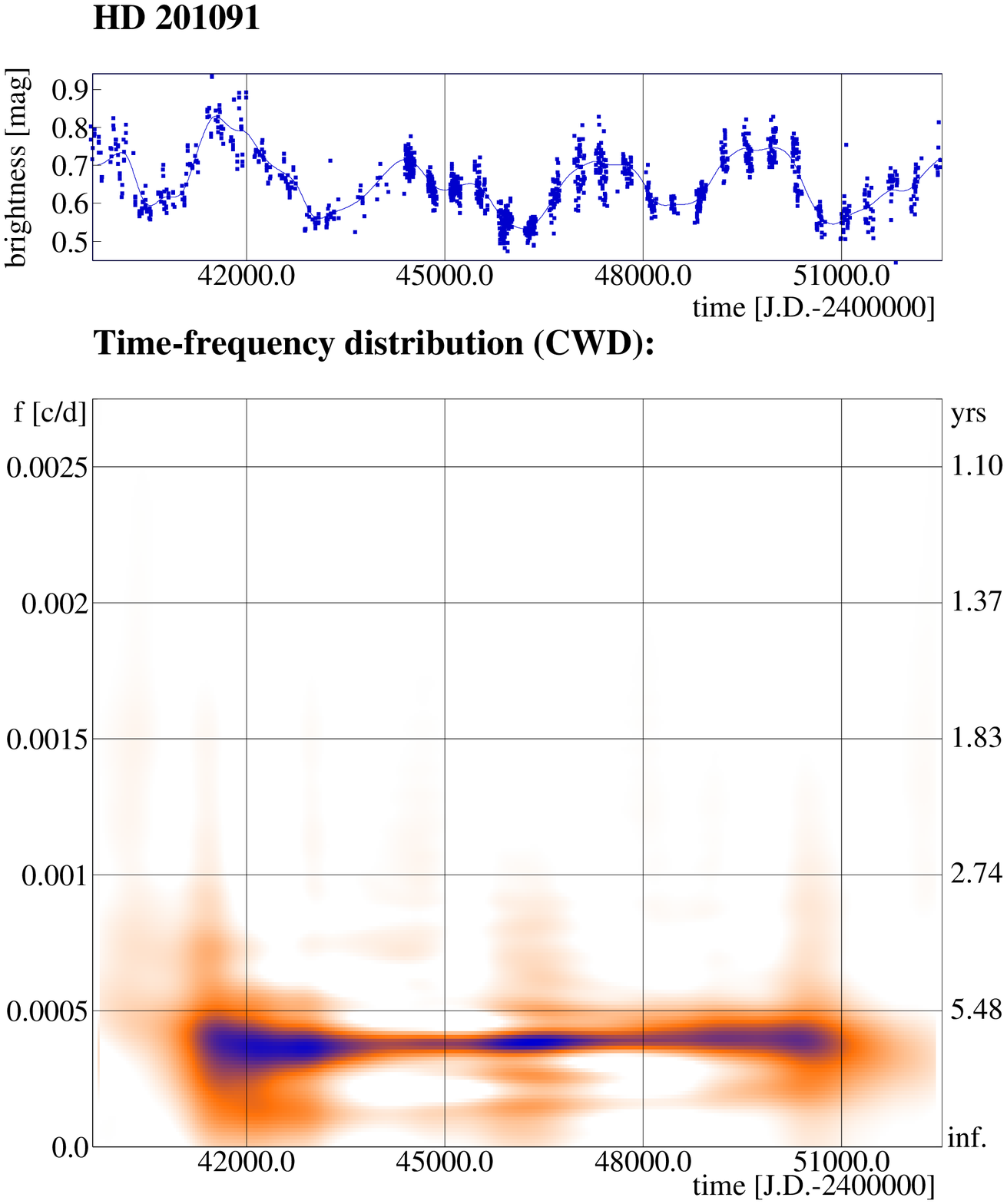}\includegraphics[width=4.3cm]{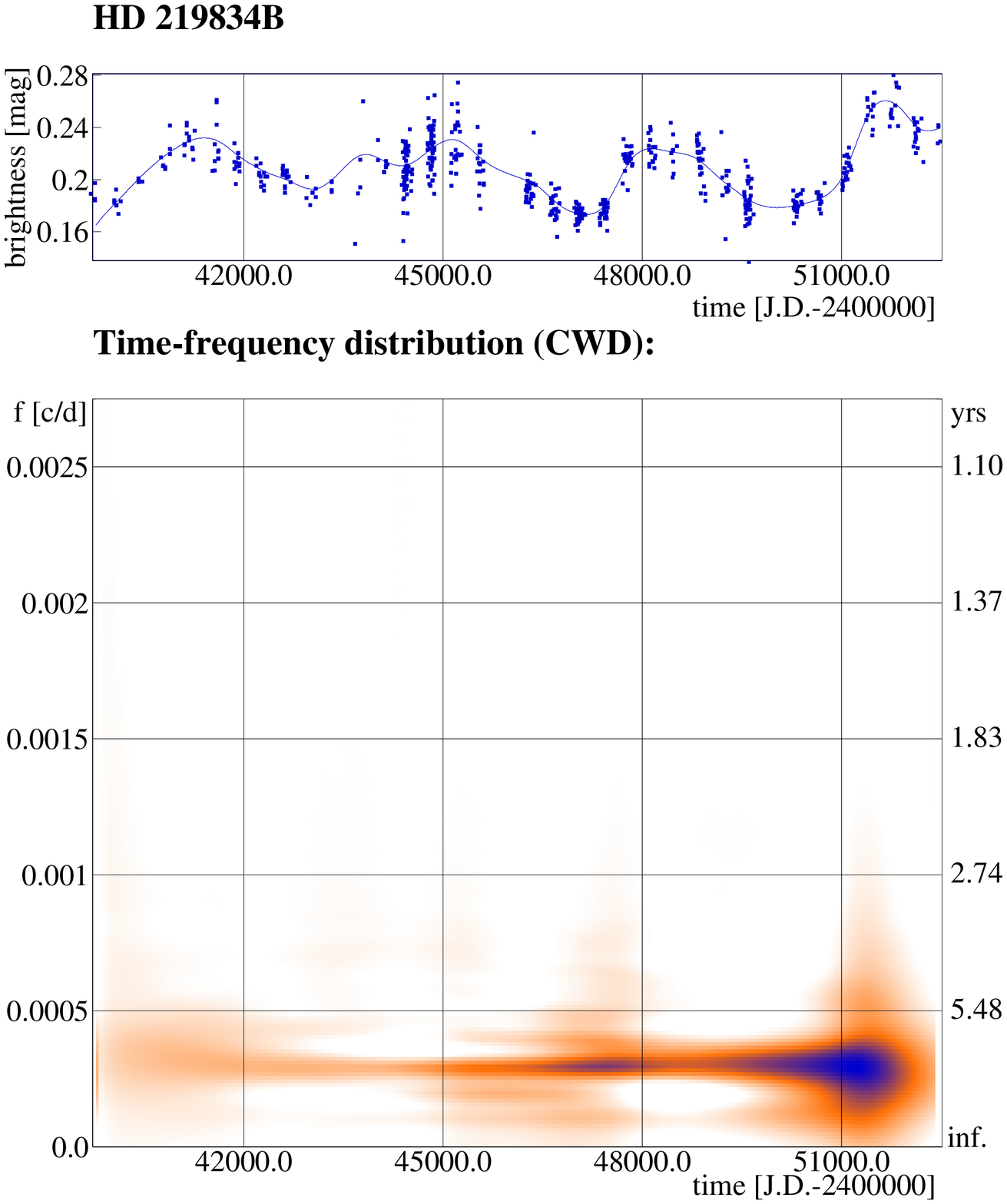}
\caption{CWD of the ``old" MW stars with simple cycles. In the upper panels the datasets and the used spline interpolation is plotted, below the time-frequency diagrams are seen.}
 \label{cwd_simple}
 \end{figure*}
 
   \begin{figure*}[tbp]
   \centering
 \includegraphics[width=4.3cm]{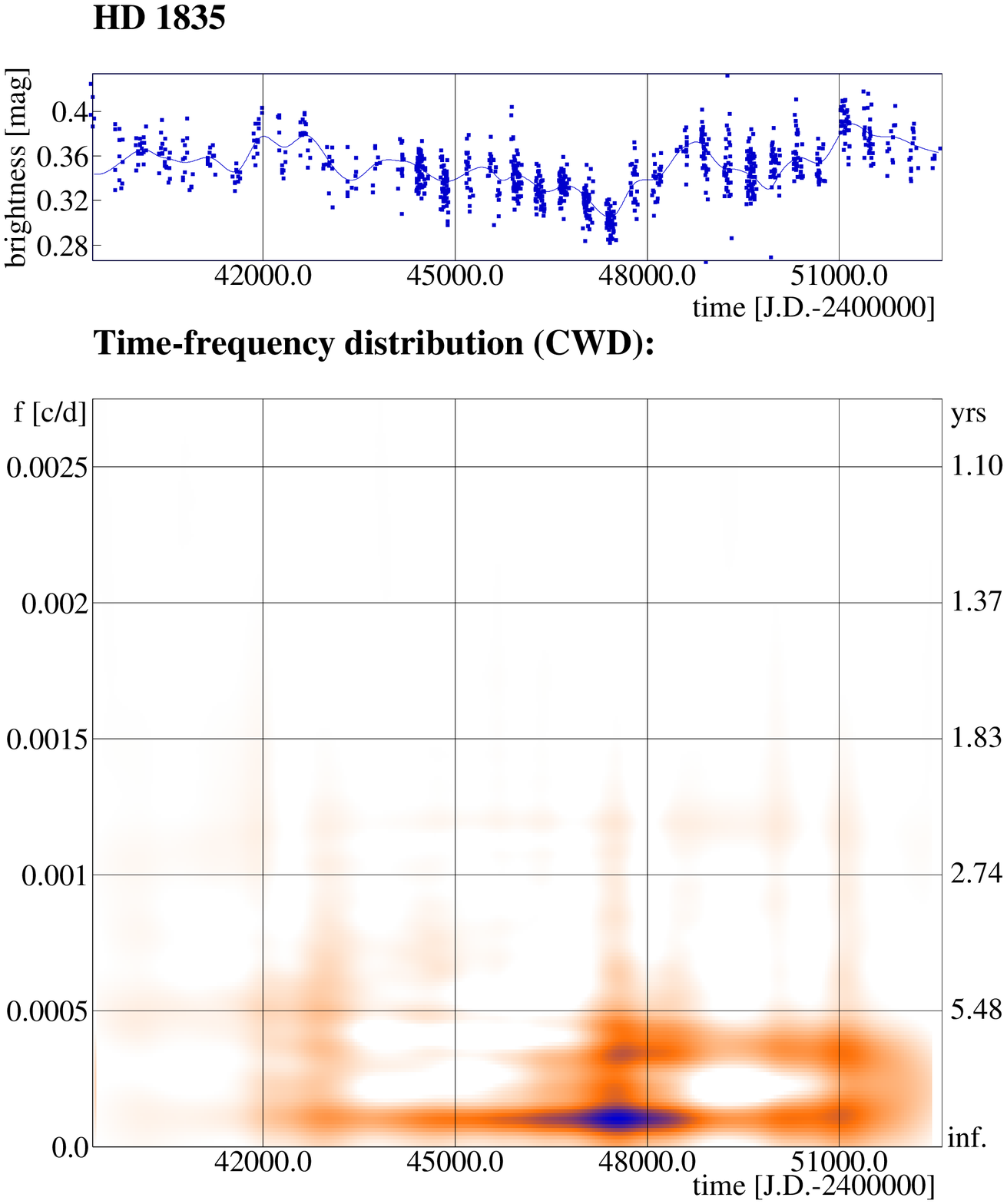}\includegraphics[width=4.3cm]{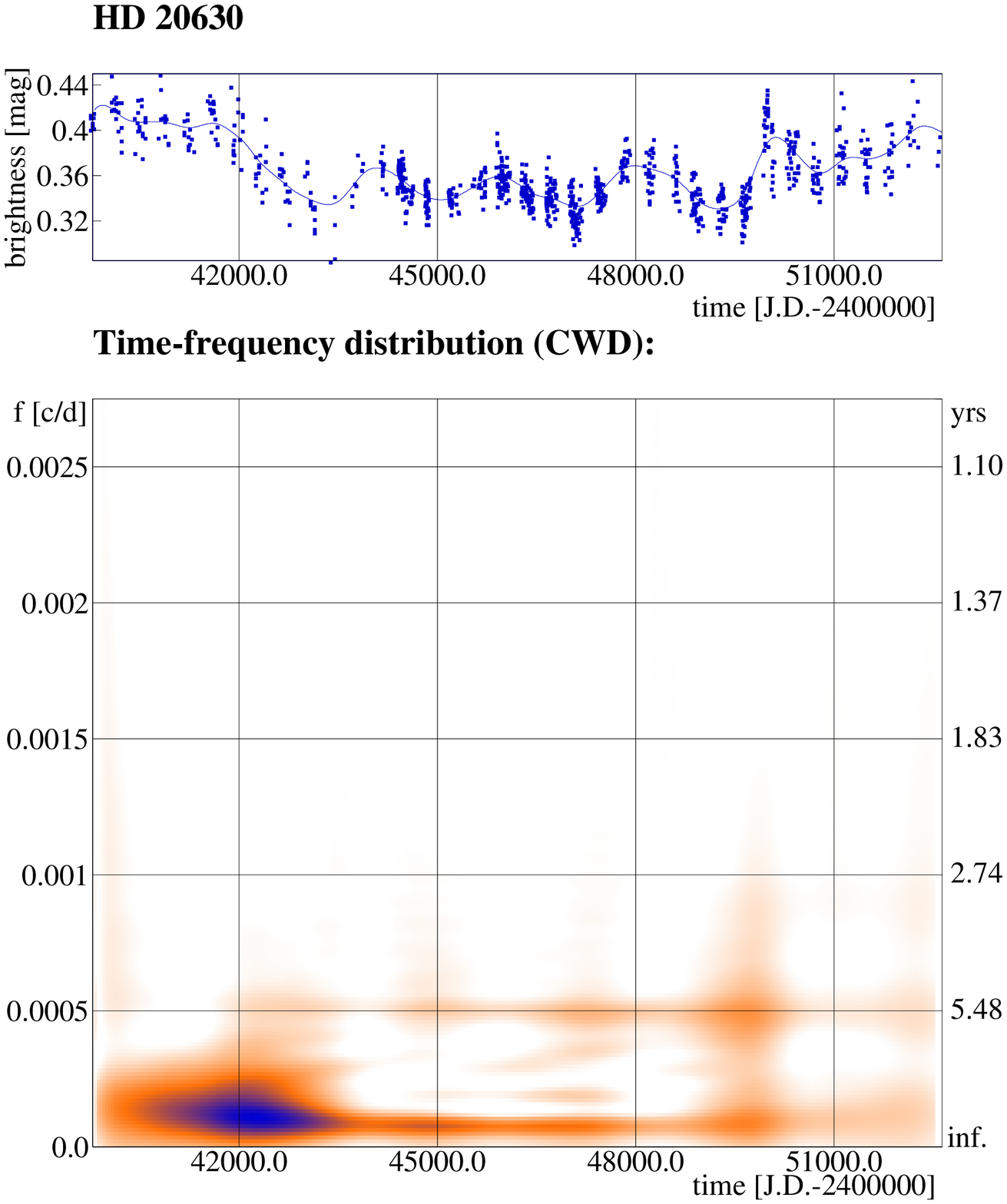}\includegraphics[width=4.3cm]{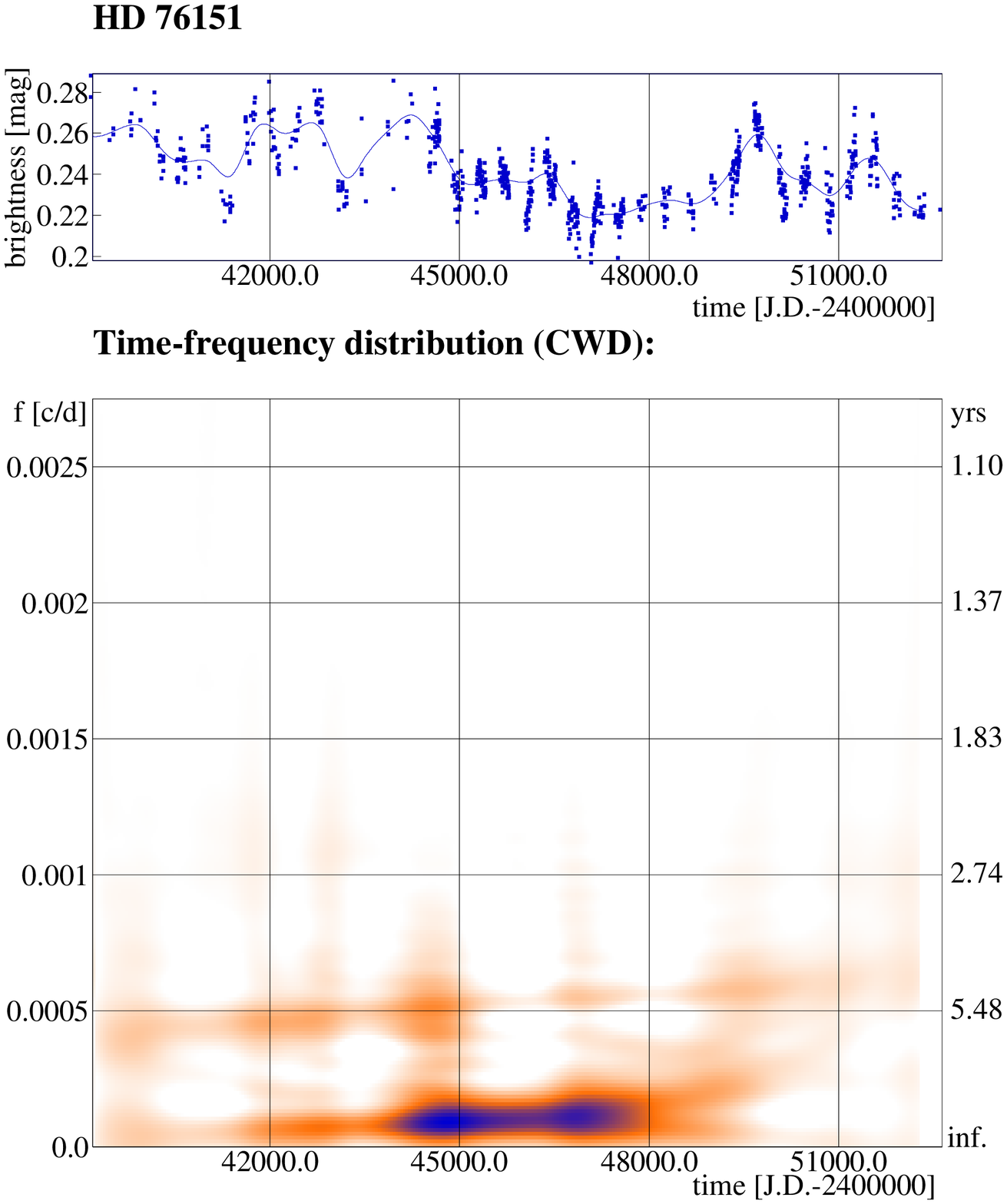}\includegraphics[width=4.3cm]{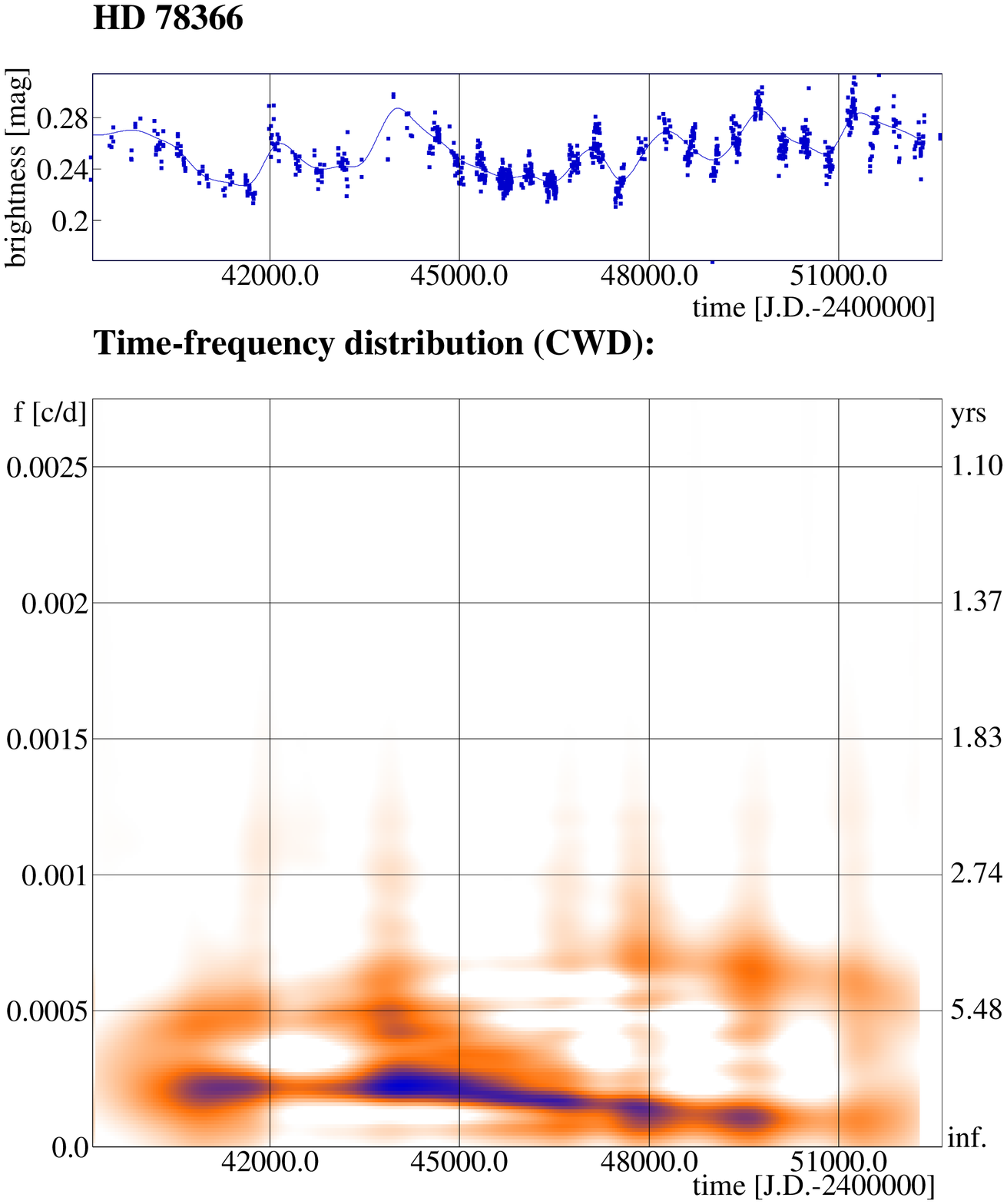}
\includegraphics[width=4.3cm]{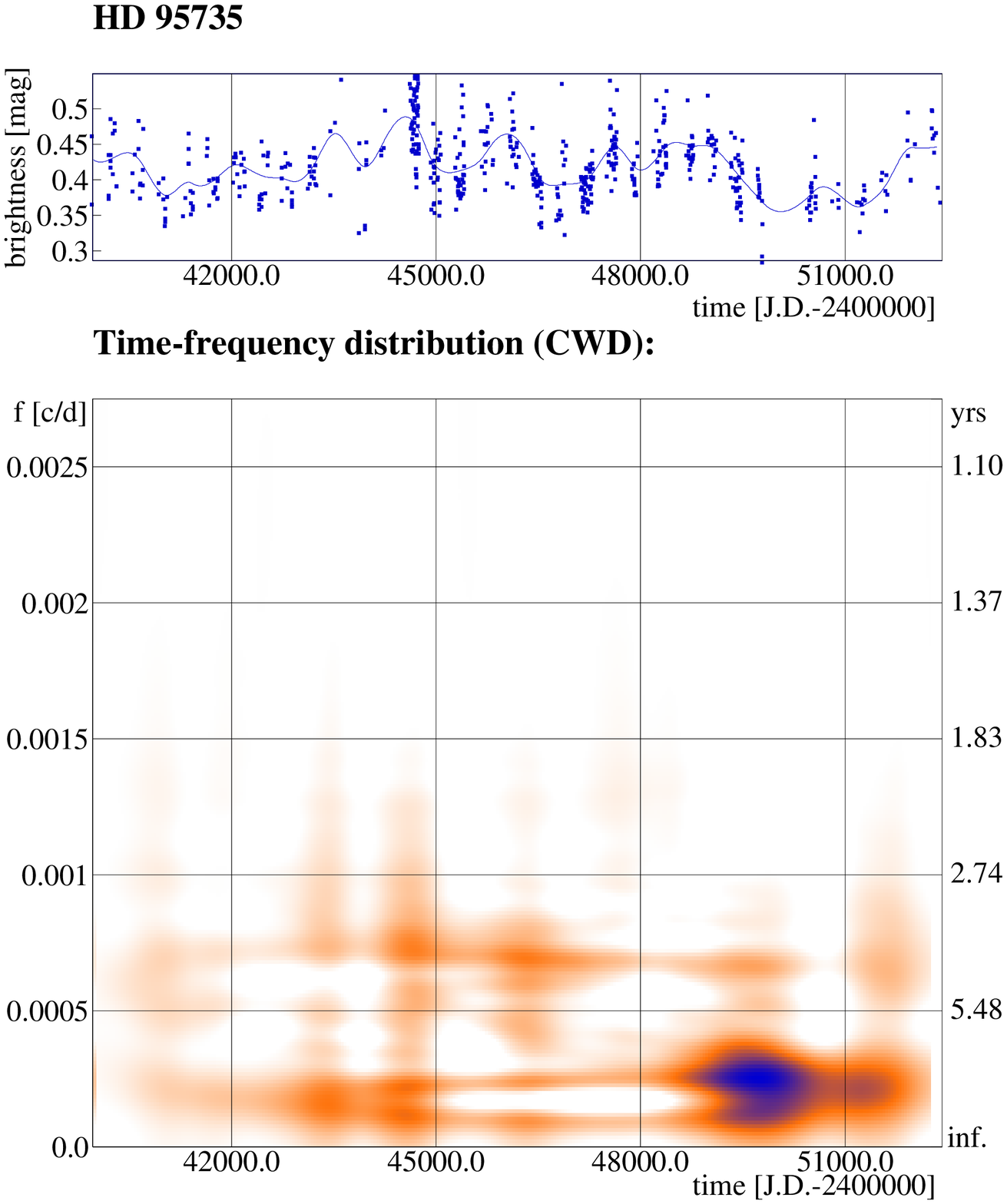}\includegraphics[width=4.3cm]{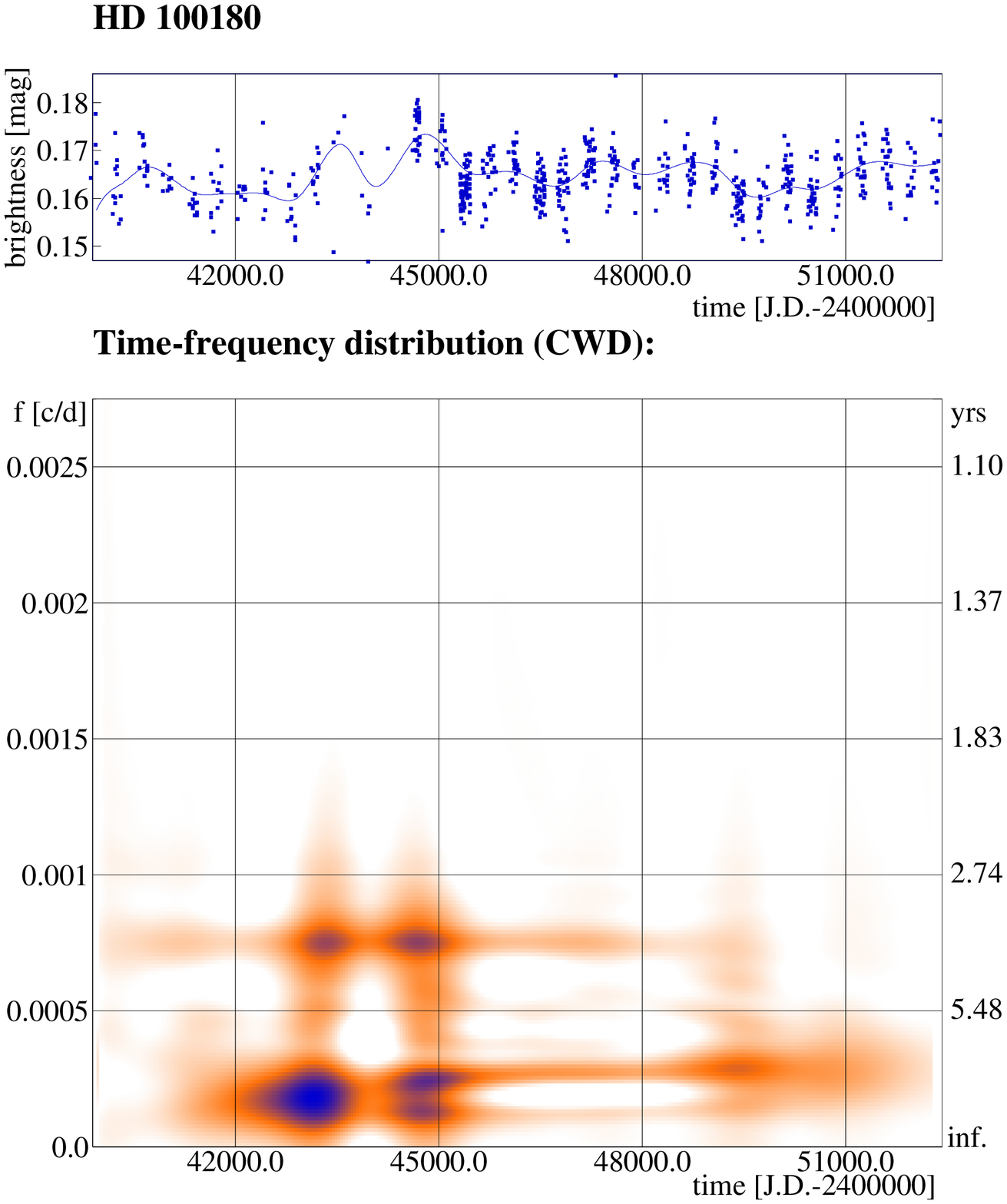}\includegraphics[width=4.3cm]{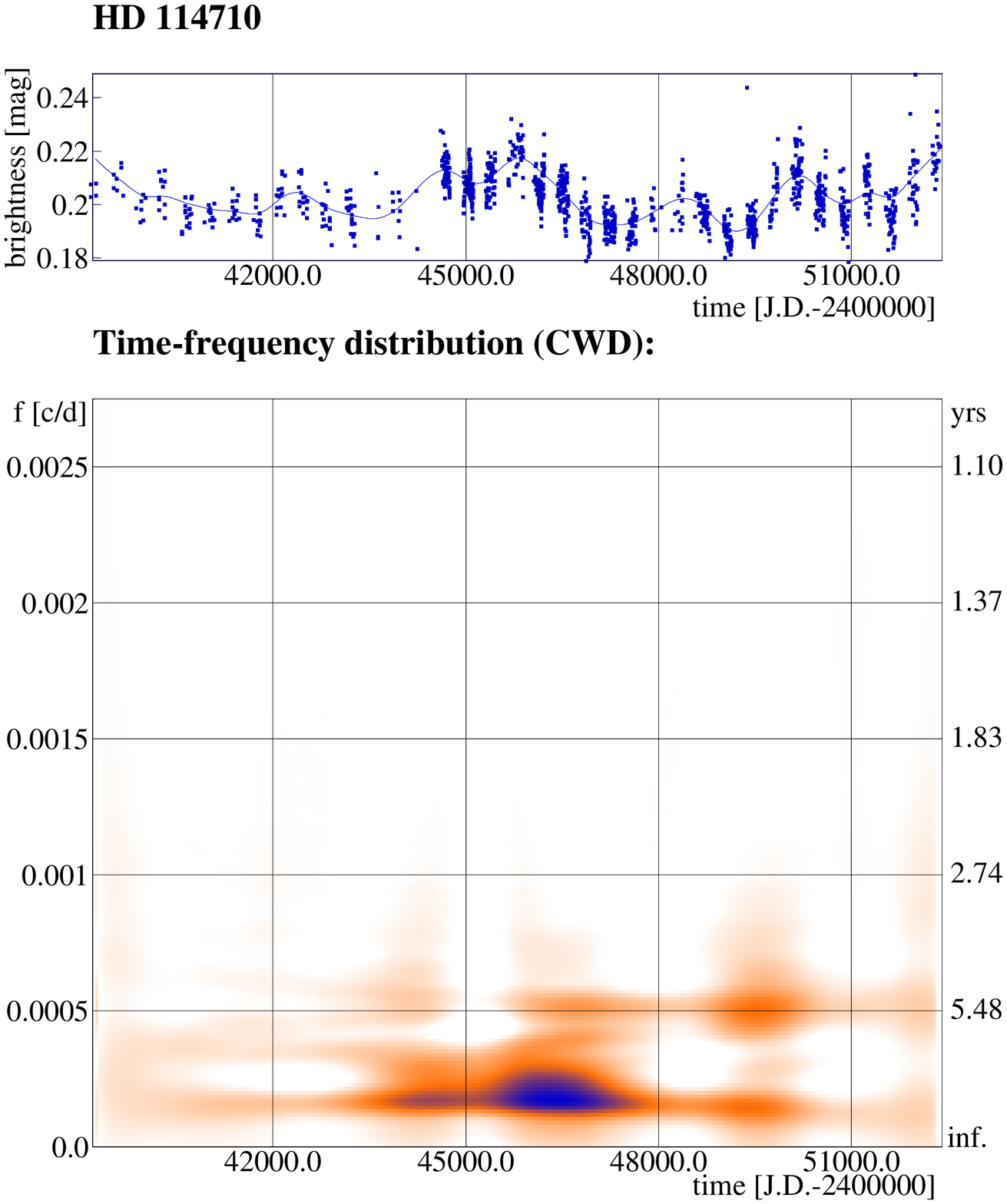}\includegraphics[width=4.3cm]{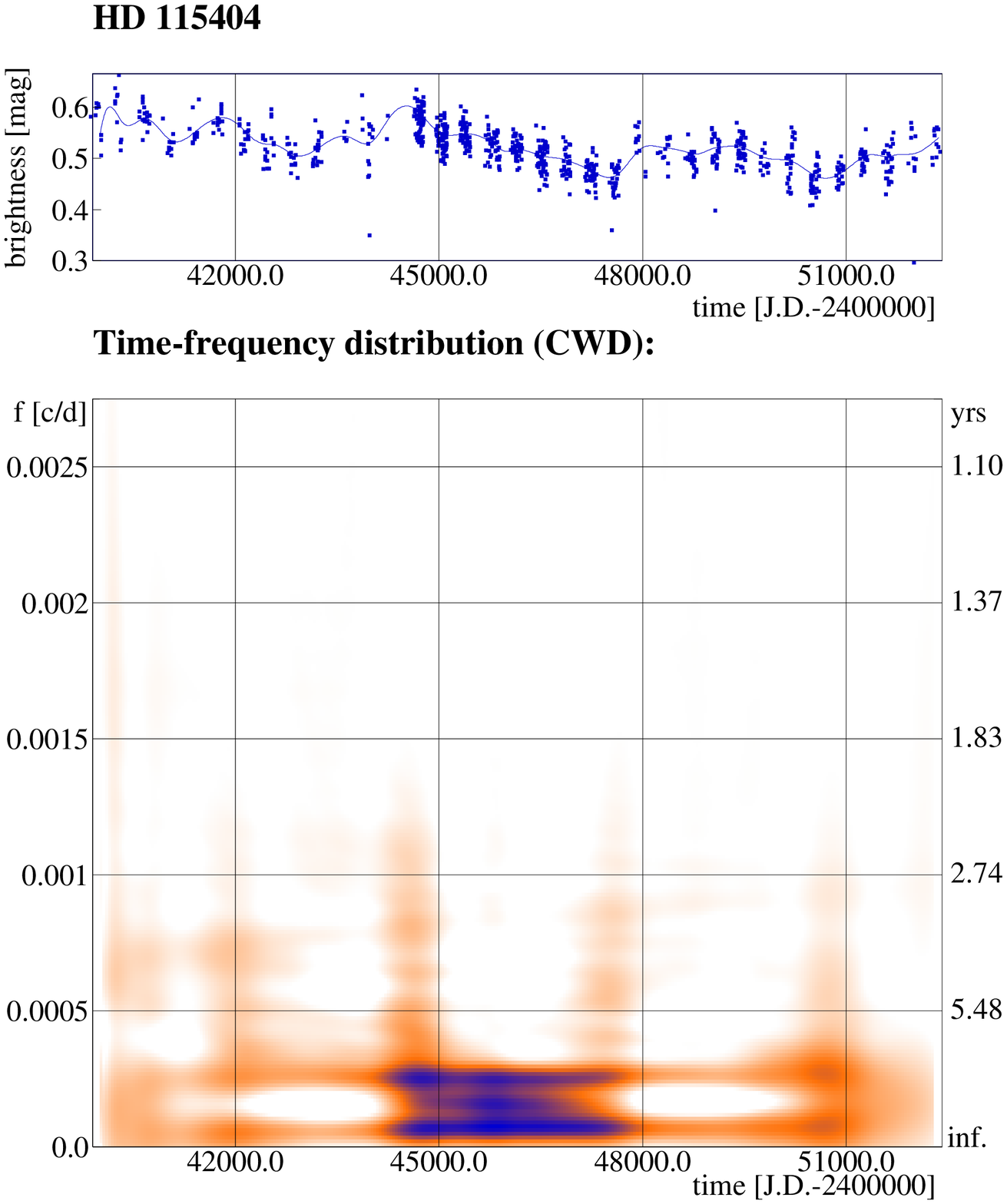}
\includegraphics[width=4.3cm]{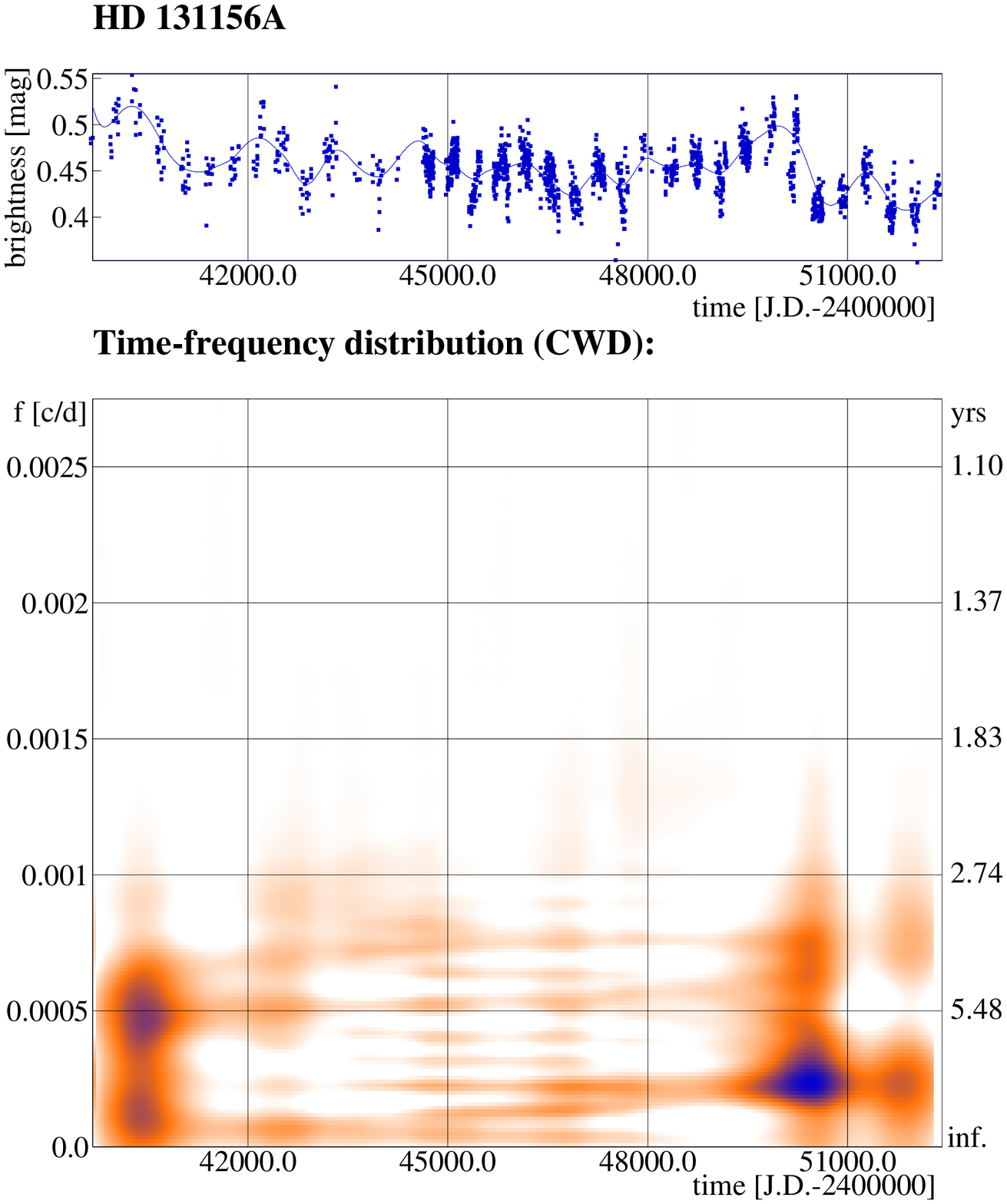}\includegraphics[width=4.3cm]{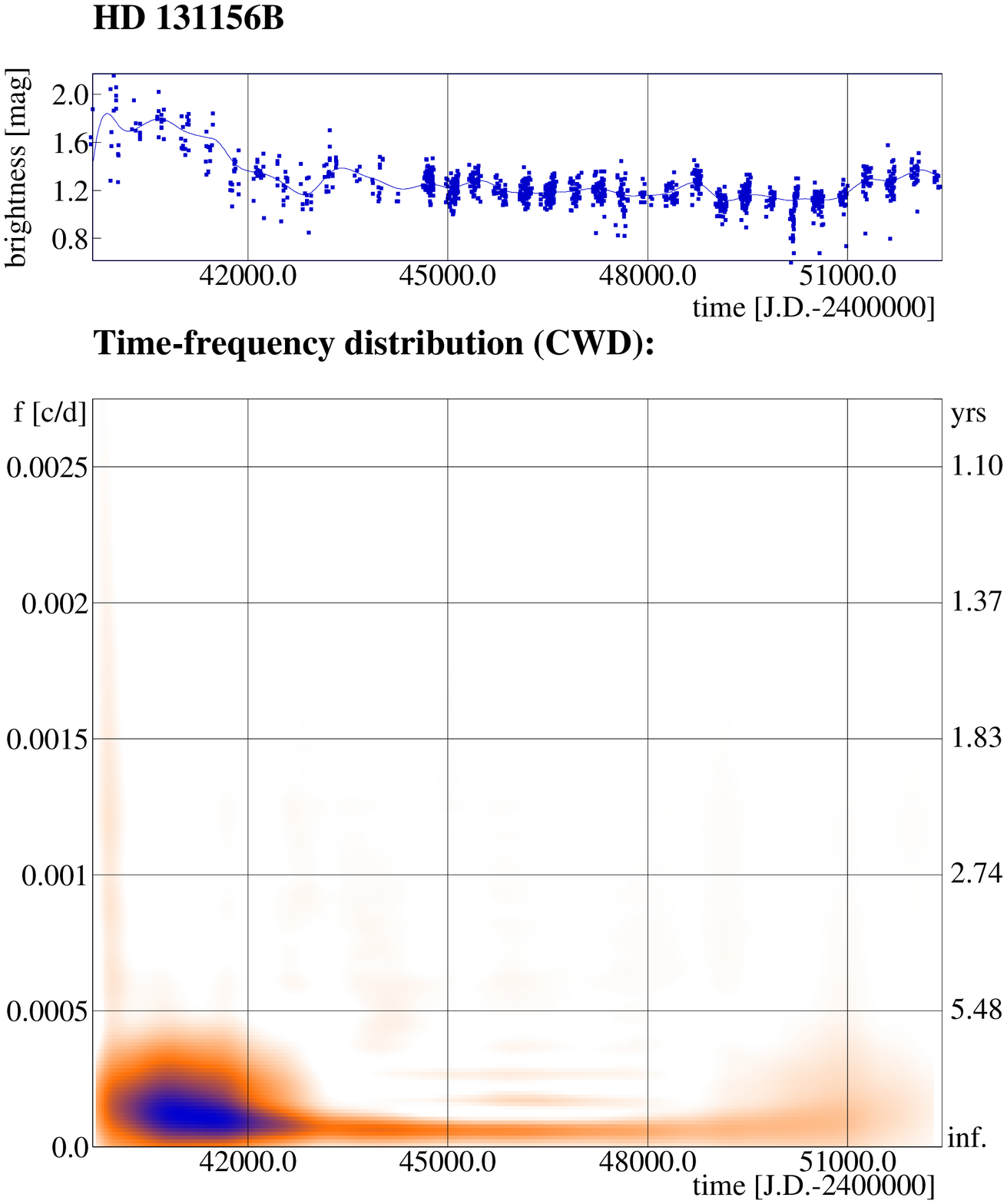}\includegraphics[width=4.3cm]{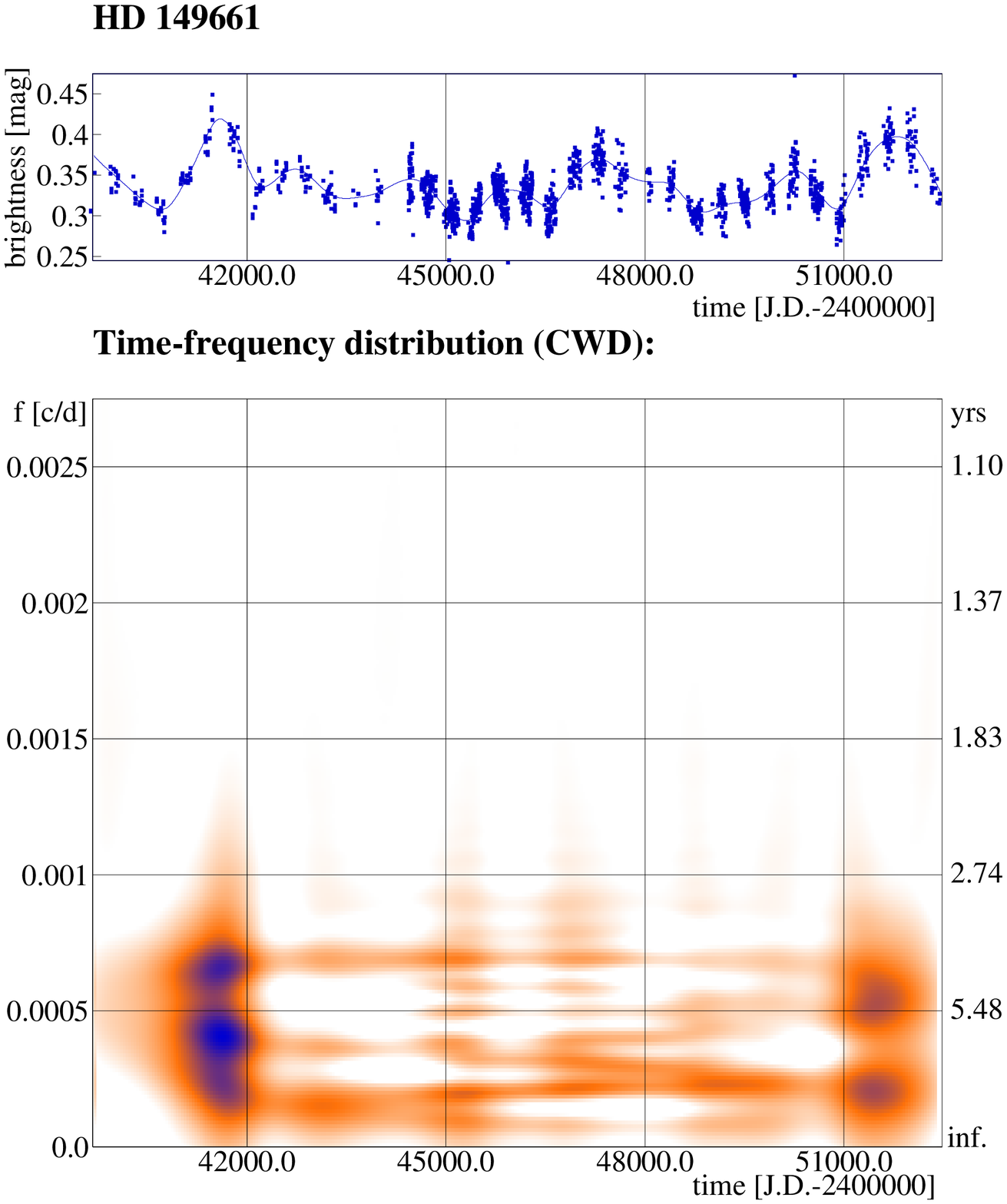}\includegraphics[width=4.3cm]{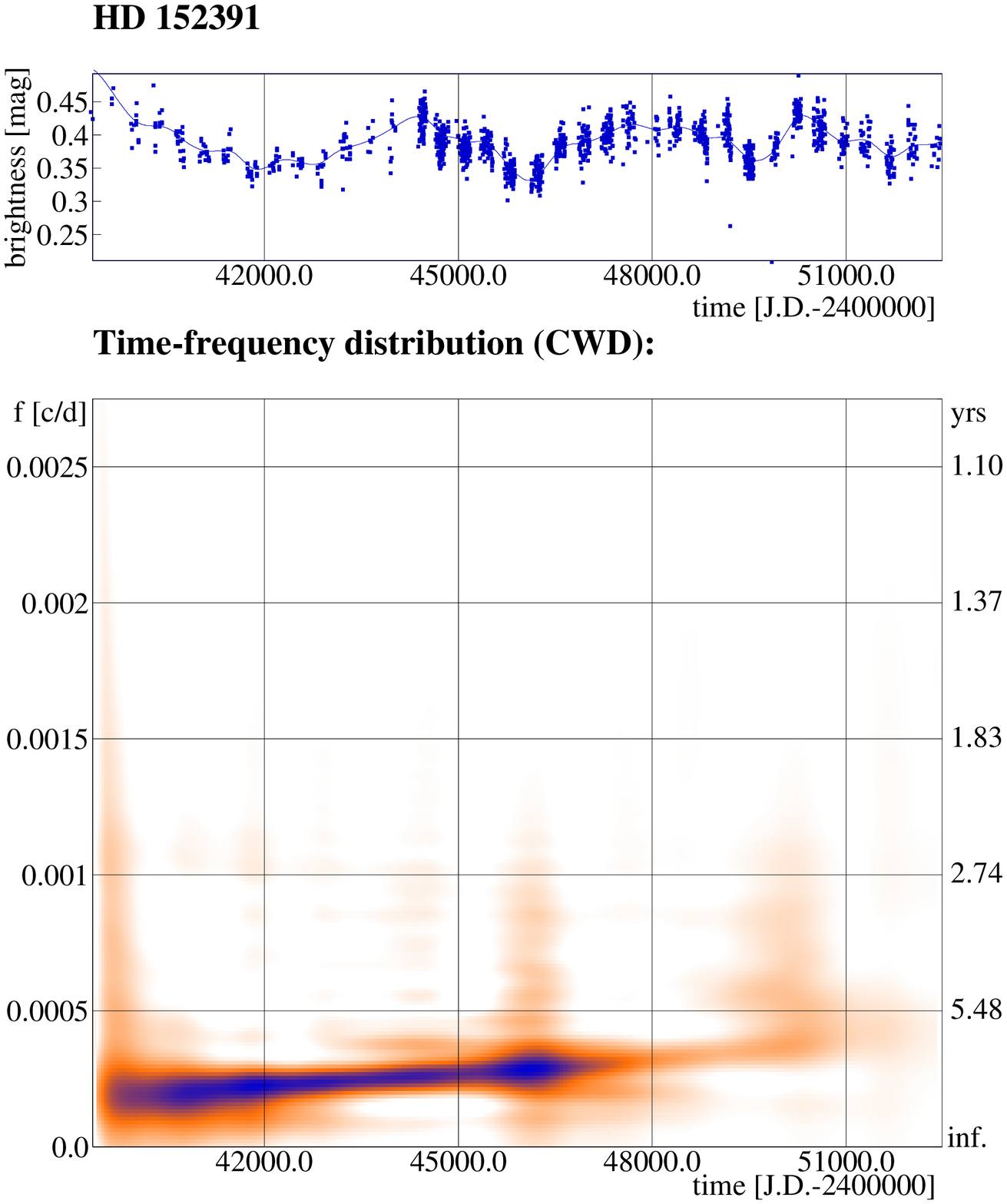}
\includegraphics[width=4.3cm]{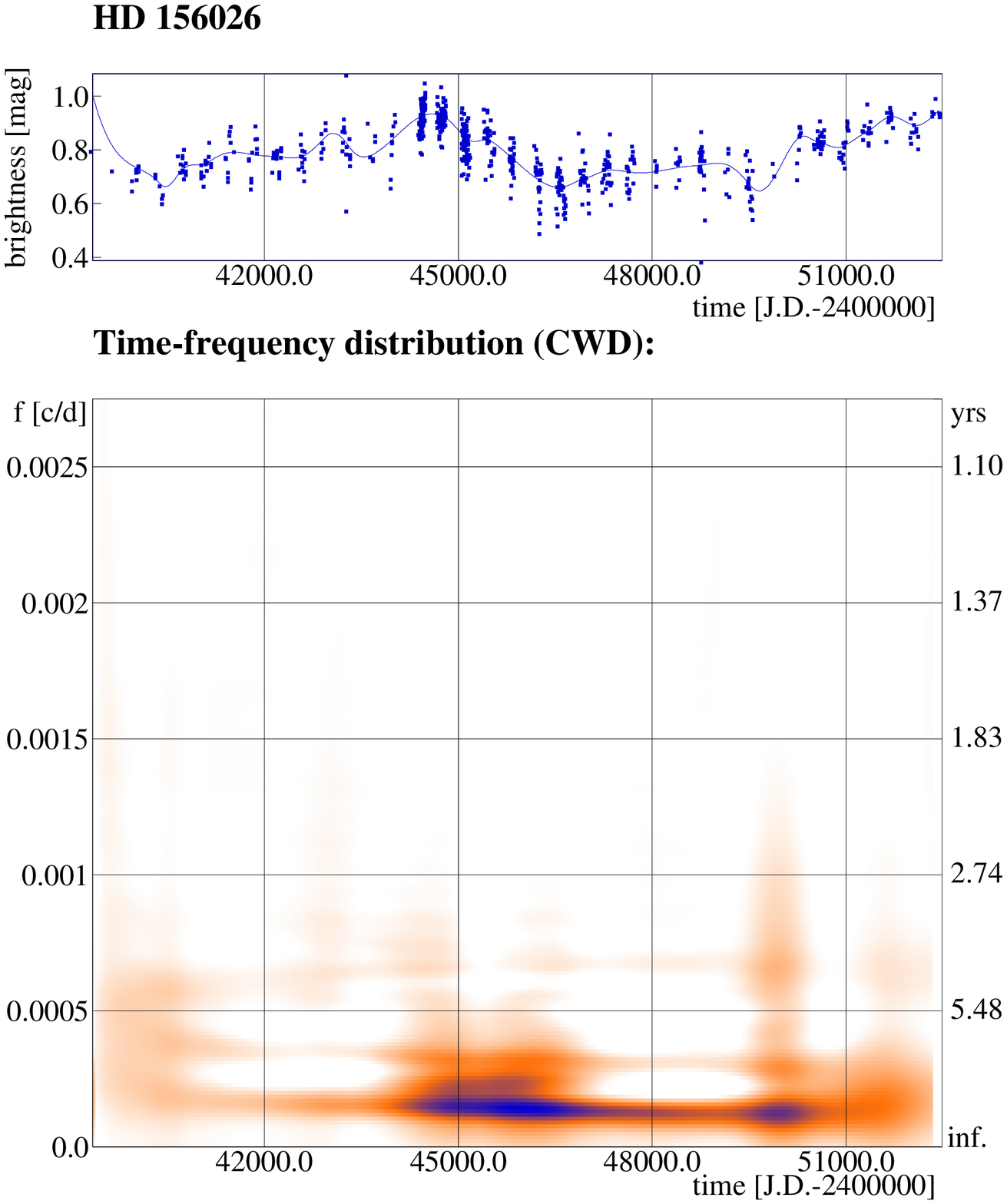}\includegraphics[width=4.3cm]{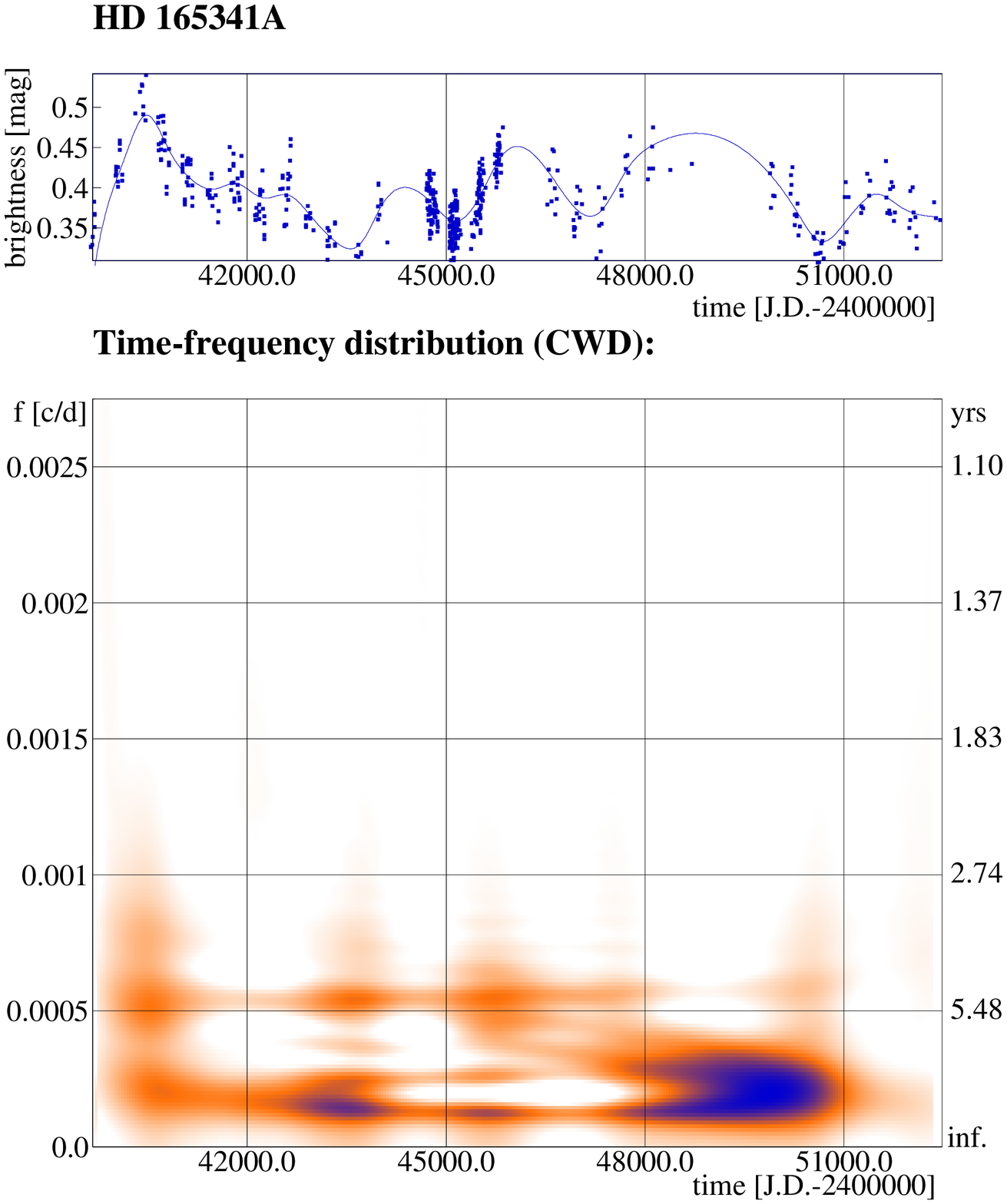}\includegraphics[width=4.3cm]{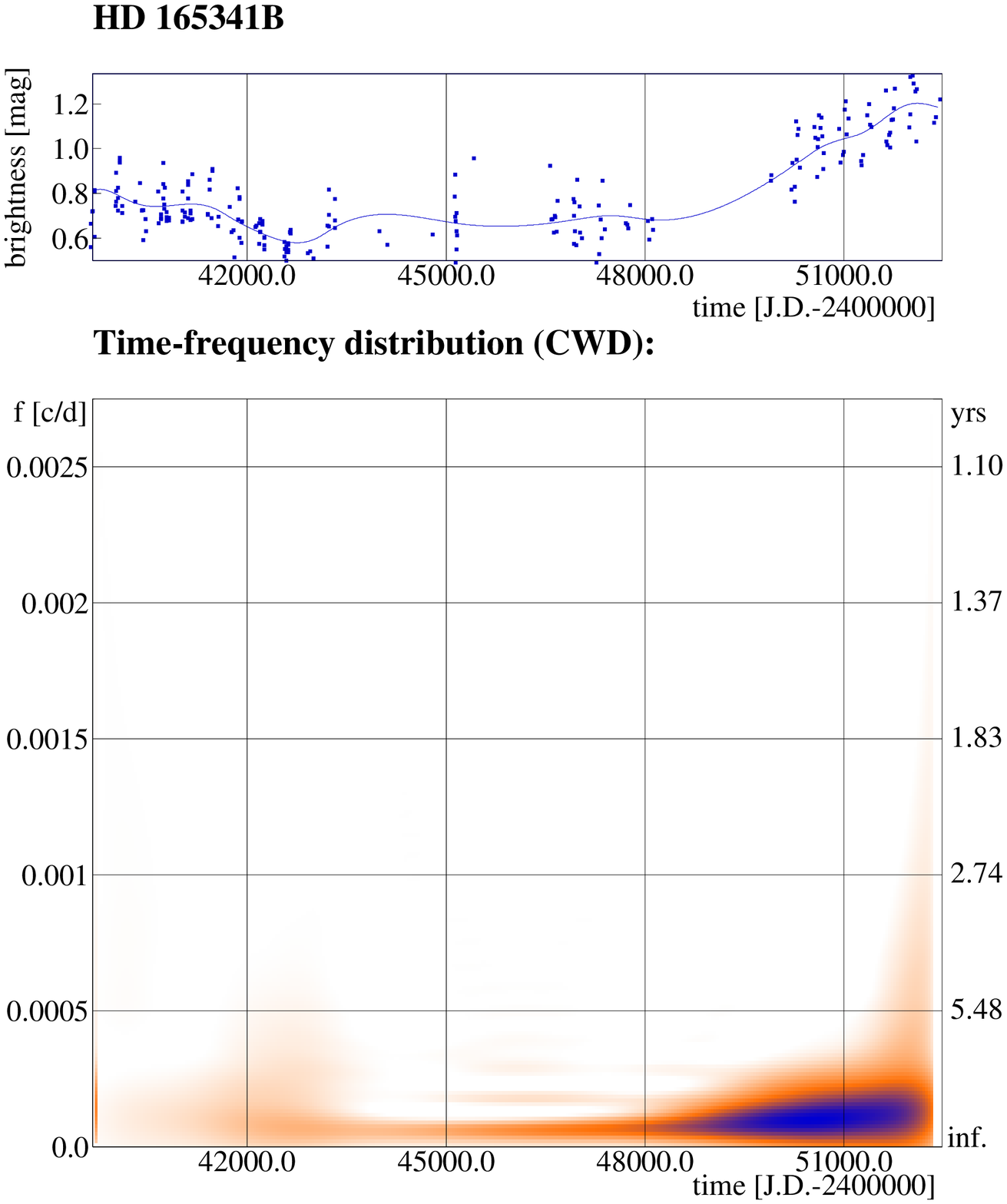}\includegraphics[width=4.3cm]{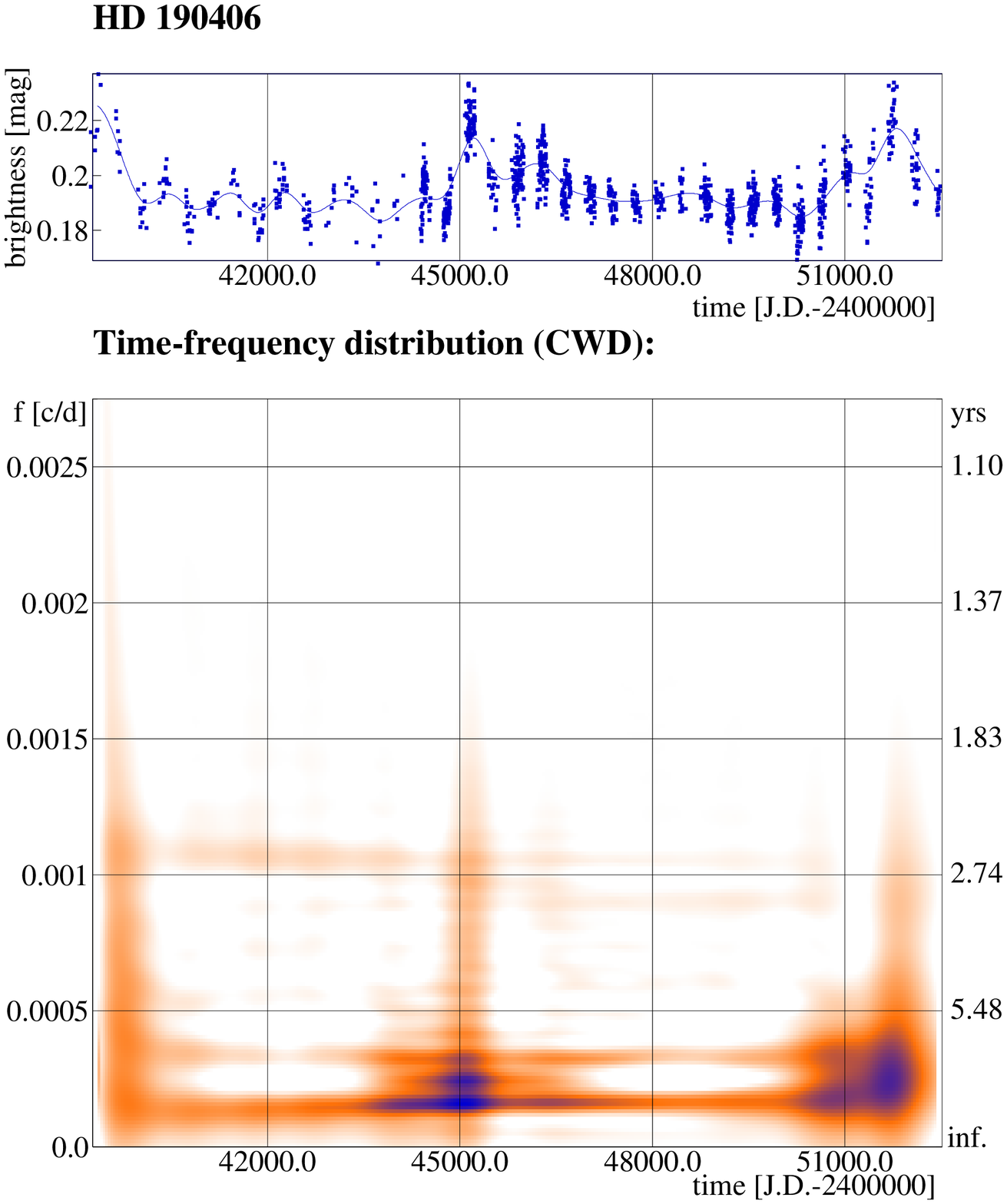}

  \hspace*{-12.8cm}\includegraphics[width=4.3cm]{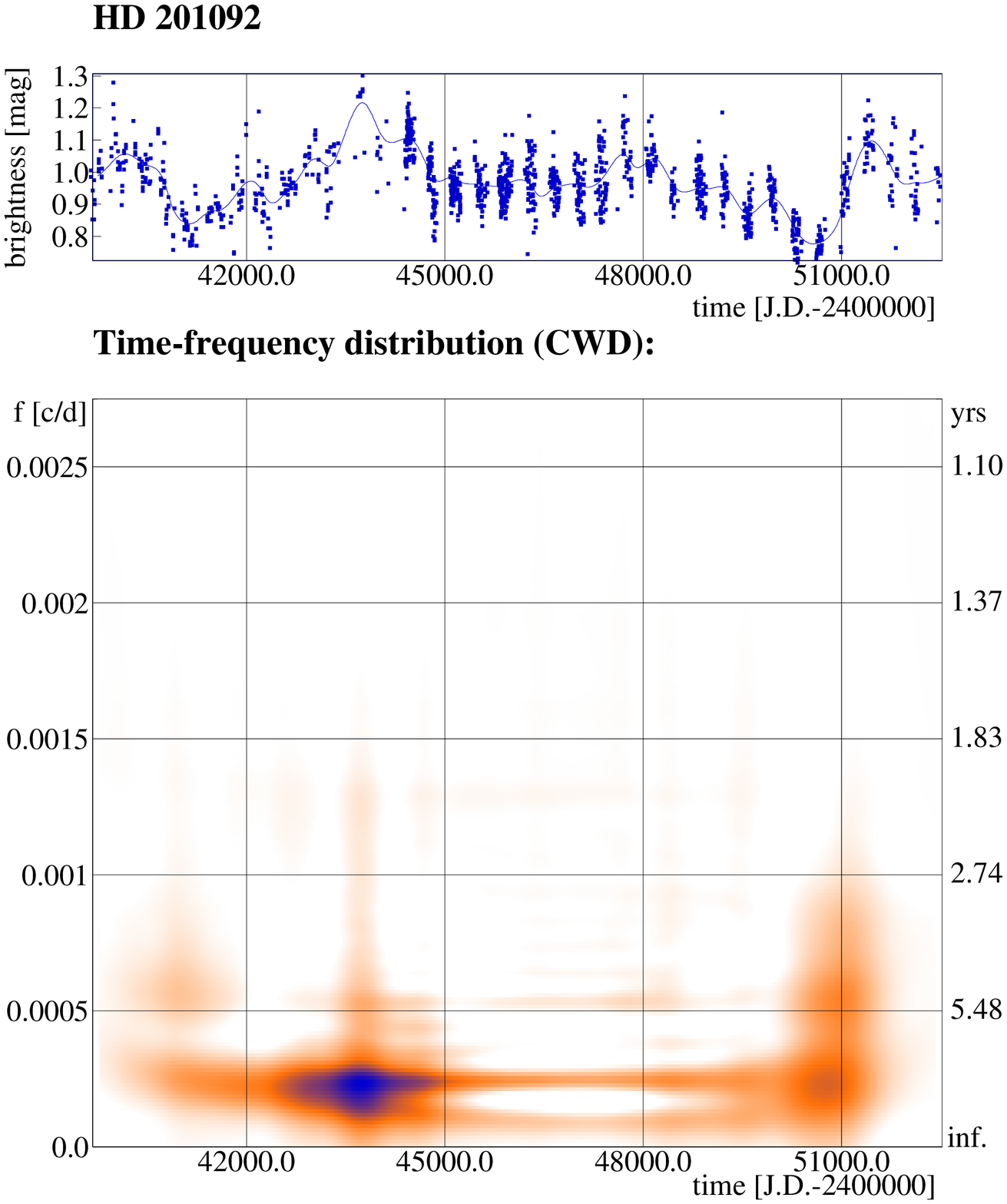}
      \caption{CWD of the ``young" MW stars with complex cycles. In the upper panels the datasets and the used spline interpolation is plotted, below the time-frequency diagrams are seen.}
    \label{cwd_compl}
    \end{figure*}  
    
           \begin{figure}[tbp]
   \centering
   \includegraphics[width=8cm]{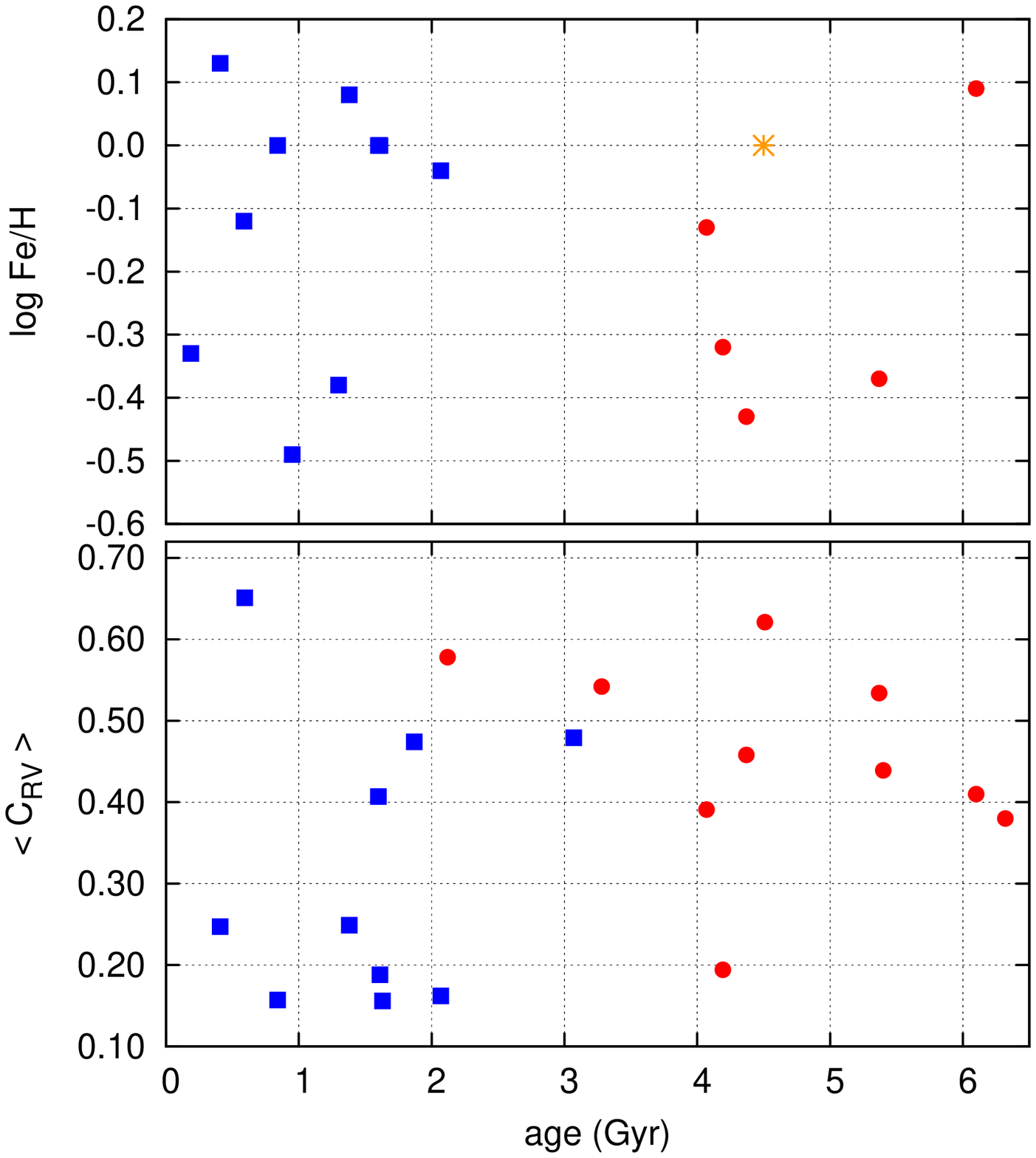}
   \caption{Metallicity vs. age of the MW stars. \emph {Top:} Log (Fe/H) from Gray et al. (\cite{gray1}, \cite{gray2}), {\emph {Bottom:} $< C_{RV} >$} from Soon et al. (\cite{soon2}).}
         \label{metal}
         \end{figure}

\end{appendix}         
\end{document}